\documentclass[
aps,
prc,
superscriptaddress,
floatfix,
nofootinbib,
twocolumn
]{revtex4-2}

\usepackage{amsmath,amssymb,amsfonts,physics,graphicx,float}
\graphicspath{{pic/}}
\usepackage[setpagesize=false]{hyperref}
\allowdisplaybreaks[4]

\usepackage[labelformat=simple]{subcaption} 

\usepackage{ulem}
\usepackage{color}
\definecolor{ar}{rgb}{1.0, 0.01, 0.24}
\definecolor{al}{rgb}{0.82, 0.1, 0.26}
\definecolor{ev}{rgb}{0.56, 0.0, 1.0}

\newcommand{\blueflag}[1]{{\color{blue} #1}}

\newcommand{\lag}{\mathcal{L}}
\newcommand{\U}[1]{\mathrm{U}(#1)}
\newcommand{\UA}[1]{\mathrm{U}(#1)_\mathrm{A}}
\newcommand{\SU}[1]{\mathrm{SU}(#1)}
\newcommand{\diag}[1]{\mathrm{diag}(#1)}
\newcommand{\rl}{\mathrm{L}}
\newcommand{\rr}{\mathrm{R}}
\newcommand{\mir}{\mathrm{mir}}

\newcommand{\GO}{Gell-Mann--Okubo mass relation}

\begin{document}

\title{
Parity doublet model for baryon octets: \\
diquark classifications and mass hierarchy based on the quark-line diagram 
}

\author{Takuya Minamikawa}
\email{minamikawa@hken.phys.nagoya-u.ac.jp}
\affiliation{Department of Physics, Nagoya University, Nagoya 464-8602, Japan}

\author{Bikai Gao}
\email{gaobikai@hken.phys.nagoya-u.ac.jp}
\affiliation{Department of Physics, Nagoya University, Nagoya 464-8602, Japan}

\author{Toru Kojo}
\email{torujj@nucl.phys.tohoku.ac.jp}
\affiliation{Department of Physics, Tohoku University, Sendai 980-8578, Japan}

\author{Masayasu Harada}
\email{harada@hken.phys.nagoya-u.ac.jp}
\affiliation{Department of Physics, Nagoya University, Nagoya 464-8602, Japan}
\affiliation{Advanced Science Research Center, Japan Atomic Energy Agency, Tokai 319-1195, Japan}
\affiliation{Kobayashi-Maskawa Institute for the Origin of Particles and the Universe, Nagoya University, Nagoya, 464-8602, Japan}

\date{\today}

\begin{abstract}
We construct $ {\rm SU(3)}_{\rm L} \otimes {\rm SU(3)}_{\rm R}$ invariant parity doublet models within the linear realization of the chiral symmetry.
Describing baryons as the superposition of linear representations should be useful description for transitions toward the chiral restoration.
The major problem in the construction is that there are much more chiral representations for baryons than in the two-flavor cases.
To reduce the number of possible baryon fields, 
we introduce a hierarchy between representations with good or bad diquarks (called soft and hard baryon representations, respectively).
We use $(3,\bar3)+(\bar3,3)$ and $(8,1)+(1,8)$ as soft to construct a chiral invariant Lagrangian, while the $(3,6)+(6,3)$ representations are assumed to be integrated out,
leaving some effective interactions.
The mass splitting associated with the strange quark mass is analyzed in the first and second order in the meson fields $M$ in $(3,\bar3)+(\bar3,3)$ representations.
We found that the chiral $ {\rm SU(3)}_L \otimes {\rm SU(3)}_R$ constraints are far more restrictive than the $ {\rm SU(3)}_V$ constraints used in conventional models for baryons. 
After extensive analyses within $(3,\bar3)+(\bar3,3)$ and $(8,1)+(1,8)$ models, we found that models in the first order of $M$ do not reproduce the mass hierarchy correctly,
although the {\GO} is satisfied.
In the second order, the masses of the positive parity channels are reproduced well up to the first radial excitations, 
while some problem in the mass ordering remains in a negative parity channel.
Apparently the baryon dynamics is not well-saturated by just $(3,\bar3)+(\bar3,3)$ and $(8,1)+(1,8)$ representations,
as indicated by the necessity of terms higher order in $M$.

\end{abstract}

\maketitle


\section{Introduction}

\begin{figure*}
\centering
\includegraphics[width=0.7\hsize]{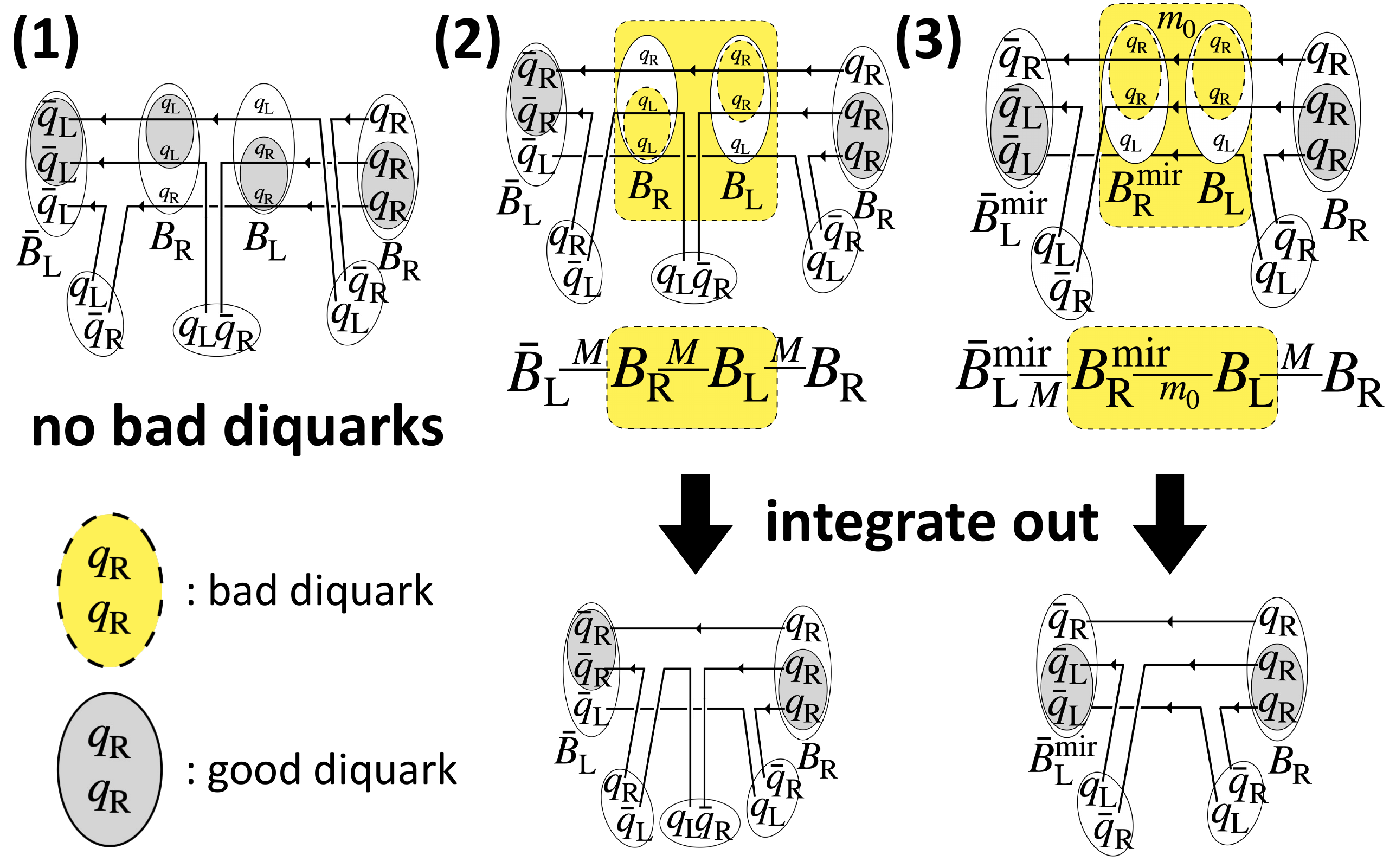}
\caption{
Higher-order quark exchange diagrams. 
(1) is just combination of the first-order interaction without the bad diquarks. 
(2) has three quark exchanges (three meson fields) through the bad diquarks, while 
(3) has two quark exchanges (two meson fields). 
In (3), there is 
a mixing between naive and mirror representations. }
\label{fig-iterations}
\end{figure*}

Chiral symmetry in quantum chromodynamics (QCD) is 
the key symmetry to describe the low energy hadron dynamics. 
Although the chiral symmetry is spontaneously broken by the formation of chiral condensates \cite{Nambu:1961tp,Nambu:1961fr,Hatsuda:1994pi},
the chiral symmetry in the underlying theory leaves a number of constraints on the low energy dynamics \cite{Weinberg:1966kf,Weinberg:1969hw,Weinberg:1990xn,Weinberg:2010bq}.
Effective Lagrangians for hadrons are constructed by grouping a set of fields in a chiral invariant way,
modulo small explicit breaking associated with the current quark masses.

The most general construction of chiral Lagrangian is based on the nonlinear realization of the chiral symmetry \cite{Coleman:1969sm,Callan:1969sn}
in which fields transform nonlinearly under chiral transformations.
The great advantage of this construction is that pions accompany space-time derivatives appearing in powers of $\sim \partial / \Lambda_\chi$, 
where $\Lambda_\chi$ is the typical chiral symmetry breaking scale
related to the pion decay constant $f_\pi$ as $\Lambda_\chi \sim 4 \pi f_\pi$~\cite{Manohar:1983md}, 
which leads to the low-energy constants of the chiral perturbation theory to be $\sim {\cal O}(10^{-3})$~\cite{Gasser:1983yg,Gasser:1984gg}.
This power counting greatly systematizes the construction of effective Lagrangians.

While the nonlinear realization has the advantage in generality and systematics, 
it also has the disadvantage when we try to address the physics at energies 
near or greater than $\sim \Lambda_\chi$.  
One simple way of improving the description is to manifestly include massive degrees of freedom.
The problem also occurs when we consider the chiral restoration in extreme environments;
there, the denominators of the derivatives, 
$\sim \Lambda_\chi$, become small,
invalidating the derivative expansion
with only pions.

A model of linear realization is less general but more suitable when we describe QCD in extreme environments (e.g., neutron stars \cite{Baym:2017whm}) with partial restoration of chiral symmetry.
Near chiral restored region the hadron spectra should recover chiral multiplets, e.g., $(\sigma, \vec{\pi})$.
Implementing candidates of chiral multiplets from the beginning should simplify our descriptions;
we do not have to dynamically generate relevant degrees of freedom.

In this work we consider a model of baryons in linear realization of chiral symmetry, aiming at its application to dense QCD.
We include the parity doublet structure which allows us to introduce 
the chiral invariant mass \cite{Detar:1988kn,Jido:1998av,Jido:1999hd,Jido:2001nt,Nagata:2008xf,Gallas:2009qp,Gallas:2013ipa}.
For increasing baryon densities, the existence of such mass has large impacts on the density dependence of baryon masses as well as baryon-meson couplings.
Previously we have analyzed models of two-flavors~\cite{Minamikawa:2020jfj,Minamikawa:2021fln,Gao:2022klm,Minamikawa:2023eky}, 
but in this work we extend the model to the three-flavor case.
This is necessary to analyze dense baryonic matter with hyperons.

The extension from two-flavors to three-flavors, however, 
drastically complicates the construction of chiral Lagrangian for baryons
since there are so many possible representations.
Combining three quarks in linear chiral representations, one can create several representations for baryons.
For two-flavors, we start with quarks in $(2_L,1_R)$ and $(1_L, 2_R)$, then
the three products yield $(2_L, 1_R)$, $(4_L, 1_R)$, $(3_L, 2_R)$ and $L\leftrightarrow R$.
When we include only nucleons, we may focus on $(2_L, 1_R)$ and $(1_L, 2_R)$,
and the number of fields is managable.
For three flavors, we start with quarks in $(3_L,1_R)$ and $(1_L, 3_R)$, and find much more representations for their products.
Although there are several studies of baryons based on the models including possible chiral representations of baryons~\cite{Chen:2009sf,Chen:2010ba,Chen:2011rh,Schramm:2015hba,Nishihara:2015fka,Dmitrasinovic:2016hup,Sasaki:2017glk,Shivam:2019cmw,Motornenko:2020vqm},
to the best of our knowledge, for three-flavors,
the construction of linearly realized chiral Lagrangian for baryons has not been established.

In order to keep the number of representations tractable,
in this work we introduce dynamical assumptions based on the quark dynamics.
We assume that baryons in representations including ``good diquarks'', the representations $(\bar{3}_L, 1_R)$ or $(1_L, \bar{3}_R)$, 
are lighter than those including ``bad diquarks'',  the representations $(6_L, 1_R)$ or $(1_L, 6_R)$ \cite{Jaffe:1976ig,Jaffe:1976ih,Rapp:1997zu,Jaffe:2003sg}.
In this paper we call baryon representations including good diquarks {\it soft baryons}, and those with bad diquark {\it hard baryons}.
In this paper the hard baryons are integrated out and do not manifestly appear,
but the consequence of such integration can be traced in effective vertices including high powers of $M$ (Fig. \ref{fig-iterations}).

Based on this idea we build the chiral Lagrangian for soft baryon fields in $(3_L, \bar{3}_R)$, $(8_L, 1_R)$ with $L\leftrightarrow R$.
We include mesonic fields and the parity doublet structure in a chiral invariant way. 
Both the spectra of positive and negative baryons as well as the first radial excitations are studied.
As usual in the linear realization, we do not have good rationals to restrict the power of mesonic fields,
so we examine how important higher order terms are.

The remarkable and unexpected finding in our construction is that
the chiral symmetry and the above dynamic assumption give very strong impacts on baryon masses, especially the SU(3) flavor breaking due to the strange quark mass.
For example, for models including only $(3_L, \bar{3}_R)$ and $(\bar{3}_L, 3_R)$, the usual baryon mass ordering based on the number of strange quarks
does not hold, at least at the order of meson fields we have worked on.
We then add $(8_L, 1_R)$ and $(1_L, 8_R)$ representations, finding them to be insufficient.
To improve the description of spectra,
we are forced to increase powers of mesonic fields up to the second order of Yukawa interactions.

We try to reproduce the ground and first radially excited states for positive and negative baryon octets.
Our modeling works for positive parity baryons,
but for negative baryons, some of mass ordering related to the strange quark appears to be inconsistent 
with the picture based on the constituent quark models \cite{DeRujula:1975qlm}.
This situation persists even after our extensive survey for parameter space.

 Some comments are in order for comparison with the previous studies.
 The textbook example of the octet mass formula \cite{georgi1982lie}       
 is based on the SU(3) symmetry with the explicit breaking as perturbation,
 but  the underlying Hamiltonian does not have the chiral symmetry.
There are some previous studies for the parity doublet model including 
hyperons~\cite{Chen:2009sf,Chen:2010ba,Chen:2011rh,Schramm:2015hba,Nishihara:2015fka,Dmitrasinovic:2016hup,Sasaki:2017glk,Shivam:2019cmw,Motornenko:2020vqm}.
For example, in Ref.\cite{Dmitrasinovic:2016hup}, current quark masses are incorporated into a parity doublet model based on the SU(3)$_{\rm L} \times$SU(3)$_{\rm R}$ chiral symmetry, 
and the pion-nucleon $\Sigma_{\pi N}$ and kaon-nucleon $\Sigma_{KN}$ terms are studied.
In Ref.~\cite{Sasaki:2017glk}, explicit breaking is effectively introduced into the masses without explicit forms of the Lagrangian terms to study the difference of behavior in hot matter.
However, to the best of our knowledge, there is no analysis of mass spectra of baryons including hyperons 
in a chiral invariant model.

This paper is structured as follows.
In Sec.\ref{sec-representation}, the chiral representations of 
$(3_L,\bar{3}_R)+(\bar{3}_L,3_R)$ and $(8_L,1_R)+(1_L,8_R)$ 
for octet baryons are defined. 
In Sec.\ref{sec-case}, we study a Lagrangian up to the first order of Yukawa interactions,
and found that the mass hierarchy of the baryon octet cannot be reproduced.  
In Sec.\ref{sec-QD-CYI}, we classify hadronic effective interactions based on quark diagrams. 
Then, in Sec.\ref{sec-2nd}, we construct the second-order Yukawa-type interactions which should be induced by integrating out hard baryons.
In Sec.\ref{sec-numerical}, we perform numerical fit of baryon spectra.
Sec.\ref{sec-summary} is devoted to the summary. 




%
\begin{figure*}[t]\centering
\begin{subfigure}{0.3\hsize}\centering
	\includegraphics[width=0.8\hsize]{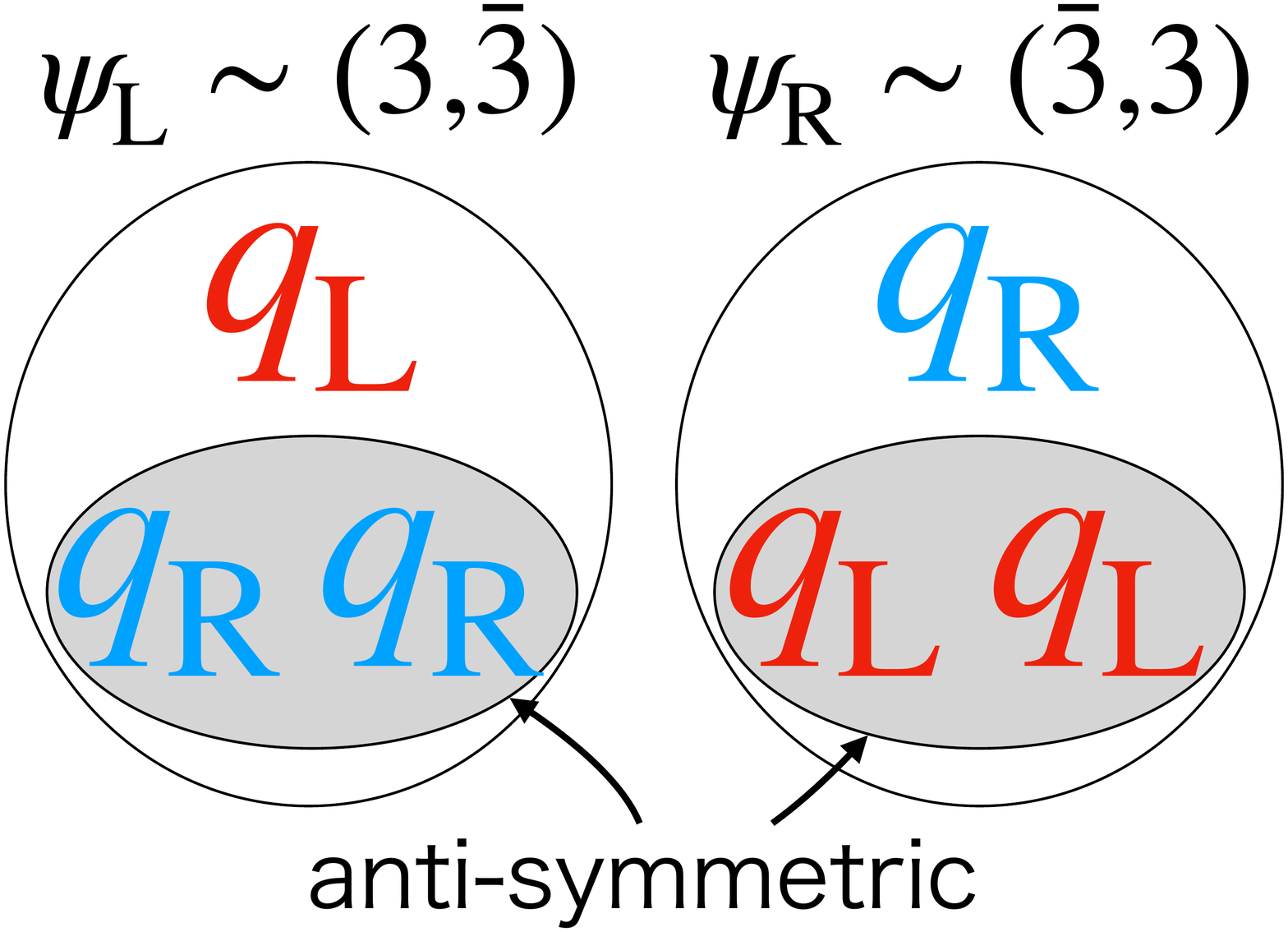}
	\caption{}
	\label{fig-baryon-psi}
\end{subfigure}
\begin{subfigure}{0.3\hsize}\centering
	\includegraphics[width=0.8\hsize]{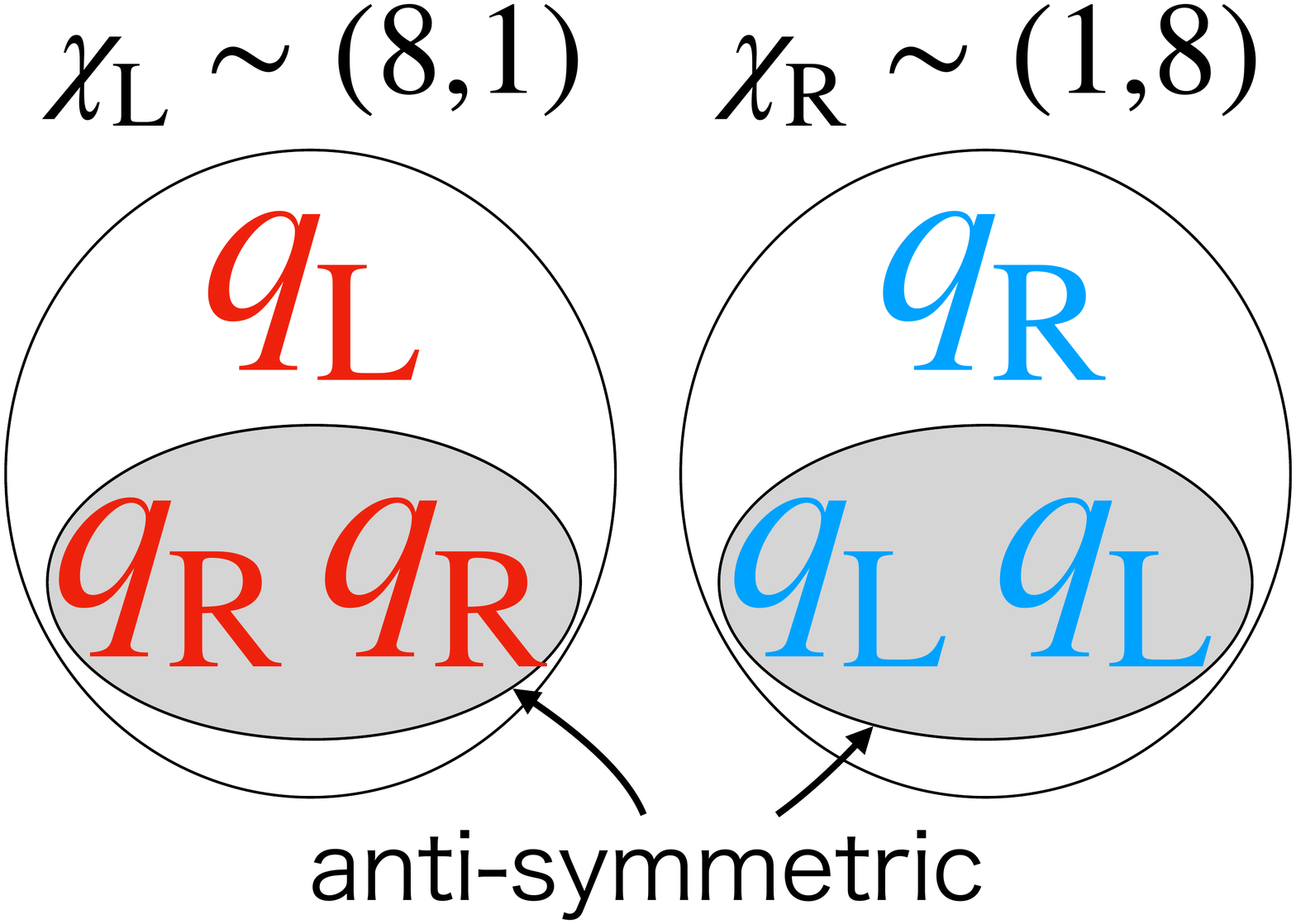}
	\caption{}
	\label{fig-baryon-chi}
\end{subfigure}
\begin{subfigure}{0.3\hsize}\centering
	\includegraphics[width=0.8\hsize]{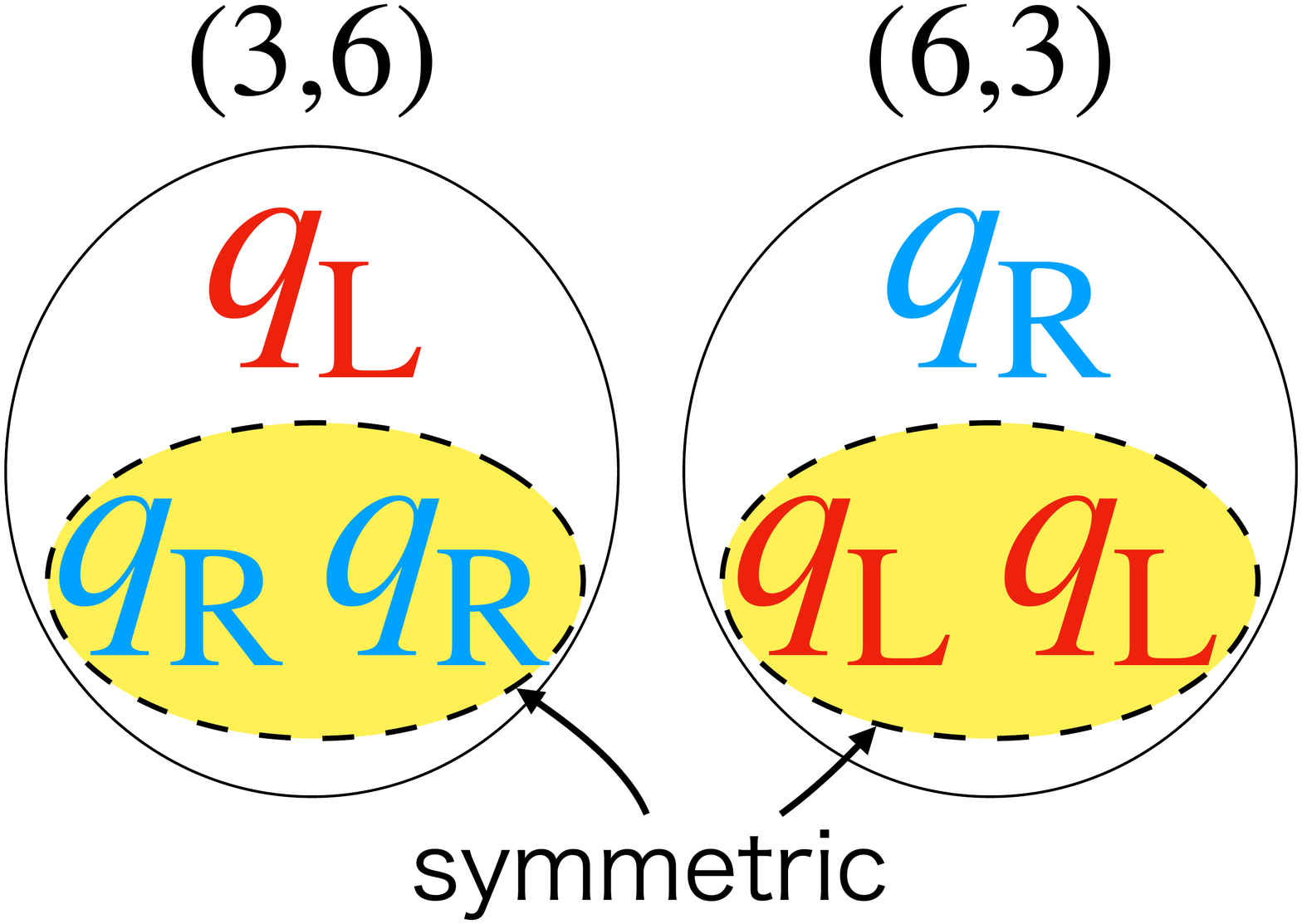}
	\caption{}
	\label{fig-baryon-eta}
\end{subfigure}
\caption[]{
Quark contents for each baryon representations: 
(a) $(3,\bar3)+(\bar3,3)$, 
(b) $(8,1)+(1,8)$, 
and 
(c) $(3,6)+(6,3)$. 
The gray shaded diquark indicates flavor antisymmetric representation, 
while the yellow indicates symmetric one. 
}
\label{fig-baryon}
\end{figure*}
%


\section{Chiral Representation for Hadron}\label{sec-representation}

In three-flavor chiral symmetry $\SU{3}_\rl\times\SU{3}_\rr$, 
quark fields are defined as the fundamental representations, 
namely left-handed $(q_\rl)^l\sim(3,1)$ and right-handed $(q_\rr)^r\sim(1,3)$, 
with upper indices $l,r=1,2,3=u,d,s$. 
The antiquark fields are defined as the dual representations 
$(\bar{q}_\rl)_l\sim(\bar3,1)$ and $(\bar{q}_\rr)_r\sim(1,\bar3)$ 
with lower indices $l,r$. 
The scalar meson field is defined as 
$(M)^l_r\sim(q_\rl)^l\otimes(\bar{q}_\rr)_r\sim(3,\bar3)$ in this paper. 

Since baryons consist of three valence quarks, 
the baryon fields are related with the tensor products of three quark fields. 
We define the left-handed baryon field 
as a product of a spectator left-handed quark 
and left- or right-handed diquark, 
while the right-handed baryon has a right-handed spectator quark. 
Taking irreducible decomposition, the left-handed baryon 
can be expressed as the following representations 
\begin{gather}
q_\rl\otimes(q_\rl\otimes q_\rl+q_\rr\otimes q_\rr)\sim\notag\\
(1,1)+(8,1)+(8,1)+(10,1)
+(3,\bar3)+(3,6)\,. 
\end{gather}
%
%
The octet baryons are included in 
$(3,\bar3)$, $(8,1)$, and $(3,6)$, 
which can be illustrated as in Fig.\ref{fig-baryon}. 
The representations $(3,\bar3)$ and $(8,1)$ contain flavor-antisymmetric diquarks $\sim\bar3$ which is called ``good'' diquark,  
while $(3,6)$ contains flavor-symmetric diquark $\sim6$ called ``bad'' diquark. 
We call baryon representations including good diquarks {\it soft baryons}, and those with bad diquark {\it hard baryons}.
In this paper the hard baryons are integrated out and do not manifestly appear,
but the consequence of such integration can be traced in effective vertices including high powers of $M$.
The detailed discussions are given in Sec.\ref{sec-2nd}.

The baryon fields denoted as $\psi$ and $\chi$ are related with the quark fields as follows: 
For example, the left-handed baryons have the relations,  
\begin{align}
(\psi_{\rl})^{l[r_1r_2]}&\sim
(q_\rl)^l\otimes(q_\rr)^{[r_1}\otimes(q_\rr)^{r_2]}\,, \label{eq-psi-qqq}\\
(\chi_{\rl})^{l_1[l_2l_3]}&\sim
(q_\rl)^{l_1}\otimes(q_\rl)^{[l_2}\otimes(q_\rl)^{l_3]}\,, \label{eq-chi-qqq}
\end{align}
where $[\,\cdot\,]$ implies that two indices in the bracket are antisymmetrized. 
The relations can be rewritten as 
\begin{align}
(\psi_{\rl})^l_r&\sim
\varepsilon_{rr_1r_2}(q_\rl)^l\otimes(q_\rr)^{r_1}\otimes(q_\rr)^{r_2}\,, \label{eq-psi-qd}\\
(\chi_{\rl})^l_{l'}&\sim
\varepsilon_{l'l_1l_2}(q_\rl)^l\otimes(q_\rl)^{l_1}\otimes(q_\rl)^{l_2}\,, \label{eq-chi-qd}
\end{align}
where $\varepsilon_{ijk}$ is the totally asymmetric tensor. 
For these baryon fields, upper indices are interpreted as the ones of quarks, 
and lower indices are as the ones of good diquarks. 
For example, 
$(\psi_\rl)^{l[r_1r_2]}$ consists of a left-handed quark with upper index $l$ and 
two antisymmetrized right-handed quarks with upper indices $r_1$ and $r_2$, 
while $(\psi_\rl)^l_r$ consists of a left-handed quark with upper index $l$ and 
a scalar right-handed diquark ($\bar3$ representation) with lower index $r$. 
The baryon fields with three indices and the ones with two indices are equivalent 
through the following relations, 
\begin{align}
(\psi_{\rl})^l_r&=
\frac12\varepsilon_{rr_1r_2}(\psi_{\rl})^{l[r_1r_2]}\,, \label{eq-psi-qqq-qd-1}\\
(\psi_{\rl})^{l[r_1r_2]}&=\varepsilon^{rr_1r_2}(\psi_{\rl})^l_r\,, \label{eq-psi-qqq-qd-2}
\end{align}
and there are also the same relations for $\chi$. 
We call the baryon fields with three indices ``three-index notation'' 
(Eqs.\eqref{eq-psi-qqq}-\eqref{eq-chi-qqq}), 
and the ones with two indices ``two-index notation'' 
(Eqs.\eqref{eq-psi-qd}-\eqref{eq-chi-qd}) in this paper. 

The two-index notation is often used to calculate as usual, 
because it is directly related with the adjoint representation matrices as 
\begin{align}
(\chi)^i_j&\sim
\begin{bmatrix}
\frac1{\sqrt2}\Sigma^0+\frac1{\sqrt6}\Lambda & \Sigma^+ & p \\
\Sigma^- & -\frac1{\sqrt2}\Sigma^0+\frac1{\sqrt6}\Lambda & n \\
\Xi^- & \Xi^0 & -\frac2{\sqrt6}\Lambda \\
\end{bmatrix}\,,\\
(\psi)^i_j&\sim
\frac1{\sqrt3}\Lambda_0 \notag\\
&+\begin{bmatrix}
\frac1{\sqrt2}\Sigma^0+\frac1{\sqrt6}\Lambda & \Sigma^+ & p \\
\Sigma^- & -\frac1{\sqrt2}\Sigma^0+\frac1{\sqrt6}\Lambda & n \\
\Xi^- & \Xi^0 & -\frac2{\sqrt6}\Lambda \\
\end{bmatrix}\,, 
\end{align}
for left-handed and right-handed respectively. 
To distinguish $\psi$ and $\chi$ explicitly, 
we treat a flavor of a baryon as an index of $\psi$ or $\chi$ fields, e.g.,
$\psi^1_3=\psi_p$, $\chi^1_3=\chi_p$, $\psi^3_2=\psi_{\Xi^0}$, 
$\psi^3_3=\psi_{\Lambda_0}/\sqrt3-2\psi_{\Lambda}/\sqrt6$, and so on. 
For simplicity, we define the isospin vectors as 
\begin{align}
\psi_N&\equiv(\psi_p,\psi_n)\\
\psi_\Sigma&\equiv(\psi_{\Sigma^-},\psi_{\Sigma^0},\psi_{\Sigma^+})\\
\psi_\Xi&\equiv(\psi_{\Xi^-},\psi_{\Xi^0})\,. 
\end{align}
We also define the same notations for $\chi$. 

In the three-index notation, 
it is easy to distinguish the antiquark ($\bar3$) and the diquark (also $\bar3$), 
because it has a one-to-one correspondence between 
the indices of the baryon field and the ones of the quark fields, 
as in Eqs.\eqref{eq-psi-qqq}-\eqref{eq-chi-qqq}. 
In addition, we can easily see the charge of U(1)$_{\rm A}$ symmetry in the three-index notation. 
For example, if one wants to make a contraction 
of the baryon field $\psi_\rr\sim(\bar3,3)$ and the meson field $M\sim(3,\bar3)$ as 
\begin{align}
(\psi_\rr)^r_l (M)^l_r = \tr(\psi_\rr M)\,, 
\end{align}
which is invariant under SU(3)$_L\times$SU(3)$_R$ but not invariant under U(1)$_A$, 
because there are three left-handed quarks but there are no left-handed antiquarks.
The transformation property is the same as
\begin{align}
(\psi_\rr)^r_l (M)^l_r \sim 
(q_\rr)^r \varepsilon_{ll_1l_2} (q_\rl)^{l_1} (q_\rl)^{l_2} (q_\rl)^l (\bar{q}_\rr)_r\,,
\end{align}
where the left and right components are SU(3) singlet but the left handed one has a finite U(1)$_{\rm A}$ charge.
We emphasize that such term actually appears 
in the form of a specific combination with other terms 
(Eq.\eqref{eq-2nd-sd-2index} in Sec.\ref{sec:psi_psi}).
This property also leads to a correspondence between 
the quark diagrams and the hadronic effective interactions, 
as will be explained in Sec.\ref{sec-QD-CYI}. 

Next, let us define the parity doubling partners $\psi^\mir$ and $\chi^\mir$ for $\psi$ and $\chi$ respectively 
as the opposite assigns for the chiral representations (``mirror'' assignment) as 
\begin{align}
(\psi_{\rl})^l_r\sim(3,\bar3)\ , \quad&\quad(\psi_{\rl}^\mir)^r_l\sim(\bar3,3)\label{eq-mir-33}\ , \\
(\chi_{\rl})^{l_1}_{l_2}\sim(8,1)\ , \quad&\quad(\chi_{\rl}^\mir)^{r_1}_{r_2}\sim(1,8)\, , \label{eq-mir-18}
\end{align}
and these fields have opposite parity respectively as, 
\begin{align}
\psi\xrightarrow{\text{parity}}+\gamma^0\psi\ , \quad&\quad
\psi^\mir\xrightarrow{\text{parity}}-\gamma^0\psi^\mir \ ,\\
\chi\xrightarrow{\text{parity}}+\gamma^0\chi\ , \quad&\quad
\chi^\mir\xrightarrow{\text{parity}}-\gamma^0\chi^\mir\ . 
\end{align}
The right-handed ones are also defined in the similar way. 

Note that, the mirror assigned fields can be interpreted as 
spatially excited states in the following sense. 
One choice of the interpolating fields for $\psi_\rl$ is 
$q_\rl(q_\rr^TCq_\rr)=P_\rl qd_\rr$, 
where $C=i\gamma^2\gamma^0$ is the charge conjugate matrix, 
$P_\rl=(1-\gamma^5)/2$ is the chiral projection matrix, and 
$d_\rr=q_\rr^TCq_\rr$ is the scalar diquark field. 
One possible choice of interpolating field for its excited state is 
$(\gamma_\mu\partial^\mu q_\rr)(q_\rl^TCq_\rl) =P_\rl(\gamma_\mu\partial^\mu q)d_\rl$, 
which has the same chirality but the opposite chiral representation.

Introducing the mirror fields $\psi^\mir$ and $\chi^\mir$, 
there are mixing terms between the naive and the mirror fields as 
\begin{align}
\lag^\mathrm{CIM}=
&- m_0^\psi \big( \bar{\psi}_{\rm L} \gamma_5 \psi^\mir_{\rm R} + \bar{\psi}_{\rm R} \gamma_5 \psi^\mir_{\rm L}  \big) +  \mathrm{h.c.} \notag \\
&- m_0^\chi \big( \bar{\chi}_{\rm L} \gamma_5 \chi^\mir_{\rm R} + \bar{\chi}_{\rm R} \gamma_5 \chi^\mir_{\rm L} \big) +  \mathrm{h.c.}\,, 
\end{align}
where the transformation properties, e.g., 
$\psi_{\rm L} \rightarrow U_{\rm L} \psi_{\rm L}$ and $\psi^\mir_{\rm R} \rightarrow U_{\rm L} \psi^\mir_{\rm R}$, with $U_{\rm L,R} \in $SU(3)$_{\rm L,R}$,
make the above mass terms chiral invariant.
%
This parameters $m_0^{\psi,\chi}$ corresponds to he chiral invariant mass, 
since this mixing terms are chiral symmetric.


\section{Models with first-order Yukawa interactions}
\label{sec-case}

In this section, 
we study the mass hierarchy of octet baryons in models with first-order Yukawa interactions.
First, in Sec.\ref{sec-GO}, we review the {\GO} for octet baryons, 
which is derived from the flavor symmetry. 
Next, in Sec.\ref{sec-1st-1}-\ref{sec-1st-2}, 
we deal with models based on the chiral U(3)$_{\rm L}\times$U(3)$_{\rm R}$ symmetry which include $\psi$s and $\chi$s given in the previous section with first-order Yukawa interactions, 
and see that these models have some problems to satisfy the {\GO}. 
Consequently, we will see that the minimal chiral model for octet baryons
must include second-order Yukawa interactions.

\subsection{Review of {\GO}}\label{sec-GO}

A basic model with flavor $\SU{3}$ symmetry is 
\begin{align}\label{eq-GO}
\lag^\mathrm{V}=
-a\tr\bar{B}B
-b\tr\bar{B}MB
-c\tr\bar{B}BM\,, 
\end{align}
where $B$ is a $3\times3$ matrix of the octet baryon fields,
and $a$, $b$, $c$ are real parameters.
We emphasize that this Lagrangian, commonly appearing in textbooks of group theories,
has the SU(3) flavor symmetry but does not possess the SU(3)$_{\rm L} \times$ SU(3)$_{\rm R}$ symmetry.

Assuming that the meson field $M$ has 
a vacuum expectation value (VEV) $\ev{M}=\diag{\alpha,\beta,\gamma}$ 
and isospin symmetry $\alpha=\beta$, 
the masses of octet baryons are obtained as
\begin{align}
m_N&=a+b\alpha+c\gamma \ , \\
m_\Sigma&=a+b\alpha+c\alpha\ , \\
m_\Xi&=a+b\gamma+c\alpha\ , \\
m_\Lambda&=a+b\frac{\alpha+2\gamma}{3}+c\frac{\alpha+2\gamma}{3}\ . 
\end{align}
Erasing the parameters $a$, $b$ and $c$, 
we have the following relation: 
\begin{align}
\frac{m_N+m_\Xi}{2}=\frac{3m_\Lambda+m_\Sigma}{4}\,, 
\end{align}
which is called the {\GO} for octet baryons. 

In a naive quark mass counting,
the {\GO} is satisfied by assuming 
$M_u \simeq M_d$, $m_N \sim 3M_u$, $m_\Xi \sim M_u + 2M_s$, $m_\Lambda \sim 2M_u + M_s$, and  $m_\Sigma \sim 2M_u + M_s$, 
where $M_q$ ($q=u,d,s$) are the constituent quark masses. 
These estimates hold for typical constituent quark models. 
On the other hand, these quark counting is sufficient but not necessary conditions;
the {\GO} is a weaker condition than that deduced from the quark counting.

\subsection{Model 1: only $(3,\bar3)+(\bar3,3)$}\label{sec-1st-1}

First we consider a model including only the $(3,\bar3)+(\bar3,3)$ representations for octet baryons, $\psi$, 
and the $(3,\bar3)$ representation of mesons, $M$, which generates the chiral variant mass of baryons through the spontaneous chiral symmetry breaking.
The chiral invariant term for Yukawa interactions at the first order in $M$ is
\begin{align}
\lag^\text{model(1)} 
=-g\big[&\varepsilon^{l_1l_2l_3}\varepsilon_{r_1r_2r_3}
(\bar\psi_\rl)^{r_1}_{l_1}(M^\dag)^{r_2}_{l_2}(\psi_\rr)^{r_3}_{l_3}\notag\\
+&\varepsilon_{l_1l_2l_3}\varepsilon^{r_1r_2r_3}
(\bar\psi_\rr)^{l_1}_{r_1}(M)^{l_2}_{r_2}(\psi_\rl)^{l_3}_{r_3}\big]\label{eq-case1}\,, 
\end{align}
where $\varepsilon_{ijk}$ is the totally asymmetric tensor. 
%

\begin{figure*}\centering
\begin{subfigure}{0.3\hsize}\centering
	\includegraphics[width=0.8\hsize]{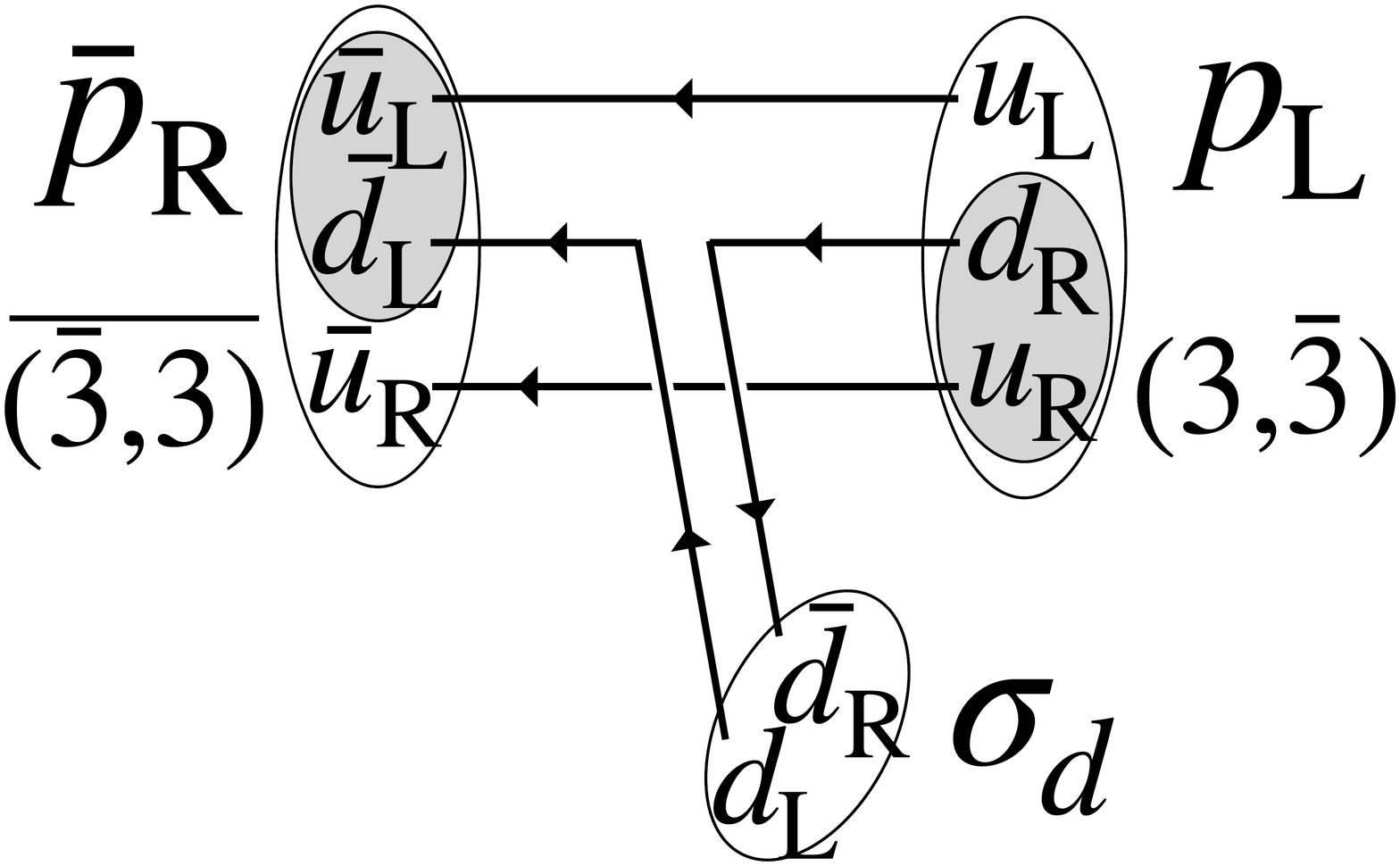}
	\caption{}
	\label{fig-quarklines-det-proton}
\end{subfigure}
\begin{subfigure}{0.3\hsize}\centering
	\includegraphics[width=0.8\hsize]{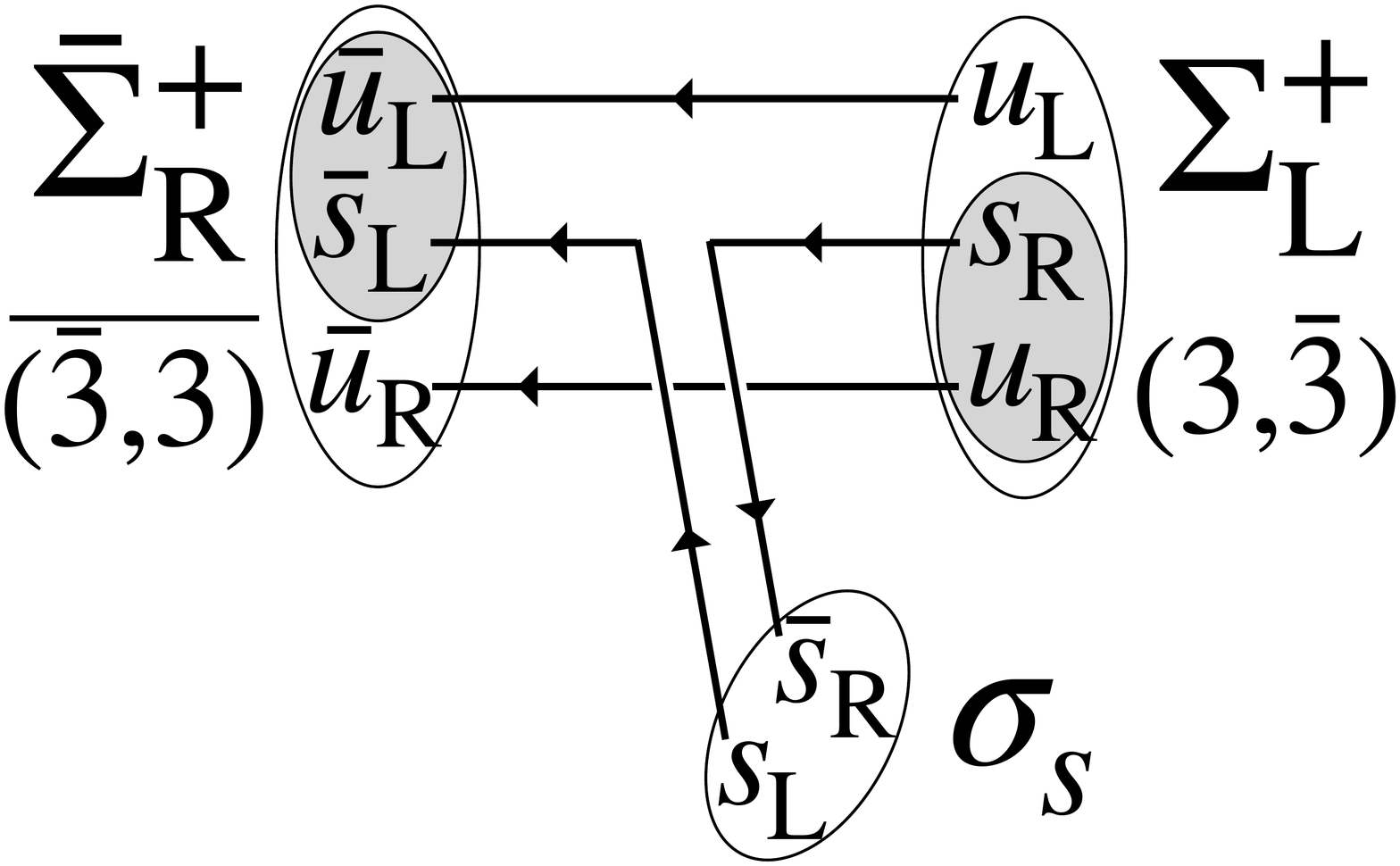}
	\caption{}
	\label{fig-quarklines-det-Sigma}
\end{subfigure}
\begin{subfigure}{0.3\hsize}\centering
	\includegraphics[width=0.8\hsize]{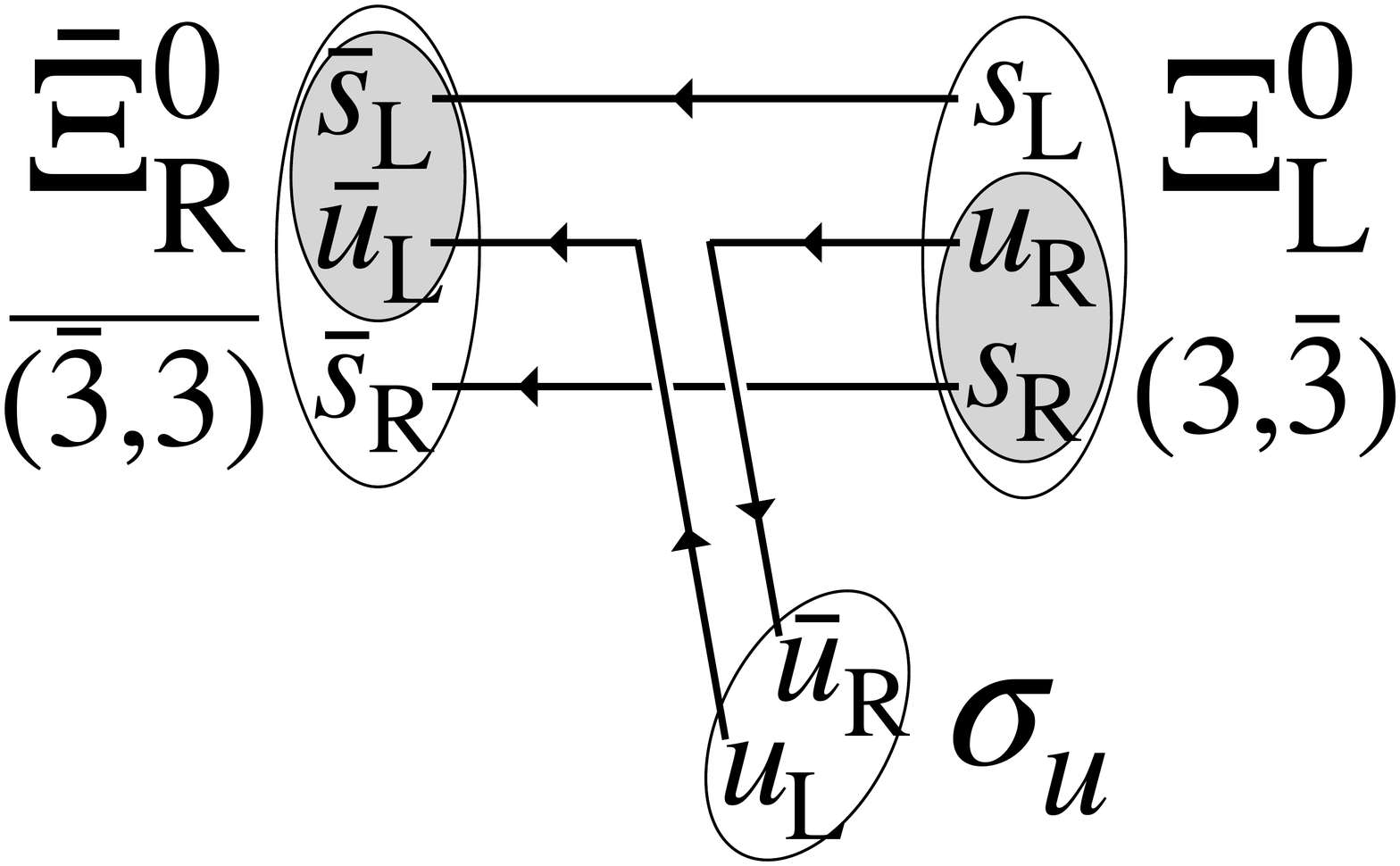}
	\caption{}
	\label{fig-quarklines-det-Xi}
\end{subfigure}
\caption{  Yukawa couplings between $(3,\bar3)$ and $(\bar3,3)$ baryon fields. }
\label{fig-quarklines-det-flavor}
\end{figure*}

Equation \eqref{eq-case1} can be represented graphically in Fig.\ref{fig-quarklines-det-flavor}.
In this model, 
$\sigma_s$ ($\propto(M)^3_3$, or $\propto\ev{\bar{s}s}$) 
contributes only to the $\Sigma$ baryons. 
This implies that the $\Xi$ baryons must be degenerated with the nucleons in this model, 
and therefore, this model cannot reproduce the octet baryon masses correctly. 

It should be instructive to mention the difference from the two-flavor case
where we have only nucleons for which we use $(2_L, 1_R)$ and $(1_L, 2_R)$ representations.
The $(2_L,1_R)$ representations may be constructed as $q_{\rm L} (q_{\rm L})^2$ or $q_{\rm L} (q_{\rm R})^2$.
Here good diquarks are the SU(2) singlet,
but in the context of three-flavor, these diquarks are $\bar{3}$ representation in SU(3).
To construct baryon octet analogous to nucleons in two-flavor models,
we need to add more representations in SU(3).

\subsection{Model 2: $(3,\bar3)+(\bar3,3)$ and $(8,1)+(1,8)$}\label{sec-1st-2}

Next we add the representation $(8,1)+(1,8)$, $\chi$, to models of $(3,\bar3)+(\bar3,3)$.
We emphasize that, at the first (and second) orders in $M$,
there are no Yukawa interactions that couple $\chi_L$ and $\chi_R$ fields.
This is because the $\chi$ contains 
three valence quarks with all left-handed or right-handed
so that Yukawa interactions with $\chi$ should include three quark exchanges
that flip the chirality of all three quarks. 
In other words, since $\UA{1}$-charges for $\chi_\rl$, $\chi_\rr$, $M$ are 
$-3$, $3$, and $-2$ respectively, 
a $\UA{1}$ symmetric term cannot be constructed unless we consider the cubic orders, $M^3$ or $(M^\dag)^3$.

\begin{figure*}\centering
\begin{subfigure}{0.3\hsize}\centering
	\includegraphics[width= 0.8\hsize]{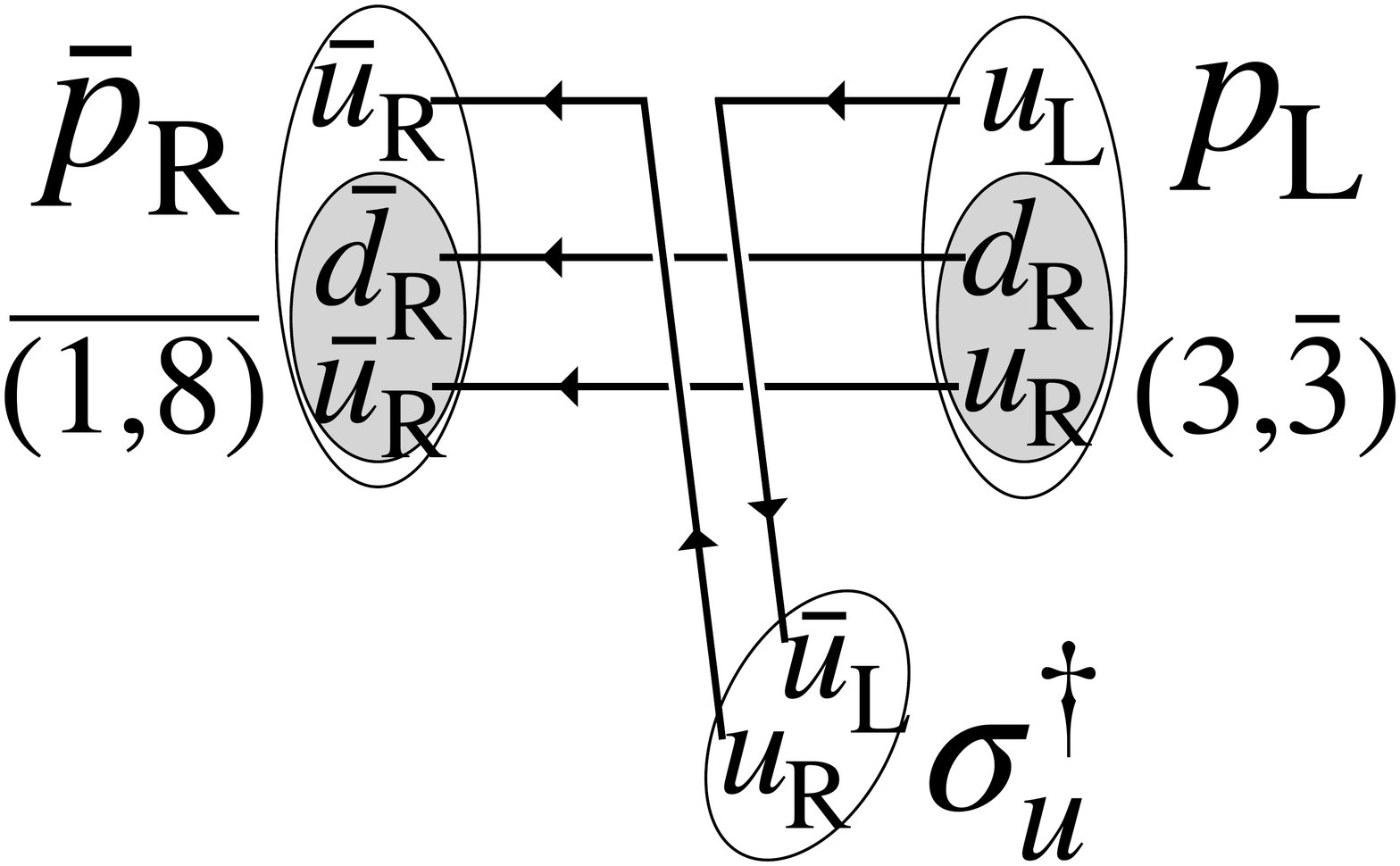}
	\caption{}
	\label{fig-quarklines-spectator-proton}
\end{subfigure}
\begin{subfigure}{0.3\hsize}\centering
	\includegraphics[width= 0.8\hsize]{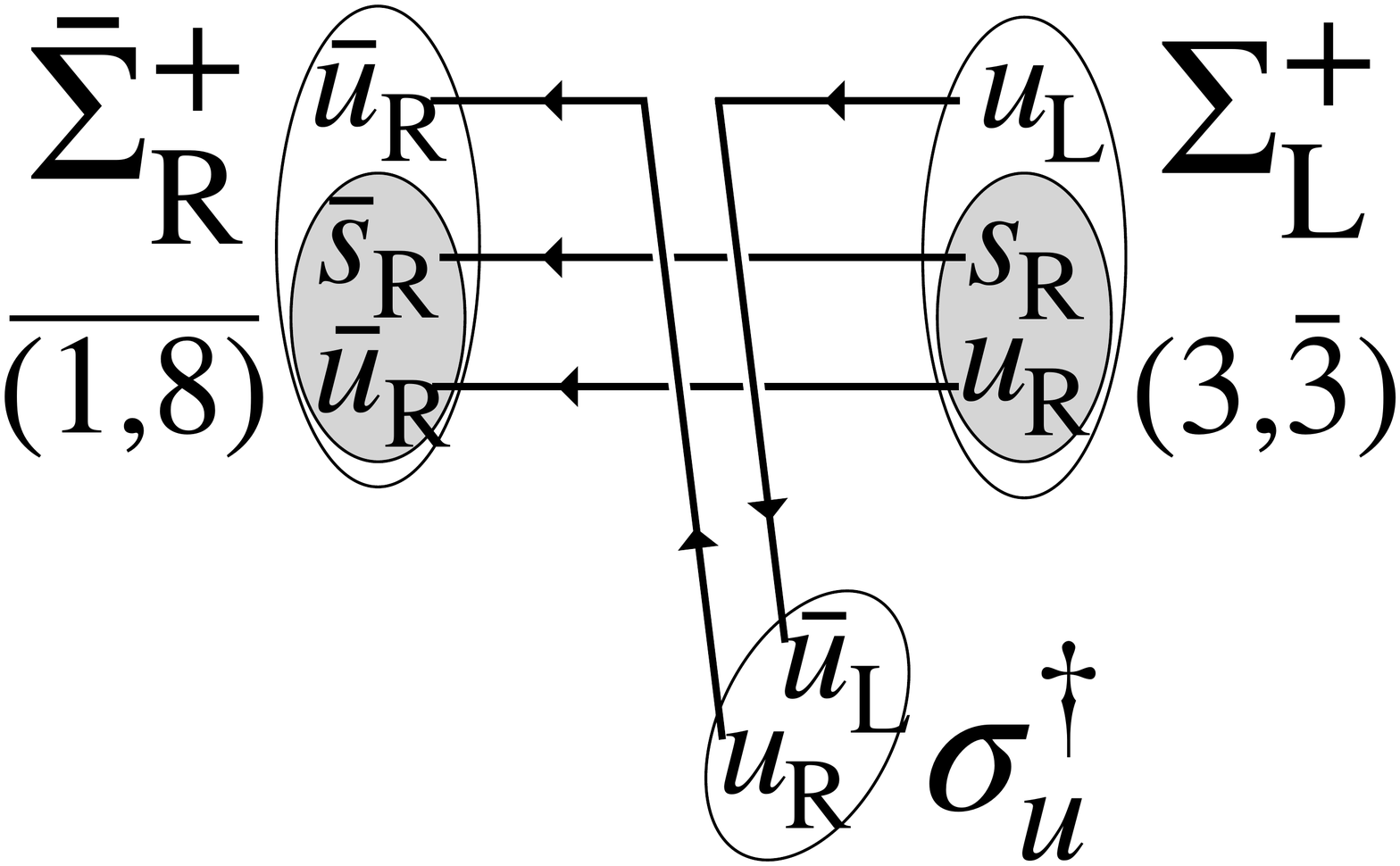}
	\caption{}
	\label{fig-quarklines-spectator-Sigma}
\end{subfigure}
\begin{subfigure}{0.3\hsize}\centering
	\includegraphics[width= 0.8\hsize]{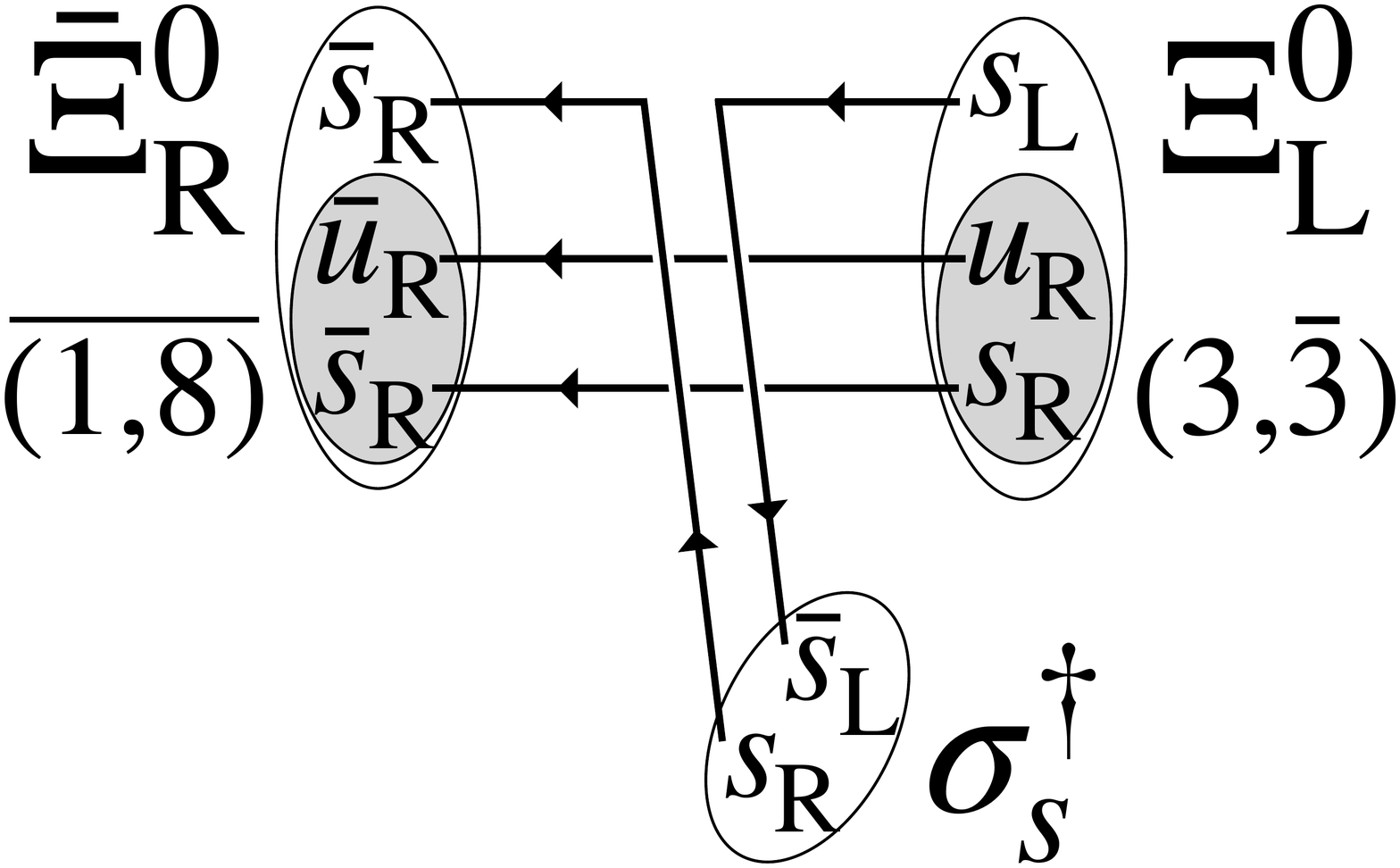}
	\caption{}
	\label{fig-quarklines-spectator-Xi}
\end{subfigure}
\caption{ Yukawa couplings between $(3,\bar{3})$ and $(1,8)$ baryon fields.
}
\label{fig-quarklines-spectator-flavor}
\end{figure*}

There are, however, the first order Yukawa interactions between $\psi$ and $\chi$.
The simplest Lagrangian, at the leading order in $M$, is
\begin{align}
\lag^\text{model(2)} 
& = \lag^\text{model(1)}
-g'\tr\big[
\bar\chi_\rl M\psi_\rr
+\bar\chi_\rr M^\dag\psi_\rl
+\mathrm{h.c.}\big]\,. \label{eq-case2}
\end{align}
This additional interaction $\tr(\bar\chi_\rr M^\dag\psi_\rl)$ 
can be interpreted as in Fig.\ref{fig-quarklines-spectator-flavor}. 
%
The strange quark contributes to $\Xi$ baryons through this interaction yields splitting between $\Sigma$ and $\Xi$. 

This model still contains problems in reproducing the spectra of octet baryons.
To see this, let us calculate the mass eigenvalues for the ground-state octet baryons in this model
by taking the VEV as $\ev{M}=\diag{\alpha, \beta, \gamma}$ with $\alpha=\beta$ as before.
We note that 
$\alpha$, $\beta$, and $\gamma$ correspond to the contribution from 
$\ev{\bar{u}u}$, $\ev{\bar{d}d}$, and $\ev{\bar{s}s}$, and 
$\alpha=\beta$ is assured by the isospin symmetry. 
According to the linear sigma model, 
when the pion and kaon decay constants are 
$f_\pi\approx93\,\mathrm{MeV}$ and $f_K\approx110\,\mathrm{MeV}$, 
its value is $\ev{M}\propto\diag{f_\pi,\, f_\pi,\, 2f_K-f_\pi} \approx\diag{93,\, 93,\,127}\, \mathrm{MeV}$. 
Using the VEV of $M$, the lagrangian can be decomposed as 
\begin{align}
\begin{aligned}
& 
\lag^\text{model(2)} =
-\begin{pmatrix} \bar\psi_N & \bar\chi_N \\ \end{pmatrix}
\hat{M}_N 
\begin{pmatrix} \psi_N \\ \chi_N \\ \end{pmatrix}
%
-\begin{pmatrix} \bar\psi_\Sigma & \bar\chi_\Sigma \\ \end{pmatrix}
\hat{M}_\Sigma
\begin{pmatrix} \psi_\Sigma \\ \chi_\Sigma \\ \end{pmatrix}\\
& \quad {} -\begin{pmatrix} \bar\psi_\Xi & \bar\chi_\Xi \\ \end{pmatrix}
\hat{M}_\Xi
\begin{pmatrix} \psi_\Xi \\ \chi_\Xi \\ \end{pmatrix}
+(\text{terms for }\Lambda\text{ baryons})
\ ,
\end{aligned}
\end{align}
where $\hat{M}_N$, $\hat{M}_\Sigma$ and $\hat{M}_\Xi$ are $2\times2$ mass matrices given by 
\begin{align}
\begin{aligned}
& \hat{M}_N = 
\begin{pmatrix}
-g\alpha & h\alpha \\
h\alpha & 0 \\
\end{pmatrix}
\ , 
\\
& \hat{M}_\Sigma = 
\begin{pmatrix}
-g\gamma & h\alpha \\
h\alpha & 0 \\
\end{pmatrix}
\ , \\
& \hat{M}_\Xi = 
\begin{pmatrix}
-g\alpha & h\gamma \\
h\gamma & 0 \\
\end{pmatrix}
\ .
\end{aligned}
\end{align}
The strange quark contributions for $\Sigma$ baryons, 
which enters the diagonal components in the mass matrix, 
corresponds to Fig.\ref{fig-quarklines-det-Sigma}. 
The one for $\Xi$ baryons, which enters the off-diagonal components, 
corresponds to Fig.\ref{fig-quarklines-spectator-Xi}. 
The mass eigenvalues for the ground-state octet members can be written as 
\begin{align}
m[N]&=m(|g\alpha|,|h\alpha|)\,, \\
m[\Sigma]&=m(|g\gamma|,|h\alpha|)\,, \\
m[\Xi]&=m(|g\alpha|,|h\gamma|)\,, 
\end{align}
where $m(x,y)\equiv\sqrt{(x/2)^2+y^2}-x/2$ is an eigenvalue of the matrix 
$\begin{pmatrix} x & y \\ y & 0 \end{pmatrix}$.

Here we note $|g\alpha|<|g\gamma|$ and $\partial_xm(x,y)<0$;
this means that this model leads to $m[N] > m[\Sigma]$.
Therefore, somewhat counter-intuitively, 
this model cannot reproduce the octet baryon masses correctly.

Note that, when we neglect the mixing with the singlet $\Lambda$ baryon, 
the mass term is expressed as
\begin{align}
-\begin{pmatrix} \bar\psi_{\Lambda} & \bar\chi_{\Lambda} \\ \end{pmatrix}
\hat{M}_{\Lambda} 
\begin{pmatrix} \psi_{\Lambda} \\ \chi_{\Lambda} \\ \end{pmatrix}
\ , 
\end{align}
with
\begin{align}
\hat{M}_{\Lambda} = \begin{pmatrix}
- \frac{g}{3} \left( 4 \alpha - \gamma\right)  & \frac{h}{3} \left( \alpha + 2 \gamma \right) \\
\frac{h}{3} \left( \alpha + 2 \gamma \right) & 0 \\
\end{pmatrix}
\ .
\end{align}
From this,
the mass of the octet $\Lambda$ baryon can be calculated as 
\begin{align}
m[\Lambda]=m\big( | g (4\alpha-\gamma)|/3, \, |h(\alpha+2\gamma)|/3 \big)\,. 
\end{align}
We stress that the mass matrices of octet baryons satisfy the {\GO} as
\begin{align}
\frac{1}{2} \left[ \hat{M}_N+ \hat{M}_\Xi \right] = \frac{1}{4} \left[ 3 \hat{M}_{\Lambda} + \hat{M}_\Sigma \right] \ .
\label{GO first order Yukawa}
\end{align}
%
(See Sec.\ref{sec-GO-matrix} for detail.) 
However, the mass eigenvalues can satisfy the {\GO} only up to first order of strange quark breaking $\mathcal{O}(\gamma-\alpha)$. 

To summarize this section,
we found that simple models based on the baryon octet with good diquarks
do not reproduce the baryon octet masses correctly, unless we go beyond the lowest order in $M$.
We need to add more representations including bad diquarks or go to higher orders in $M$.

\section{Quark Diagram and Chiral Yukawa Interaction}\label{sec-QD-CYI}

In the previous section we have found that the simplest version of 
the $(3,\bar{3}) + (\bar{3},3)$ and $(8,1) + (1,8)$ 
model
does not work well.
We have to go beyond the leading order,
in other word, we need to include two or more $M$ fields in Yukawa interaction terms.
Since there are many possible terms for Yukawa interactions at higher order, we need 
to set up some rules for systematic treatments.

In this paper, we propose to use quark diagrams to classify Yukawa interactions.
Quark fields in baryon and meson fields are connected 
to manifestly conserve the quantum numbers.
At the first order of $M$ (the first order Yukawa interactions),
 there are only two types of chiral Yukawa interaction.
%
Here, 
although the first order Yukawa terms were treated in the last section,
we repeat the analysis of the graph in terms of 
meson-diquark and meson-spectator couplings.
The coupling constants are expressed as $g_{1,2}^\mathrm{a}$ and $g_{1,2}^\mathrm{s}$, respectively.
Here, the subscript $1,2$ refers to two baryon fields in the parity doublet model for a given representation, $\psi_{1,2}$ and $\chi_{1,2}$.
In the next section (Sec.\ref{sec-2nd}), 
we will deal with ``second-order'' Yukawa interactions, which include 
two quark exchanges (two meson fields). 

\subsection{Correspondence between quark diagrams 
and hadronic effective interactions   }
\label{sec: graphs_int}

We explain how to find the hadronic effective interaction corresponding to a given quark diagram. 
To find the correspondence, the three-index notation 
(Eqs.\eqref{eq-psi-qqq}-\eqref{eq-chi-qqq}) for baryons is more convenient than the two-index notation. 
The two-index notation is useful for notational simplicity, though.

\begin{figure}\centering
\includegraphics[width=0.4\hsize]{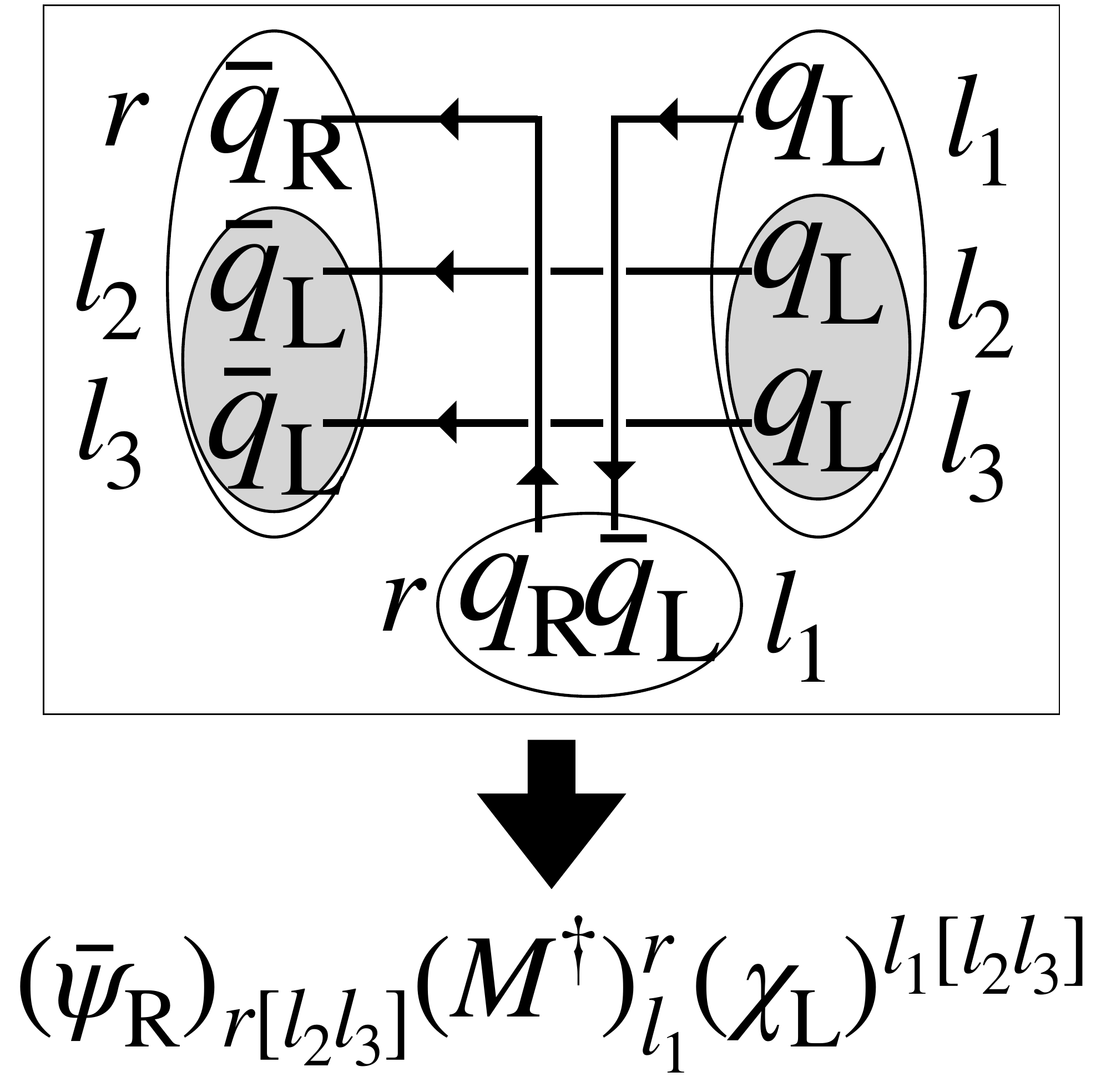}
\caption{
Correspondence between a quark diagram and 
a hadronic effective interaction. }
\label{fig-howtodraw}
\end{figure}

As shown in Fig.\ref{fig-howtodraw}, 
one draws a quark diagram 
in which 
each pair of $\bar{q}_i$ and $q_i$ ($i=\rl,\rr$) is connected though a quark line. 
Along quark lines, charges in the $\U{3}_\rl\times \U{3}_\rr$ symmetry are conserved.
Baryonic fields with different chirality are connected by inserting mesonic fields. 

According to our dynamical assumption based on diquarks,
the chirality flipping processes involving bad diquarks are assumed to be suppressed.
Integrating such intermediate states involves at least the second order in $M$.
The second-order contribution to the baryon mass is about 
$\sim \langle M \rangle^2 /\big( M_{\rm hard} - M_{\rm soft} \big) $
where $\langle M \rangle$ is the VEV of the meson field and 
$M_{\rm hard}$ and $M_{\rm soft}$ are the masses for hard- and soft-baryons, respectively.
The mass scale $M_{\rm hard} - M_{\rm soft} $ is the order of $M_\Delta - M_N \sim 300$ MeV.
The major assumption in this paper is that, 
the approximation $ \langle M \rangle /\big( M_{\rm hard} - M_{\rm soft} \big) \ll 1$,
which should become increasing valid toward the chiral restoration with $\langle M \rangle \rightarrow 0$, 
also sheds light on baryons in the vacuum.
Under this assumption 
 the second-order contributions in $\langle M \rangle$ 
 are suppressed compared with the first-order.

Meanwhile, the direct coupling between soft baryons (baryons with good diquarks) 
yields soft intermediate states
which cannot be treated perturbatively;
the Hamiltonian for soft baryonic fields must be fully diagonalized.
The full diagonalization involves  iterations of soft baryon graphs;
to avoid the double counting, from the list of higher order terms in $M$ 
we must pick up terms in which only hard baryons (baryons with bad-diquark) appear in the intermediate states.
In the following we dictate how to organize interactions
between $\psi$ and $\chi$ fields.
%

%

\subsection{``First-order'' Yukawa interaction}\label{sec-1st}


We begin with the first order Yukawa interactions. 
For soft baryon fields there are only two possible processes:
\begin{itemize}
\item {\it Coupling to diquarks}--- The Yukawa interaction couples to a quark $q_\rr$ in $(3_L, \bar{3}_R)$ representation, $\psi_\rl\sim q_\rl[q_\rr q_\rr]$.
In other words, the matrix $M$ couples to one of quarks forming a good diquark.
After the chiral flipping, the $(\bar{3}_L, 3_R)$ representation, $\psi_\rr \sim [q_{\rm L} q_{\rm L}] q_{\rm R}$ is formed.
This case was discussed in Sec.~\ref{sec-1st-1}. 
The same is true after exchanges of $L$ and $R$.

\item  {\it Coupling to a spectator quark}--- A spectator quark $q_\rl$ in $(3_L,\bar{3}_R)$ representation, $\psi_\rl\sim q_\rl[q_\rr q_\rr]$,
 couples to $M$ and flips the chirality.
The resulting representation is $(1_L,8_R)$, $\psi_\rr\sim q_\rr[q_\rr q_\rr]$. 
This case was discussed in Sec.~\ref{sec-1st-2}.
The same is true after exchanges of $L$ and $R$.
\end{itemize} 
The same arguments are applied to the mirror representations.
Below we write down the effective Lagrangian for these couplings.

\subsubsection{Diquark interaction ($g_{1,2}^\mathrm{a}$)}\label{sec-1st-det}

\begin{figure}[H]\centering
\includegraphics[width=0.5\hsize]{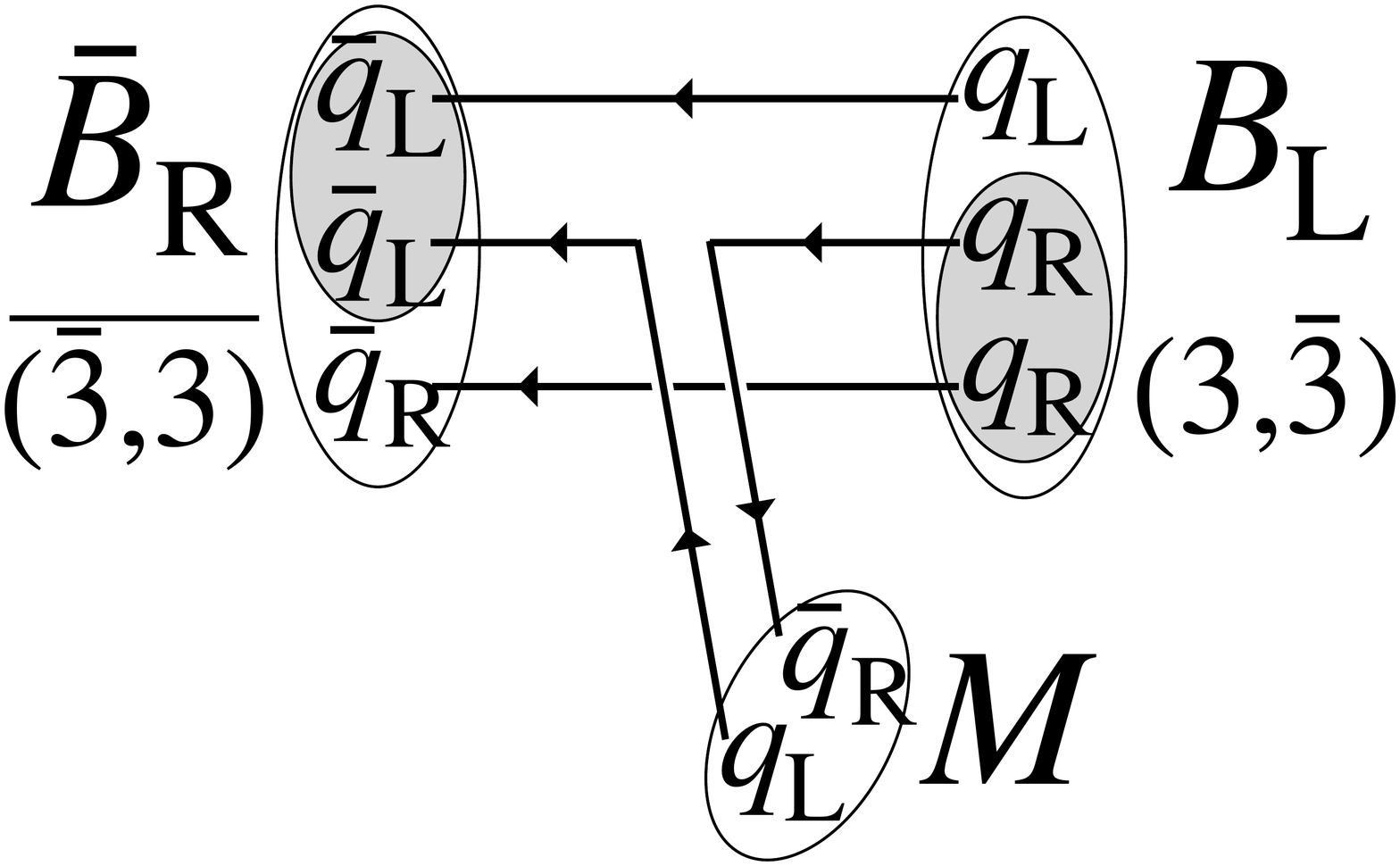}
\caption{First-order chiral Yukawa interaction 
between $(\bar3,3)$ and $(3,\bar3)$. }
\label{fig-quarklines-det}
\end{figure}
The first-order chiral Yukawa interaction corresponding to a diagram in Fig.\ref{fig-quarklines-det}
is written as
\begin{align}
(\bar\psi_\rr)_{r_1[l_1l_2]}(M)^{l_2}_{r_2}(\psi_\rl)^{l_1[r_1r_2]}\,. 
\end{align}
In the two-index notation, this is equivalent to 
\begin{align}
\varepsilon_{l_1l_2l_3}\varepsilon^{r_1r_2r_3}
(\bar\psi_\rr)^{l_3}_{r_1}(M)^{l_2}_{r_2}(\psi_\rl)^{l_1}_{r_3}\,, 
\label{g1a org 3}
\end{align}
which was treated in Sec.\ref{sec-1st-1}. 
This expression is also equivalent to the following contribution: 
\begin{align}
& \tr(\bar\psi M\psi) - \tr\big[\, \bar\psi\psi \big(\tr(M)-M \big) \, \big]\notag\\
& {} +\tr(\bar\psi)\tr(M)\tr(\psi)-\tr(\bar\psi)\tr(M\psi)-\tr(\bar\psi M)\tr(\psi) \,.
 \label{eq-det-decompose}
\end{align}
The traced baryonic fields $\tr \psi$ or $\tr \bar{\psi}$ represent the $\Lambda_0$ (flavor-singlet $\Lambda$ baryon). 
Terms without $\Lambda_0$ 
are summarized in the first line of Eq.(\ref{eq-det-decompose})
which takes the same form as Eq.\eqref{eq-GO}, 
so that the flavor-octet baryons satisfy the {\GO}.

In the following sections, the coupling constants of the Yukawa interaction of the form given in Eq.~(\ref{g1a org 3}) for naive representation 
is denoted as $g_1^\mathrm{a}$ and that for mirror representation is as $g_2^\mathrm{a}$.

\subsubsection{Spectator interaction ($g_{1,2}^\mathrm{s}$)}\label{sec-1st-spectator}

\begin{figure}\centering
\includegraphics[width=0.5\hsize]{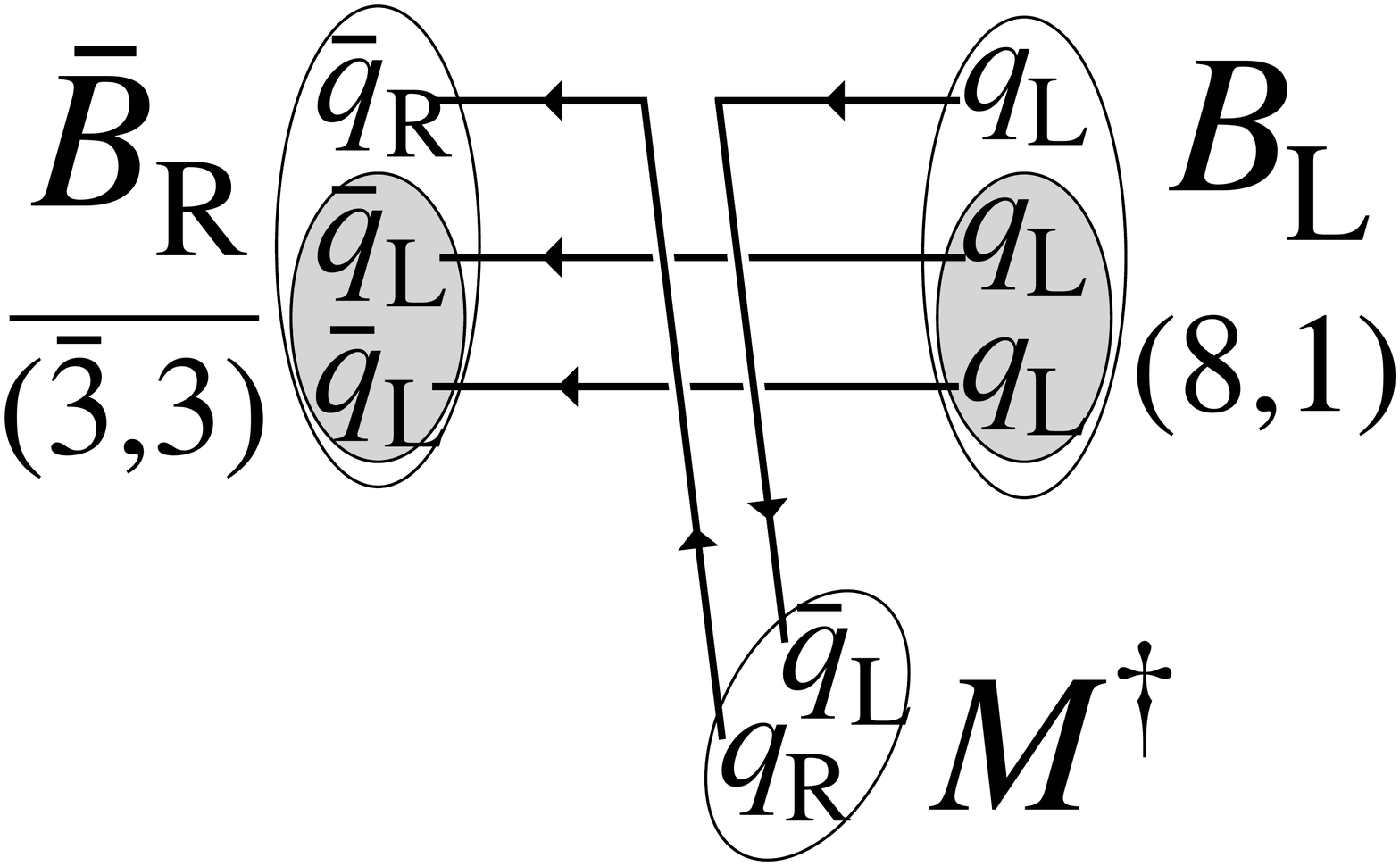}
\includegraphics[width=0.5\hsize]{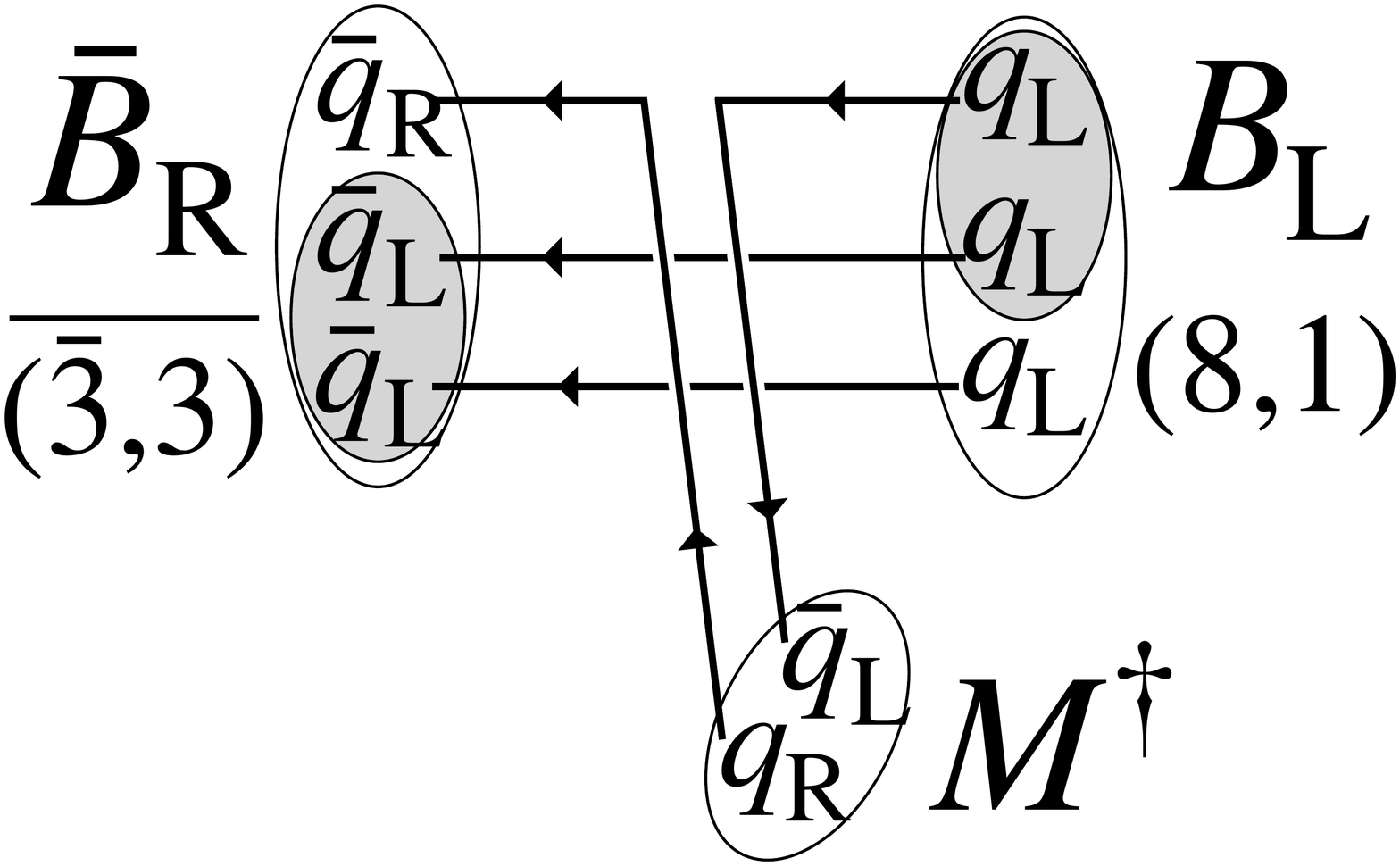}
\includegraphics[width=0.5\hsize]{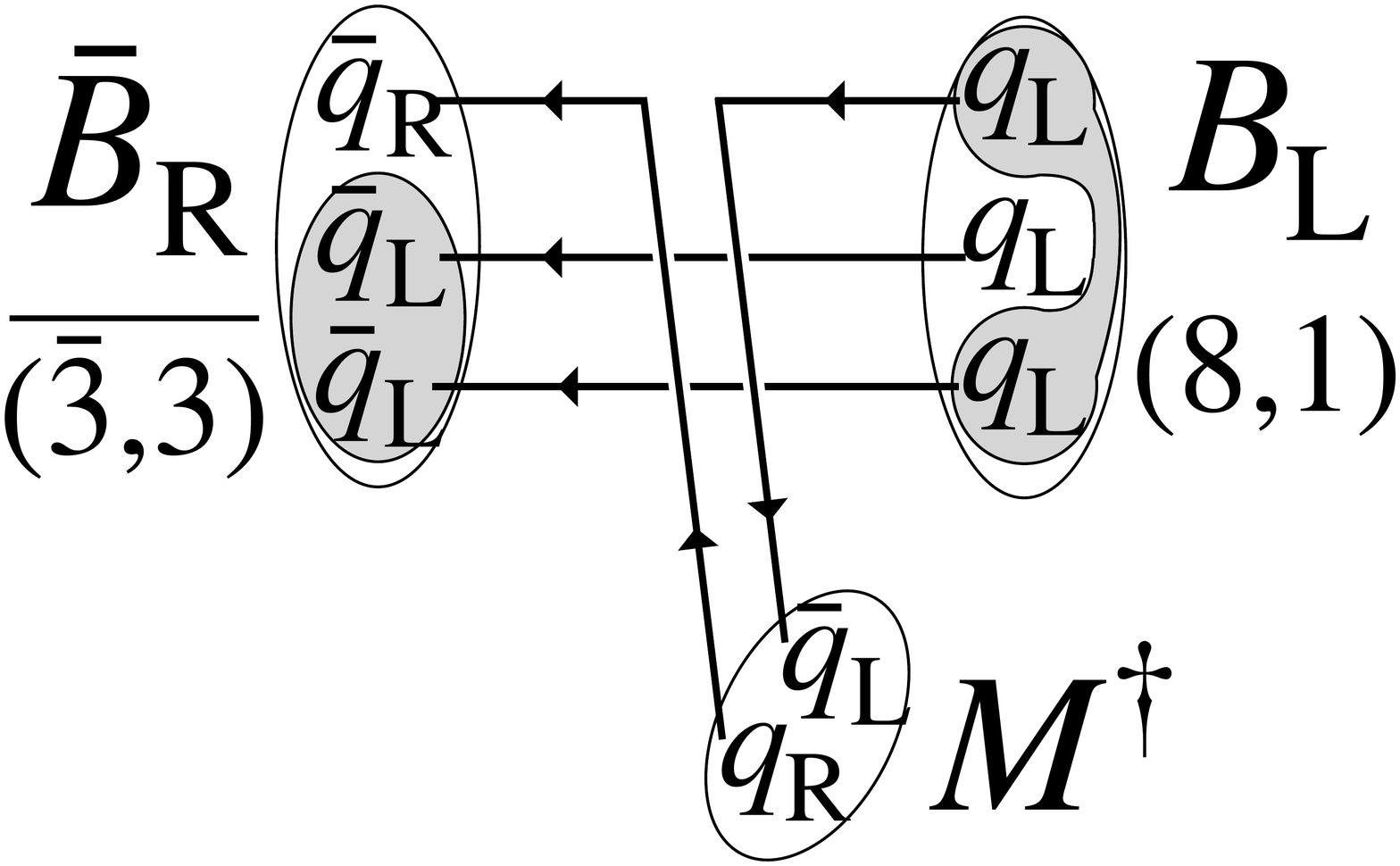}
\caption{First-order chiral Yukawa interactions connecting from $(8_L,1_R)$ to $(\bar{3}_L,3_R)$ representations.
Although there are three patterns, 
all of them correspond to the same effective interaction, due to the traceless of $\chi$. 
}
\label{fig-quarklines-spectator}
\end{figure}

Figure~\ref{fig-quarklines-spectator} shows that 
one of three quarks $q_L$ included in the $(8_L,1_R)$ representation flips its chirality to $q_R$.
The corresponding effective interaction is written as 
\begin{align}
\label{eq-1st-spectator}
(\bar\chi_\rr)_{r[r_1r_2]}(M^\dag)^{r}_{l}(\psi_\rl)^{l[r_1r_2]}\,. 
\end{align}
In the two-index notation, this can be written as 
\begin{align}
\tr(\bar\chi_\rr M^\dag \psi_\rl)\ .
\end{align}
We note that this term generates the contributions to the masses of octet baryons which satisfy the {\GO} as shown in Eq.~(\ref{GO first order Yukawa}).
We would like to stress that, 
even if any pair of quarks in $\chi$ forms a diquark, 
as seen in Fig.\ref{fig-quarklines-spectator}, 
the corresponding effective interaction is expressed by the term given in Eq.~\eqref{eq-1st-spectator}. 
This property is because of the traceless property of $\chi$, or equivalently, 
\begin{align}\label{eq-traceless}
(\chi)^{i[jk]}+(\chi)^{j[ki]}+(\chi)^{k[ij]}=0\, .
\end{align}
In the following sections, 
the coupling constants of the Yukawa interaction connecting $\chi$ and $\psi$ are 
denoted as $g_1^\mathrm{s}$ and $g_2^\mathrm{s}$.

%
%
%
%
%
%

\section{Integrating out baryons including a bad diquark: Second-Order Yukawa Interaction}
\label{sec-2nd}

In this section, we construct a minimal set of the second-order Yukawa interactions based on the quark diagram introduced in the previous section.

We omit the flavor singlet $\Lambda$ baryons for simplicity, 
which may be heavier than the octet baryons due to 
the U(1)$_{\rm A}$ anomaly. 
(See Sec.\ref{sec-2nd-singlet} for the singlet baryons.) 

As mentioned in Sec.\ref{sec: graphs_int},
it is important to omit terms which generate soft intermediate states.
In this section we carefully pick out terms yielding only hard intermediate states.
\subsection{Classification of the processes: overview}\label{sec-2nd-allpossibilities}

We consider two sets of representations $\psi_{1,2}$ and $\chi_{1,2}$ for the parity doublet
and examine how to combine them to generate the second order in $M$.
Single insertion of a meson line flips the chirality and change the chiral representation of baryon fields.
As we have mentioned, we have to remove graphs which are simply iterations of the first order graph.
For this purpose, the representations generated by the chirality flipping process must belong to hard baryons 
which include a bad diquark. 
As we stated in the previous section, the transition from soft to hard baryon intermediate states 
effectively introduce a factor
$ \langle M \rangle /\big( M_{\rm hard} - M_{\rm soft} \big) $
as an expansion parameter.

In this paper we focus on 
the Yukawa interactions concerning the scalar and pseudoscalar mesons only.
Then, from the structure of the Dirac spinor,
the chirality of the baryon must flip at the interaction point. 
As we will show below, this is impossible without mirror representations.
Thanks to the availability of the mirror representations in our framework,
Yukawa interaction terms can be made ${\rm SU(3)}$ chiral invariant by using the mirror representation for one of baryon fields.

\subsection{Transition $\psi_{1,2} \rightarrow \psi_{2,1}$ }\label{sec:psi_psi}

Let us consider the transition between the same chiral representations.
First we examine $\psi_{1,2} \rightarrow \psi_{2,1}$.
There are three possible processes.

\subsubsection{Double spectator-meson interactions ($\psi$-$\psi$) }\label{sec-2nd-s2_psi-psi}
\begin{figure}[H]\centering
\vspace{-0.3cm}
\includegraphics[width=0.5\hsize]{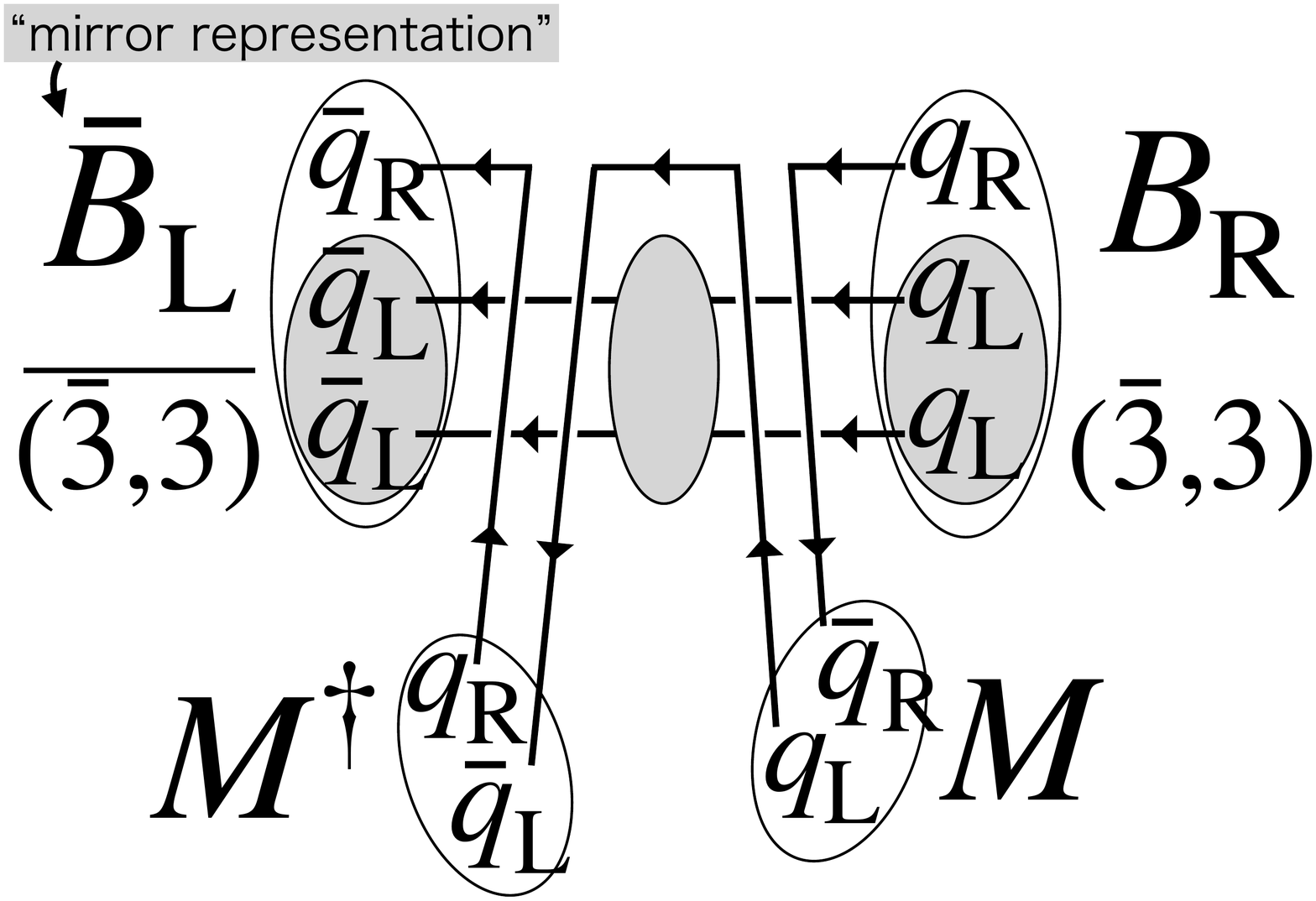}
\vspace{-0.3cm}
\caption{
Yukawa coupling between $(\bar{3},3)$ and $(\bar{3},3)$ where a spectator quark flips the chirality.
}
\label{fig-quarklines-2nd-s2_psi-psi}
\end{figure}
A spectator quark $q_\rr$ of $\psi_\rr \sim q_\rr [q_\rl q_\rl]$ flips the chirality twice
(Fig.\ref{fig-quarklines-2nd-s2_psi-psi}).
In this process hard baryons do not appear in the intermediate states, we must omit them to avoid the double counting.

\subsubsection{Spectator-meson and diquark-meson interactions ($h_1$, $\psi$-$\psi$) }\label{sec-2nd-s2}

\begin{figure}[H]\centering
\vspace{-0.3cm}
\includegraphics[width=0.5\hsize]{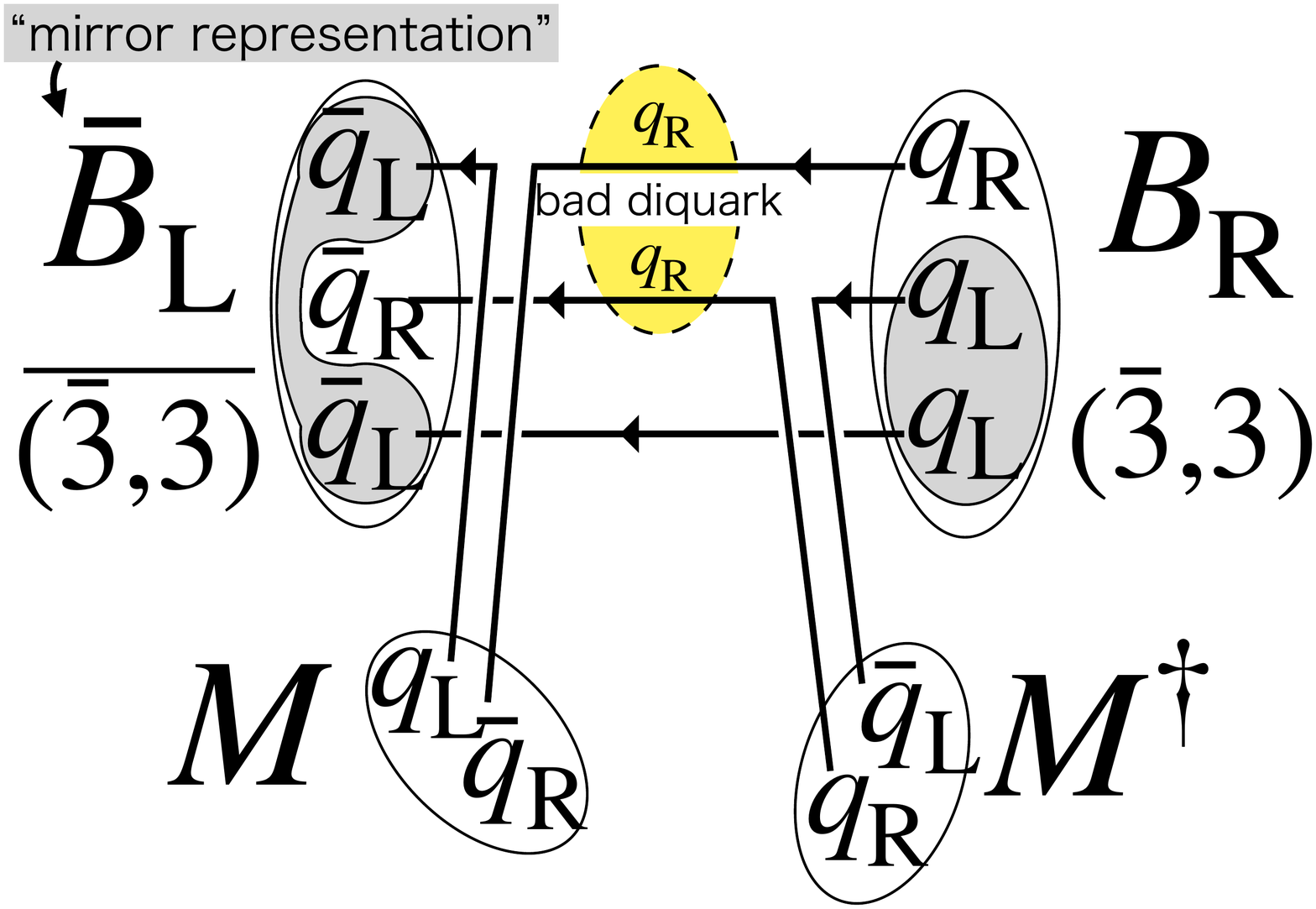}
\vspace{-0.3cm}
\caption{
Yukawa coupling between $(\bar{3},3)$ and $(\bar{3},3)$ where a spectator quark and a quark in the good-diquark flip the chirality.
}
\label{fig-quarklines-2nd-sd_psi-psi}
\end{figure}
Both a spectator quark and a quark forming a diquark flip the chirality once.
The chirality flipping in a diquark destroys a good diquark in the initial state and generates a hard baryon in the intermediate states.  
(Fig.\ref{fig-quarklines-2nd-sd_psi-psi}).
This interaction can be written as 
\begin{align}
(\bar\psi_\rr)_{r_1[l_1l']}(M)^{l_1}_{r_2}(M^\dag)^{r_1}_{l_2}(\psi_\rl^\mir)^{r_2[l_2l']} \ .
\end{align}
In terms of the two-index notation, this is written as 
\begin{align}\label{eq-2nd-sd-2index}
\tr(\bar\psi_\rr M^\dag M\psi_\rl^\mir)
-\tr(\bar\psi_\rr M^\dag)\tr(M\psi_\rl^\mir)\,. 
\end{align}
The first term $\tr(\bar\psi_\rr M^\dag M\psi_\rl^\mir)$ satisfies the {\GO}, 
while the second term $\tr(\bar\psi_\rr M^\dag)\tr(M\psi_\rl^\mir)$ breaks it, 
because it is not the form of Eq.\eqref{eq-GO}. 
However, when the flavor singlet $\Lambda$ is omitted, 
this term contributes only to the octet member of $\Lambda$ baryon and 
the breaking contribution is proportional to 
$(\gamma-\alpha)^2$.
Therefore, this interaction satisfies the {\GO} up to first-order of strange quark mass perturbation. 
We should stress that this term is possible only when there exists a mirror representation $\psi_{\rm L}^{\rm mir}$.

\subsubsection{Double meson insertions into a single quark in a diquark ($h_2$, $\psi$-$\psi$)}\label{sec-2nd-d2_psi-psi}

\begin{figure}[H]\centering
\vspace{-0.3cm}
\includegraphics[width=0.5\hsize]{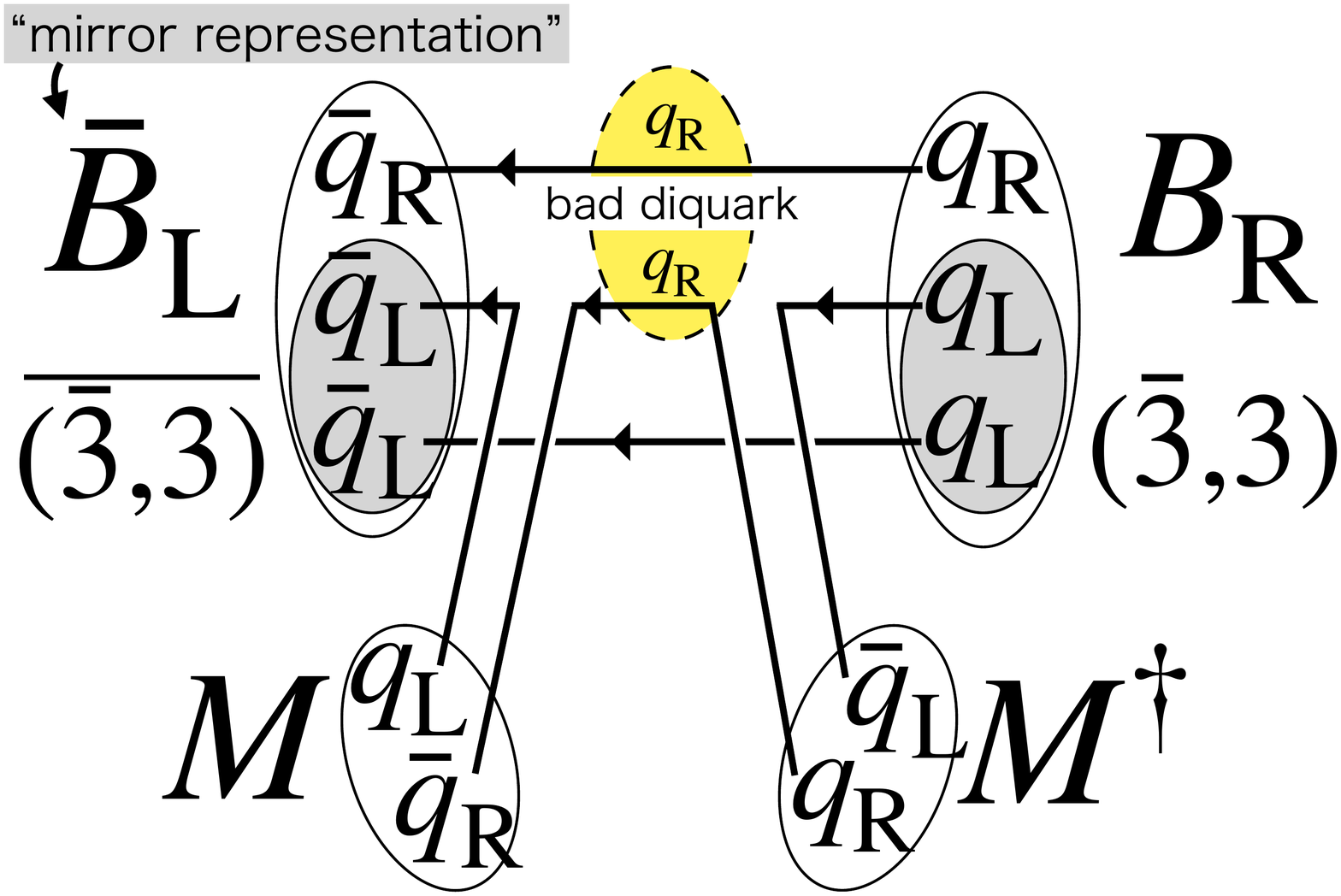}
\vspace{-0.3cm}
\caption{
Yukawa coupling between $(\bar{3},3)$ and $(\bar{3},3)$ where a spectator quark in the good-diquark flips the chirality.
}
\label{fig-quarklines-2nd-d2_psi-psi}
\end{figure}
Figure~\ref{fig-quarklines-2nd-d2_psi-psi} 
shows the diagram with double chirality flipping in a quark belonging to a diquark. 
The intermediate states are hard. 
This interaction can be written as 
\begin{align}
(\bar\psi_\rr)_{r[l_1l]}(MM^\dag)^{l_1}_{l_2}(\psi_\rl^\mir)^{r[l_2l]} \ ,
\end{align}
or in the index contracted notation,
\begin{align}
\tr[\bar\psi_\rr\psi_\rl^\mir(\tr(MM^\dag)-MM^\dag)] \ ,
\end{align}
which satisfies the {\GO}.

\subsection{Transition $\chi_{1,2} \rightarrow \chi_{2,1}$ }\label{sec:psi_psi}

Next, we examine $\chi_{1,2} \rightarrow \chi_{2,1}$.
There are three possible processes.
The processes are similar to the $\psi_{1,2} \rightarrow \psi_{2,1}$ transitions but the microphysics is not quite identical.

\subsubsection{Double spectator-meson interactions ($\chi$-$\chi$)}\label{sec-2nd-s2_chi-chi}

\begin{figure}[H]\centering
\vspace{-0.3cm}
\includegraphics[width=0.5\hsize]{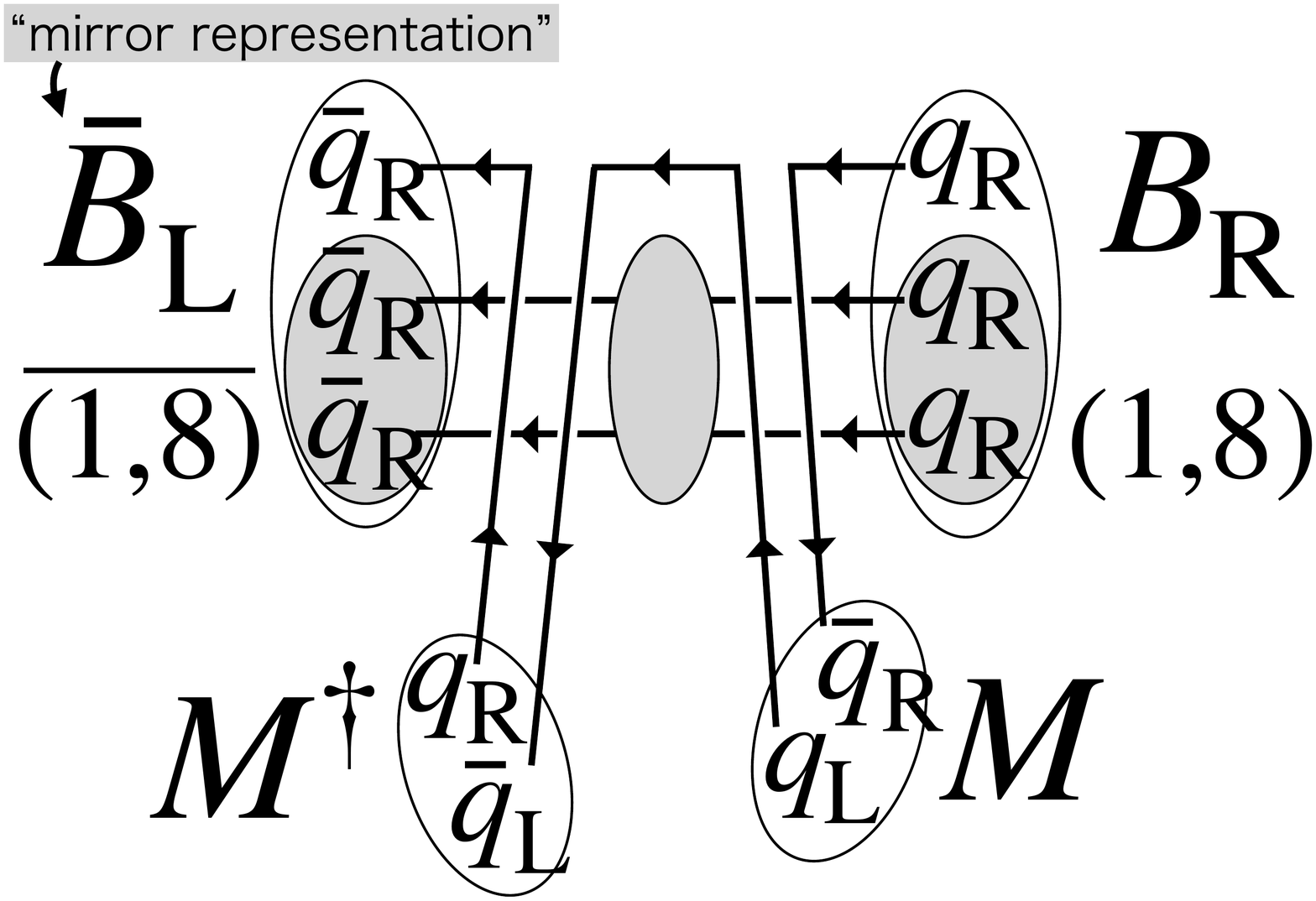}
\vspace{-0.3cm}
\caption{
Yukawa coupling between $(1,8)$ and $(1,8)$ where a spectator quark flips the chirality.
}
\label{fig-quarklines-2nd-s2_chi-chi}
\end{figure}
As in the $\psi$ cases, double meson insertions into a spectator quark 
$q_\rr$ of $\chi_\rr\sim q_\rr [q_\rr q_\rr]$
(Fig.\ref{fig-quarklines-2nd-s2_chi-chi})
does not contain any hard baryons and we must omit them to avoid the double counting.

\subsubsection{Double meson insertions into a single quark in a diquark ($\chi$-$\chi$)}\label{sec-2nd-d2_chi-chi}

%
\begin{figure}[H]\centering
\vspace{-0.3cm}
\includegraphics[width=0.5\hsize]{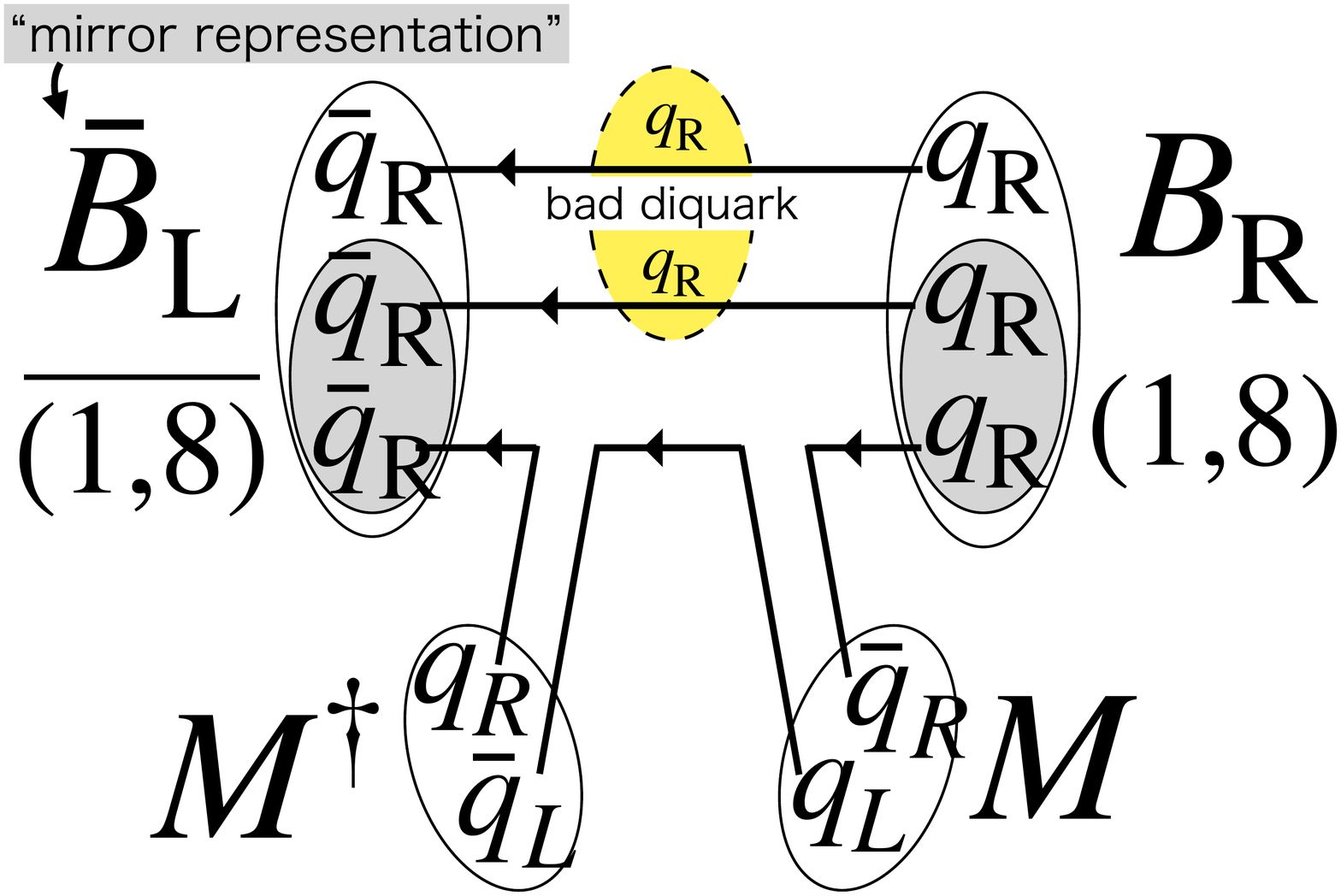}
\includegraphics[width=0.5\hsize]{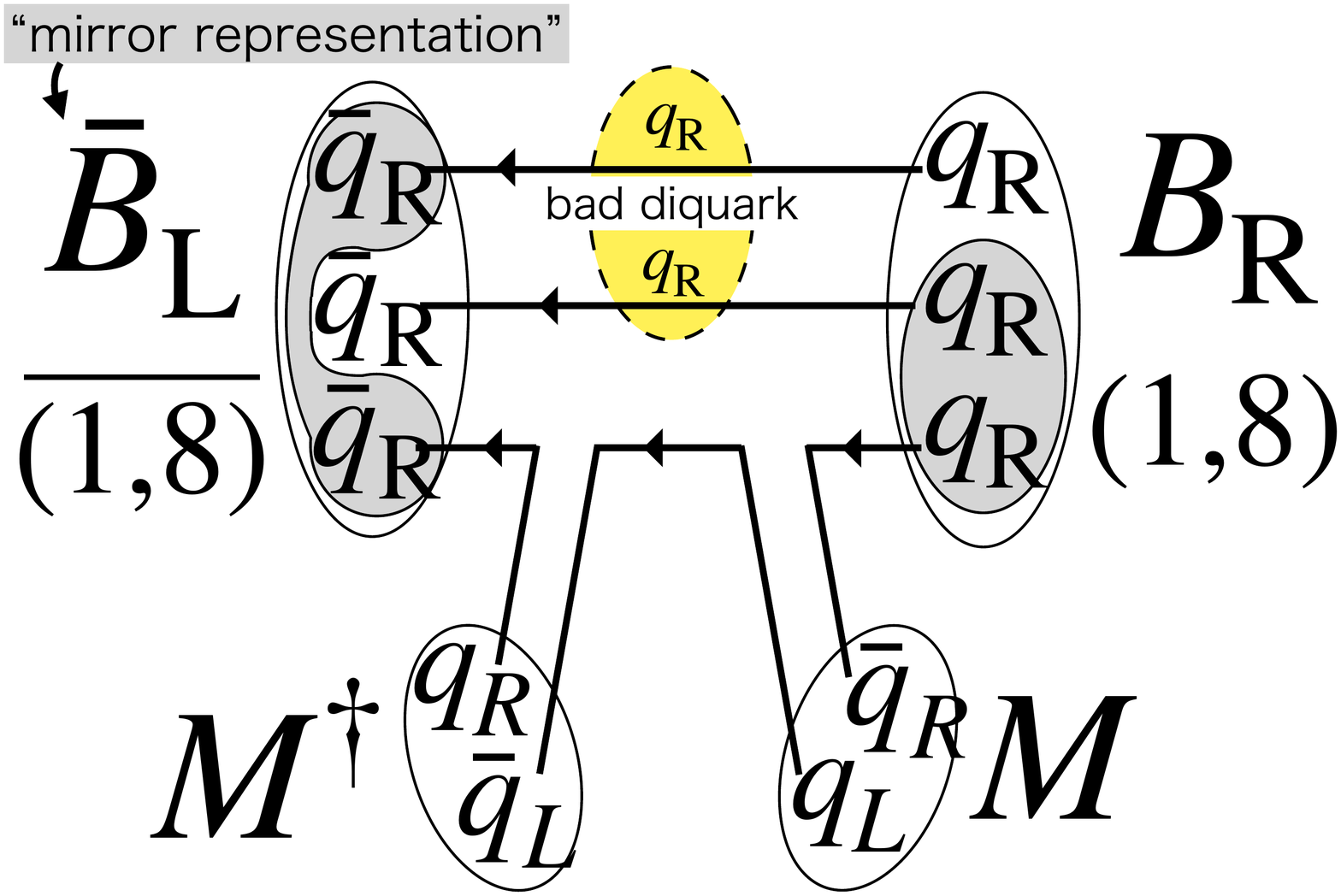}
\vspace{-0.3cm}
\caption{}
\label{fig-quarklines-2nd-chi12}
\end{figure}
Figure \ref{fig-quarklines-2nd-chi12}
shows the diagram with double chirality flipping in a quark belonging to a diquark. 
The intermediate states are hard. 
The difference between the upper and lower panels 
are the constituents forming the diquark.
In the former this interaction can be written as 
\begin{align}
(\bar\chi_\rr)_{r_1[r_2r_3]}(M^\dag M)^{r_3}_{r_4}(\chi_\rl^\mir)^{r_1[r_2r_4]} \ ,
\end{align}
or in the index contracted notation,
\begin{align}
\tr[\bar\chi_\rr\chi_\rl^\mir(\tr(M^\dag M)-M^\dag M)] \ .
\end{align}
In the latter, there is reformation of a diquark. The corresponding interaction term can be written as
\begin{align}
(\bar\chi_\rr)_{r_1[r_2r_3]}(M^\dag M)^{r_3}_{r_4}(\chi_\rl^\mir)^{r_2[r_1r_4]} \ ,
\end{align}
or  in the index contracted notation,
\begin{gather}
\tr(\bar\chi_\rr M^\dag M\chi_\rl^\mir)
-\tr[\bar\chi_\rr\chi_\rl^\mir(\tr(M^\dag M)-M^\dag M)] \ .
\end{gather}
Both terms separately satisfy the {\GO}.

\subsection{Transition $\psi_{1,2} \rightarrow \chi_{1,2} ~{\rm or}~\chi_{2,1}$ }\label{sec:psi_psi}

\begin{figure}[H]\centering
\includegraphics[width=0.5\hsize]{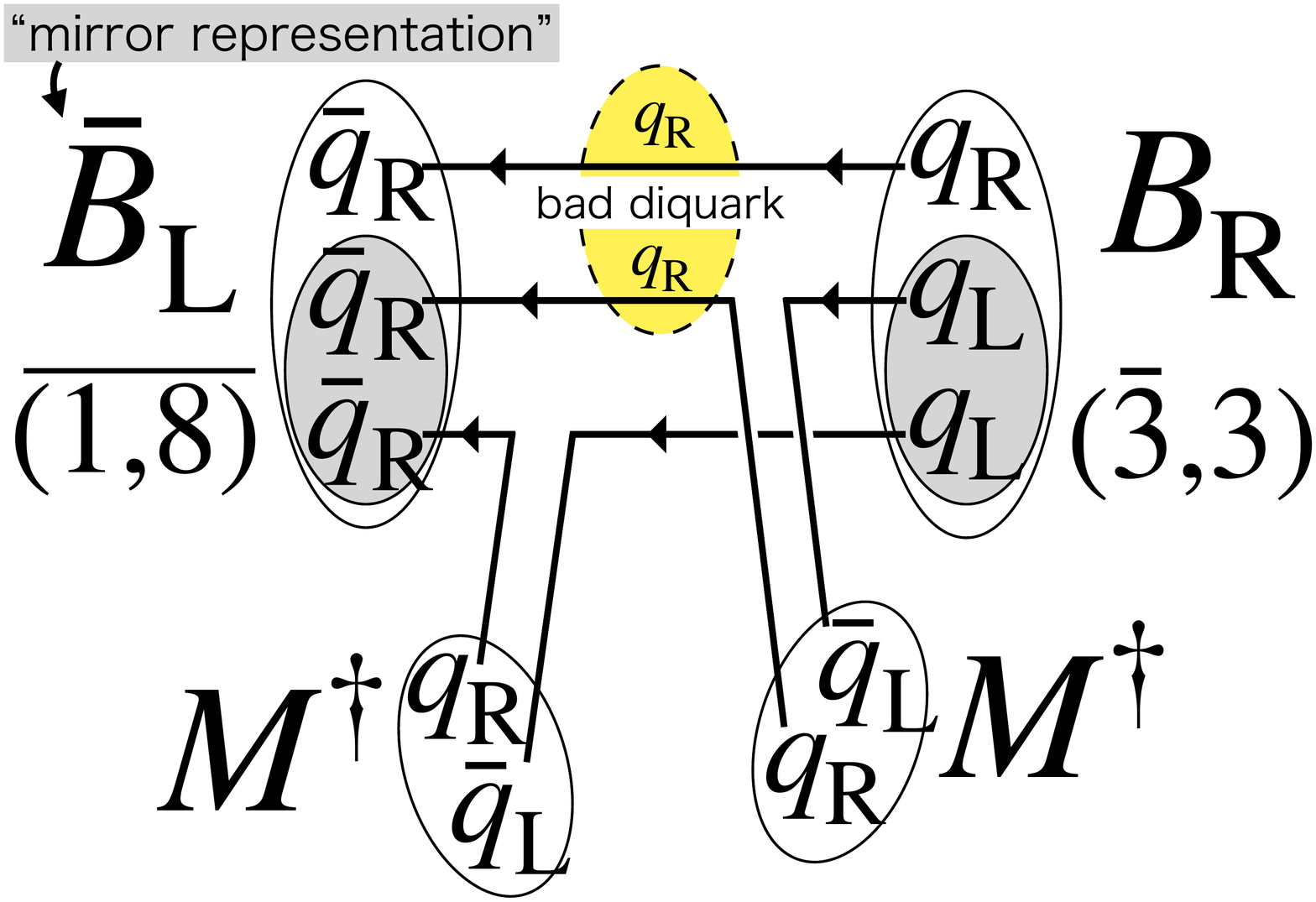}
\includegraphics[width=0.5\hsize]{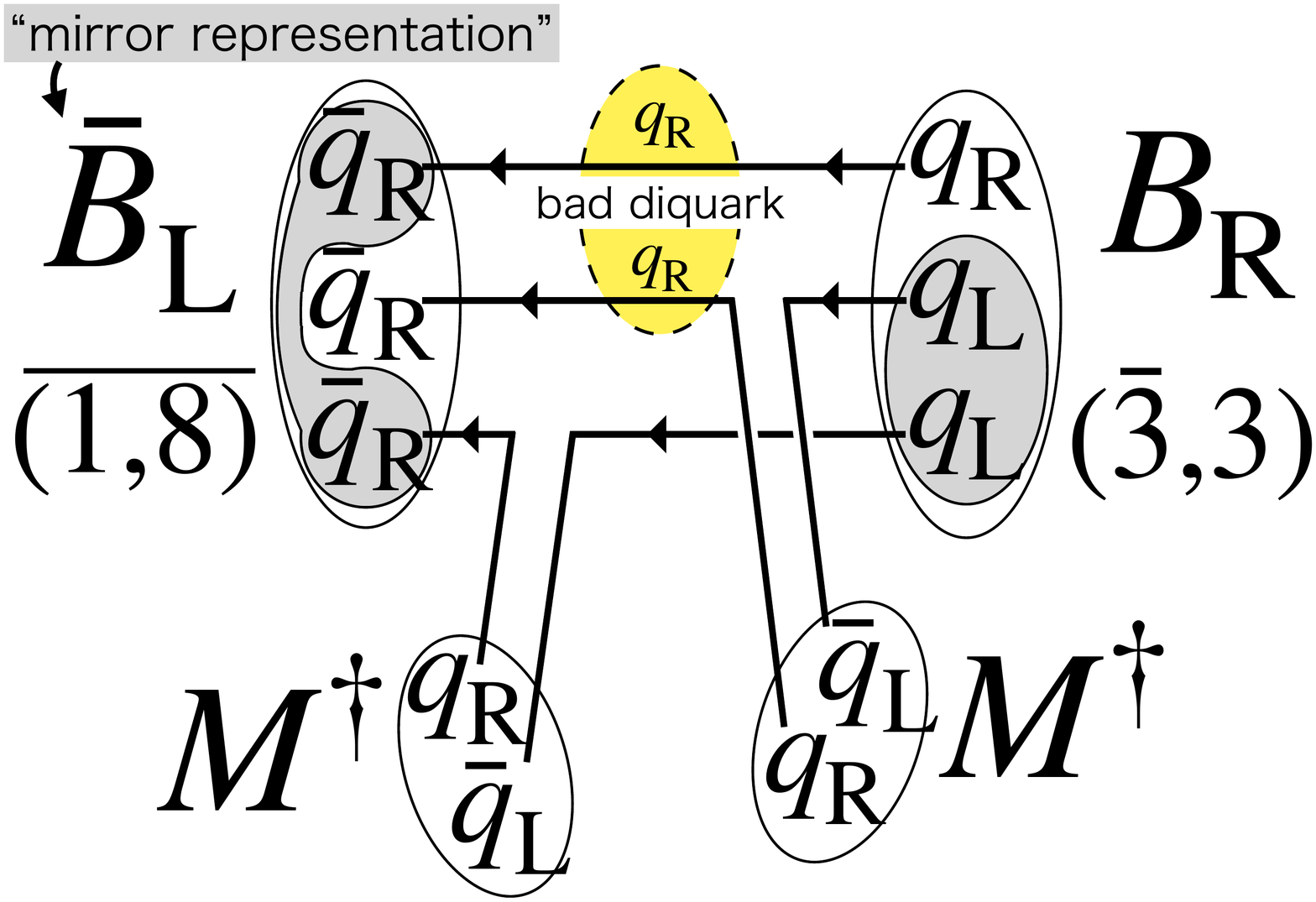}
\caption{Although there are two patterns, 
they correspond to the same effective interaction. }
\label{fig-quarklines-2nd-dd}
\end{figure}

Finally we examine the off-diagonal transitions between different chiral representations,
the $\psi_{1,2} \rightarrow \chi_{2,1}$ processes.
It turns out that the only nonzero processes are two meson insertions to quarks belonging to good diquarks.
Figure \ref{fig-quarklines-2nd-dd} 
shows two diagrams, but they can be expressed by a single term in the Lagrangian,
\begin{align}\label{eq-diq-res-1}
(\bar\psi_\rr)_{r[l_1l_2]}(M)^{l_1}_{r_1}(M)^{l_2}_{r_2}(\chi_\rl^\mir)^{r[r_1r_2]} \ ,
\end{align}
or
\begin{align}\label{eq-diq-res-2}
(\bar\psi_\rr)_{r[l_1l_2]}(M)^{l_1}_{r_1}(M)^{l_2}_{r_2}(\chi_\rl^\mir)^{r_1[rr_2]}\ . 
\end{align}
Eq.\eqref{eq-diq-res-1} and Eq.\eqref{eq-diq-res-2} 
are equivalent due to the traceless of $\chi$ given in Eq.~\eqref{eq-traceless}. 
This can be also written as 
\begin{align}
\tr(\bar\psi_\rr\chi_\rl^\mir\hat{O}) \ ,
\label{psi chi O}
\end{align}
where 
\begin{align}
\hat{O}^{r_3}_{l_3}\equiv
\varepsilon_{l_1l_2l_3}\varepsilon^{r_1r_2r_3}(M)^{l_1}_{r_1}(M)^{l_2}_{r_2} \ .
\label{def O op}
\end{align}
From the expression in Eq.~(\ref{psi chi O}), one can easily see that this term satisfies the {\GO}.

\subsection{Singlet $\Lambda$ baryon}\label{sec-2nd-singlet}

\begin{figure}[H]\centering
\includegraphics[width=0.5\hsize]{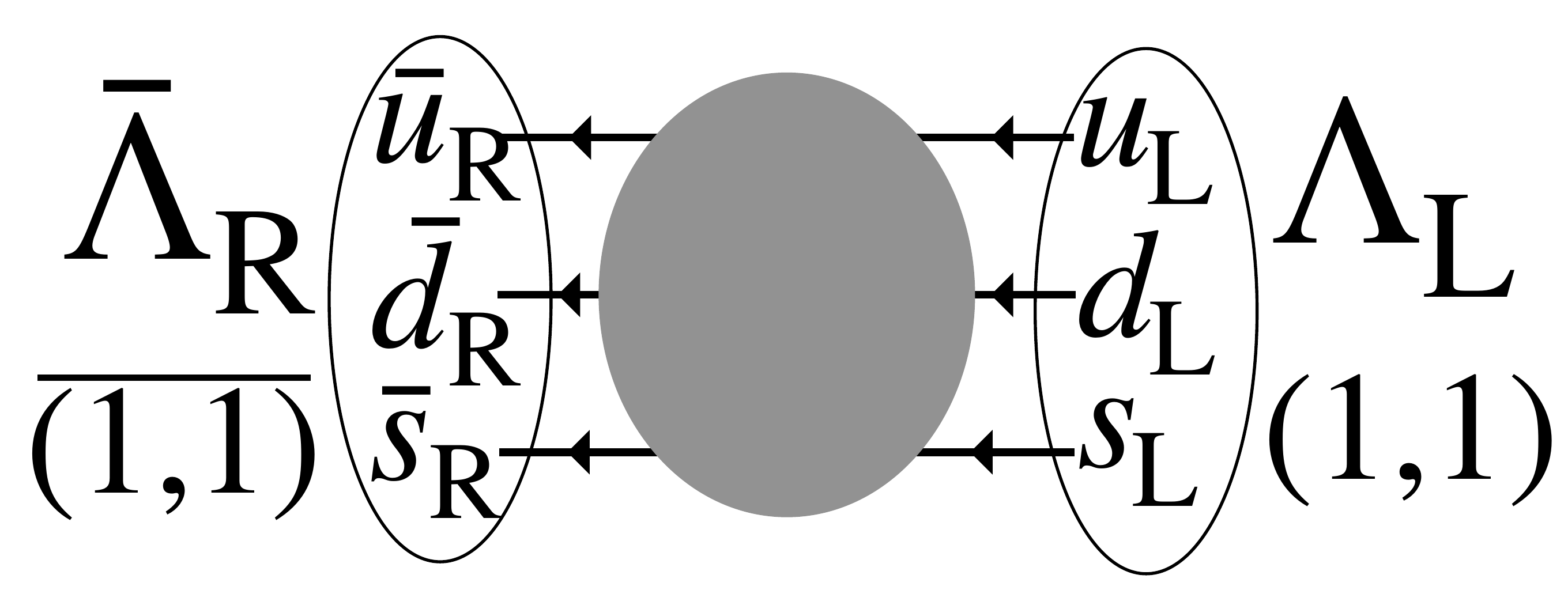}
\caption{Anomalous interaction between the chiral singlet baryons $\Lambda\sim(1,1)$. }
\label{fig-2nd-singlet}
\end{figure}
In this paper, 
we omit the contribution from the flavor singlet $\Lambda$ baryons, \blueflag{$\Lambda_0$}.
In the quark model, 
wave functions of three quarks in a flavor singlet baryon are totally antisymmetric in the flavor space as well as in the color space.
Since the spin wave functions cannot be totally antisymmetric, the space part of the wave functions should be in the excited level.
This implies that $\Lambda_0$ cannot be a groundstate.

In the present model a pair of $\Lambda_0$s 
is included in the $(3,\bar{3})$ and $(\bar{3},3)$ representations, which may mix with some $\Lambda$ baryons of the octet members when the flavor symmetry breaking is included.
However, we note that $\Lambda_0$ 
baryon appears from chiral singlet $\Lambda$ baryons of $(1,1)$ representation, 
for which there exists a chiral symmetic mass term given by
\begin{align}
-m_\Lambda(
\bar\Lambda^{(1,1)}_\rl\Lambda^{(1,1)}_\rr
+\bar\Lambda^{(1,1)}_\rr\Lambda^{(1,1)}_\rl
)\ ,
\end{align}
corresponding to the quark diagram including $\UA{1}$ anomaly as in Fig.\ref{fig-2nd-singlet}. 
The $\Lambda$ baryon of $(1,1)$ representation can be made heavy even before the spontaneous chiral symmetry breaking.
When the chiral symmetry is spontaneously broken, this mixes with the flavor singlet $\Lambda$ baryon belonging to $(3,\bar{3})$ and $(\bar{3},3)$ representations.
Thus, we naturally expect that the flavor singlet $\Lambda$ baryons are heavier than the flavor-octet $\Lambda$ baryons.

\section{Numerical fit to mass spectra } 
\label{sec-numerical}

\subsection{ Model } 

The entire Lagrangian which we use in this work consists of the following sectors: 
\begin{align}
\lag_\mathrm{total}&=
\lag_\mathrm{kin}
+\lag_\mathrm{CIM}
+\lag_\mathrm{Yukawa}
+\lag_\mathrm{2nd}\,. 
\end{align}
The kinetic term is just the ordinal one for $\psi$, $\psi^\mir$, $\chi$, and $\chi^\mir$. 
The chiral invariant mass terms are expressed as 
\begin{align}
\lag_\mathrm{CIM}=
-m_0(\bar\psi\gamma_5\psi^\mir) -m_0(\bar\chi\gamma_5\chi^\mir) +\text{h.c.}\,, 
\end{align}
where we suppose the chiral invariant masses for $\psi$ and $\chi$ are the same for simplicity. 

The first-order Yukawa interactions are given by 
\begin{align}
&\lag_\mathrm{Yukawa}= \notag \\
&-g_1^\mathrm{a}\qty[-
\varepsilon_{r_1r_2r_3}\varepsilon^{l_1l_2l_3}
(\bar\psi_\rl)^{r_1}_{l_1}(M^\dag)^{r_2}_{l_2}(\psi_\rr)^{r_3}_{l_3}
+\mathrm{h.c.}] \notag \\
&-g_2^\mathrm{a}\qty[-
\varepsilon_{l_1l_2l_3}\varepsilon^{r_1r_2r_3}
(\bar\psi_\rl^\mir)^{l_1}_{r_1}(M)^{l_2}_{r_2}(\psi_\rr^\mir)^{l_3}_{r_3}
+\mathrm{h.c.}] \notag \\
&-g_1^\mathrm{s}\qty[\tr(\bar\psi_\rl M\chi_\rr+\bar\psi_\rr M^\dag\chi_\rl+\mathrm{h.c.})] \notag \\
&-g_2^\mathrm{s}\qty[\tr(\bar\psi_\rl^\mir M^\dag\chi_\rr^\mir+\bar\psi_\rr^\mir M\chi_\rl^\mir+\mathrm{h.c.})]\,. 
\end{align}
The second-order terms introduced in the previous section are summarized as
\begin{align}
\lag_\mathrm{2nd}=
-\frac{g_1^\mathrm{d}}{f_\pi}\big[
&\tr(\bar\psi_\rl\chi_\rr^\mir\hat{O}^\dag)
-\tr(\bar\psi_\rr\chi_\rl^\mir\hat{O})+\mathrm{h.c.}\big] \notag \\
-\frac{g_2^\mathrm{d}}{f_\pi}\big[
&\tr(\bar\chi_\rl\psi_\rr^\mir\hat{O})
-\tr(\bar\chi_\rr\psi_\rl^\mir\hat{O}^\dag)+\mathrm{h.c.}\big] \notag \\
-\frac{h_1}{f_\pi}\big\{
&\tr(\bar\psi_\rl MM^\dag\psi_\rr^\mir)-\tr(\bar\psi_\rl M)\tr(M^\dag\psi_\rr^\mir)  \notag\\
-&\tr(\bar\psi_\rr M^\dag M\psi_\rl^\mir)+\tr(\bar\psi_\rr M^\dag)\tr(M\psi_\rl^\mir)\notag\\
+&\mathrm{h.c.}\big\} \notag \\
-\frac{h_2}{f_\pi}\big\{
&\tr[\bar\psi_\rl\psi_\rr^\mir(\tr(MM^\dag)-M^\dag M)]\notag\\
-&\tr[\bar\psi_\rr\psi_\rl^\mir(\tr(MM^\dag)-MM^\dag)]\notag\\
+&\mathrm{h.c.}\big\} \notag \\
-\frac{h_3}{f_\pi}\big\{
&\tr[\bar\chi_\rl\chi_\rr^\mir(\tr(M^\dag M)-MM^\dag)]\notag\\
-&\tr[\bar\chi_\rr\chi_\rl^\mir(\tr(M^\dag M)-M^\dag M)]\notag\\
+&\mathrm{h.c.}\big\} \notag \\
-\frac{h_4}{f_\pi}\big\{
&-\tr(\bar\chi_\rl MM^\dag\chi_\rr^\mir)\notag\\
&+\tr[\bar\chi_\rl\chi_\rr^\mir(\tr(MM^\dag)-MM^\dag)]\notag\\
&+\tr(\bar\chi_\rr M^\dag M\chi_\rl^\mir)\notag \notag \\
&-\tr[\bar\chi_\rr\chi_\rl^\mir(\tr(M^\dag M)-M^\dag M)]\notag\\
&+\mathrm{h.c.}\big\}\ ,
\end{align}
with $\hat{O}$ defined in Eq.~(\ref{def O op}).
We take the mean field approximation for the scalar meson 
$\ev{M}=\diag{\alpha,\beta,\gamma}$, assuming the isospin symmetry $\alpha=\beta$.
It is convenient to introduce a unified notation for the chiral representations of baryons as 
$\Psi_i=(\psi_i,\chi_i,\gamma_5\psi^\mir_i,\gamma_5\chi^\mir_i)^T$ ($i = N,\Lambda,\Sigma,\Xi$).
By using this, mass terms of baryons are written as
\begin{align}
\tilde\lag=&-\sum_{i=N,\Lambda,\Sigma,\Xi}\bar{\Psi}_i\hat{M}_i\Psi_i \ , 
\end{align}
where the mass matrices $\hat{M}_i$  ($i = N,\Lambda,\Sigma,\Xi$) is defined as 
\begin{align}
&\hat{M}(
x^\mathrm{a},
x^\mathrm{s},
x^\mathrm{d},
x^\mathrm{h1},
x^\mathrm{h23},
x^\mathrm{h4}
)
\equiv\notag\\
&\begin{pmatrix}
g^\mathrm{a}_1x^\mathrm{a} & g^\mathrm{s}_1x^\mathrm{s} & 
m_0+\frac{h_1}{f_\pi}x^\mathrm{h1}+\frac{h_2}{f_\pi}x^\mathrm{h23} & 
\frac{g^\mathrm{d}_1}{f_\pi}x^\mathrm{d} \\
 & 0 & \frac{g^\mathrm{d}_2}{f_\pi}x^\mathrm{d} & 
m_0+\frac{h_3}{f_\pi}x^\mathrm{h23}+\frac{h_4}{f_\pi}x^\mathrm{h4} \\
 & & g^\mathrm{a}_2x^\mathrm{a} & g^\mathrm{s}_2x^\mathrm{s} \\
 & & & 0 \\
\end{pmatrix}
\label{mass matrices}
\end{align}
with $x^a$, $\cdots$, $x^{h_4}$ defined in Table~\ref{tab-mass-matrices}.
We note that the matrix $\hat{M}$ is symmetric, 
so we omitted some components in Eq.~(\ref{mass matrices}).

Diagonalizing the $4\times4$ matrix $\hat{M}_i$ in Eq.~(\ref{mass matrices}), 
we obtain four mass eigenvalues, $m_i^{\rm g.s.}$, $m_i^{(1)}$, $-m_i^{(2)}$ and $-m_i^{(3)}$, 
and the corresponding mass eigenstates, $B_i^{\rm g.s.}$, $B_i^{(1)}$, $\gamma_5 B_i^{(2)}$ and $\gamma_5 B_i^{(3)}$, 
where $B_i^{(2)}$ and $B_i^{(3)}$ are negative parity baryons.
As a result, the mass term is rewritten as 
\begin{align}
\tilde{\mathcal L}
=&-\sum_i\big[
m_i^\mathrm{g.s.}\bar{B}^\mathrm{g.s.}_iB^\mathrm{g.s.}_i
+m_i^{(1)}\bar{B}^{(1)}_iB^{(1)}_i\notag\\
&\quad+m_i^{(2)}\bar{B}^{(2)}_iB^{(2)}_i
+m_i^{(3)}\bar{B}^{(3)}_iB^{(3)}_i
\big] \ .
\end{align}
\begin{table*}
\caption{
Mass matrices for the nucleons, 
the $\Sigma$ baryons, 
the $\Xi$ baryons, 
and the $\Lambda$ baryons. 
See Tab.\ref{tab-condensate-input} for $\alpha$ and $\gamma$. 
}
\label{tab-mass-matrices}
\begin{tabular}{ccccccccccccccc}
\hline\hline
&&& $g^\mathrm{a}_{1,2}$ && $g^\mathrm{s}_{1,2}$ && $g^\mathrm{d}_{1,2}$ && 
$h_1$ && $h_2,h_3$ && $h_4$ & \\
\hline
$M_N$ & $=$ & 
$M($ & $\alpha$ &,& $\alpha$ &,& $2\alpha^2$ &,& 
$\alpha^2$ &,& $2\alpha^2$ &,& $\alpha^2$ & $)$ \\
$M_\Sigma$ & $=$ & 
$M($ & $\gamma$ &,& $\alpha$ &,& $2\alpha\gamma$ &,& 
$\alpha^2$ &,& $\alpha^2+\gamma^2$ &,& $\gamma^2$ & $)$ \\
$M_\Xi$ & $=$ & 
$M($ & $\alpha$ &,& $\gamma$ &,& $2\alpha\gamma$ &,& 
$\gamma^2$ &,& $\alpha^2+\gamma^2$ &,& $\alpha^2$ & $)$ \\
$M_\Lambda$ & $=$ & 
$M($ & $\frac{4\alpha-\gamma}{3}$ &,& $\frac{\alpha+2\gamma}{3}$ &,& 
$2\frac{\alpha\gamma+2\alpha^2}{3}$ &,&
$\frac{4\alpha\gamma-\alpha^2}{3}$ &,&
$\frac{5\alpha^2+\gamma^2}{3}$ &,& 
$\frac{4\alpha^2-\gamma^2}{3}$ & $)$ \\
\hline\hline
\end{tabular}
\end{table*}

\subsection{{\GO} for mass matrices}\label{sec-GO-matrix}

As seen in Sec.\ref{sec-2nd-allpossibilities}, 
all interactions except the $h_1$ term satisfy the {\GO}. 
The breaking term is proportional to $\epsilon^2$, 
where $\epsilon$ is defined in the VEV of the meson field as 
$\ev{M}=\diag{\alpha,\alpha,\alpha+\epsilon}$. 
Therefore, assuming $\epsilon\ll\alpha$, 
the Gell-Mann--Okubo mass relations among the mass matrices 
for $N$, $\Lambda$, $\Sigma$, and $\Xi$ are approximately satisfied as
\begin{align}\label{eq-GO-matrix}
\frac{\hat{M}_N+\hat{M}_\Xi}{2}
-\frac{3\hat{M}_\Lambda+\hat{M}_\Sigma}{4}
=\mathcal{O}\left( \left(\epsilon/\alpha\right)^2\right)\ . 
\end{align}
Let $\hat{D}_i$ be 
the diagonalized matrices of $\hat{M}_i$, 
and let $\hat{U}_N$ be the unitary matrix which diagonalizes $\hat{M}_N$. 
Then the perturbations for $\epsilon$ of the mass eigenvalues are 
\begin{align}
\hat{D}_f&=\hat{D}_N
+\mbox{diag}\left[ \hat{U}_N^\dag \frac{d\hat{M}_f}{d\epsilon}\bigg|_{\epsilon=0}\hat{U}_N\right]\epsilon
+\mathcal{O}(\epsilon^2)
\end{align}
where $f=\Lambda,\Sigma,\Xi$, and
$\mbox{diag}\big[ \hat{X} \big]$ is the diagonal part of $\hat{X}$. 
Therefore, Eq.\eqref{eq-GO-matrix} implies that the mass eigenvalues in this model 
can satisfy the {\GO} for small $\epsilon$ as
\begin{align}
\frac{\hat{D}_N+\hat{D}_\Xi}{2}
-\frac{3\hat{D}_\Lambda+\hat{D}_\Sigma}{4}
=\mathcal{O}\left( \left(\epsilon/\alpha\right)^2\right)\ . 
\end{align}
This argument implies that the {\GO} is satisfied for small $\epsilon/\alpha$ in this model.
However, as seen in the previous sections, it is rather difficult to reproduce the mass ordering for hyperons. 

\subsection{Traces of mass matrices}

\begin{table}
\caption{
Physical inputs of the decay constants for pion and kaon
\cite{Workman:2022ynf},
and the VEV of the meson field $\ev{M}=\diag{\alpha,\beta,\gamma}$ 
with assuming isospin symmetry $\alpha=\beta$. 
}
\label{tab-condensate-input}
\begin{tabular}{c|c}
\hline\hline
$f_\pi$ & 93 MeV \\
$f_K$ & 110 MeV \\
$\alpha$ & $f_\pi(=93\,\mathrm{MeV})$ \\
$\gamma$ & $2f_K-f_\pi(=127\,\mathrm{MeV})$ \\
\hline\hline
\end{tabular}
\end{table}

In this section, we would like to note that there are non-trivial relations among traces of the mass matrices.
The explicit forms of traces are shown 
as follows:\footnote{
We note that these matrices satisfy the Gell-Mann--Okubo mass relation as
\begin{align}
\frac{{\tr}(\hat{M}_N)+{\tr}(\hat{M}_\Xi)}{2}
=\frac{3{\tr}(\hat{M}_\Lambda)+{\tr}(\hat{M}_\Sigma)}{4}\,. 
\end{align}
}
\begin{align}
\hspace{-0.5cm}
{\tr}(\hat{M}_N)&=(g_1^\mathrm{a}+g_2^\mathrm{a})\alpha \ , \\
{\tr}(\hat{M}_\Sigma)&=(g_1^\mathrm{a}+g_2^\mathrm{a})\gamma
={\tr}(\hat{M}_N)\frac{\gamma}{\alpha} \ , \\
{\tr}(\hat{M}_\Xi)&=(g_1^\mathrm{a}+g_2^\mathrm{a})\alpha
={\tr}(\hat{M}_N) \ , \\
{\tr}(\hat{M}_\Lambda)&=(g_1^\mathrm{a}+g_2^\mathrm{a})\frac{4\alpha-\gamma}{3}
={\tr}(\hat{M}_N)\frac{(4\alpha-\gamma)/3}{\alpha} \ .
\end{align}
\blueflag{ 
}
%
We determine the VEVs of the meson field $M$ from the decay constants of pion and kaon as
\begin{align}
\alpha = f_\pi \ , \quad \gamma = 2 f_K - f_\pi \ .
\end{align}
In Table~\ref{tab-condensate-input}, 
the input values of $f_\pi$ and $f_K$ are shown together with the determined values of $\alpha$ and $\gamma$.
As for the baryon masses, we use the values listed in Table~\ref{tab-mass-inputs} picked up from the PDG table~\cite{Workman:2022ynf}.
\begin{table*}
\caption{
Physical inputs for the four octet masses. 
}
\label{tab-mass-inputs}
\begin{tabular}{c||c|c|c|c}
 & \multicolumn{4}{c||}{Mass inputs for octet members [MeV]} \\
\hline\hline
$J^P$ & $N$ & $\Lambda$ & $\Sigma$ & $\Xi$  \\
\hline
$m_1: 1/2^+$(G.S.) & 
$N(939)$: $939$ & 
$\Lambda(1116)$: $1116$ & 
$\Sigma(1193)$: $1193$ & 
$\Xi(1318)$: $1318$  \\
$m_2: 1/2^+$ & 
$N(1440)$: $1440$ & 
$\Lambda(1600)$: $1600$ & 
$\Sigma(1660)$: $1660$ & 
$\Xi(?)$:   \\
$m_3: 1/2^-$ & 
$N(1535)$: $1530$ & 
$\Lambda(1670)$: $1674$ & 
$\Sigma(?)$:  & 
$\Xi(?)$:   \\
$m_4: 1/2^-$ & 
$N(1650)$: $1650$ & 
$\Lambda(1800)$: $1800$ & 
$\Sigma(1750)$: $1750$ & 
$\Xi(?)$:  \\
\hline
\end{tabular}
%
%
%
%
%
%
\end{table*}
%
The value of the $\mbox{tr} \big[ \hat{M}_N \big] $ is determined as
\begin{align}
{\tr}(\hat{M}_N)&=(939+1440-1530-1650)\,\mathrm{MeV}\notag\\
&=-801\,\mathrm{MeV}\,. 
\end{align}
Then, the trace values for the other flavors are also determined as the following:
\begin{align}
{\tr}(\hat{M}_\Sigma)
&={\tr}(\hat{M}_N)\frac{\gamma}{\alpha} =1094\,\mathrm{MeV} \ , 
\label{trace Sigma}\\
{\tr}(\hat{M}_\Xi)
&=-801\,\mathrm{MeV} \ , \\
{\tr}(\hat{M}_\Lambda)\label{eq-trace-Lambda}
&={\tr}(\hat{M}_N)\frac{(4\alpha-\gamma)/3}{\alpha}  =-703\,\mathrm{MeV}\ . 
\end{align}

We note that not all of four masses in a given baryon flavor, except $N$ and $\Lambda$, 
are well-established as can be seen from Table~\ref{tab-mass-inputs}.
The trace $\tr(\hat{M}_\Lambda)$ with the experimental values 
%
%
%
\begin{align}
(1116+1600-1674-1800)\,\mathrm{MeV}=-758\,\mathrm{MeV}\ .
\end{align}
This value is close to the value of $-703\,\mathrm{MeV}$ in Eq.~\eqref{eq-trace-Lambda}, with $\tr(\hat{M}_N)$ and $\langle M \rangle$ as inputs. 
The agreement seems reasonably good.
Actually the octet $\Lambda$ mass has ambiguities related to the mixing with the singlet,
so the saturation of the equality including only the octet $\Lambda$ implies the mixing between the singlet and octet is not very large.

On the other hand, 
the trace of $\Sigma$ and $\Xi$ masses is not established experimentally. 
Hence we need extra discussions about the usage of the trace formula.
This is presented in the next section.
%


\subsection{Numerical results}
\label{sec: numerical}

In this subsection, we numerically fit the model parameters to known mass spectra of light baryons, and also give some predictions.

Using twelve mass values in Table~\ref{tab-mass-inputs}, we fit the ten Yukawa couplings 
$g_1^\mathrm{a}$, $g_2^\mathrm{a}$, $g_1^\mathrm{s}$, $g_2^\mathrm{s}$, $g_1^\mathrm{d}$, $g_2^\mathrm{d}$, $h_1$, $h_2$, $h_3$ and $h_4$ by 
minimizing the following function:
\begin{align}
f_\mathrm{min}&=
\sum_{i=1}^{12}\left(
\frac{m_i^\mathrm{theory}-m_i^\mathrm{input}}{\delta m_i}
\right)^2\ , 
\end{align}
where errors $\delta m_i$ are taken as
$\delta m_i=10\,\mathrm{MeV}$ for the ground-state baryons and $\delta m_i=100\,\mathrm{MeV}$ for the excited baryons.
The difference in $\delta m_i$ is used since the masses for excited states generally contain more errors.

We select certain sets of parameters which provide reasonably good fit satisfying $f_\mathrm{min}<1$.
Since there still ramain many sets of parameters, we further restrict parameters by requiring
\begin{align}
\sum_i \left\vert \Delta_{\mathrm{GO},i} \right\vert <100\,\mathrm{MeV} \ ,
\end{align}
where
\begin{align}
\Delta_{\mathrm{GO},i}\equiv\frac{m_i[N]+m_i[\Xi]}{2}-\frac{3m_i[\Lambda]+m_i[\Sigma]}{4} \ ,
\end{align}
with $i$ indicating the octet members.
\begin{figure*}\centering
\includegraphics[width=0.7\hsize]{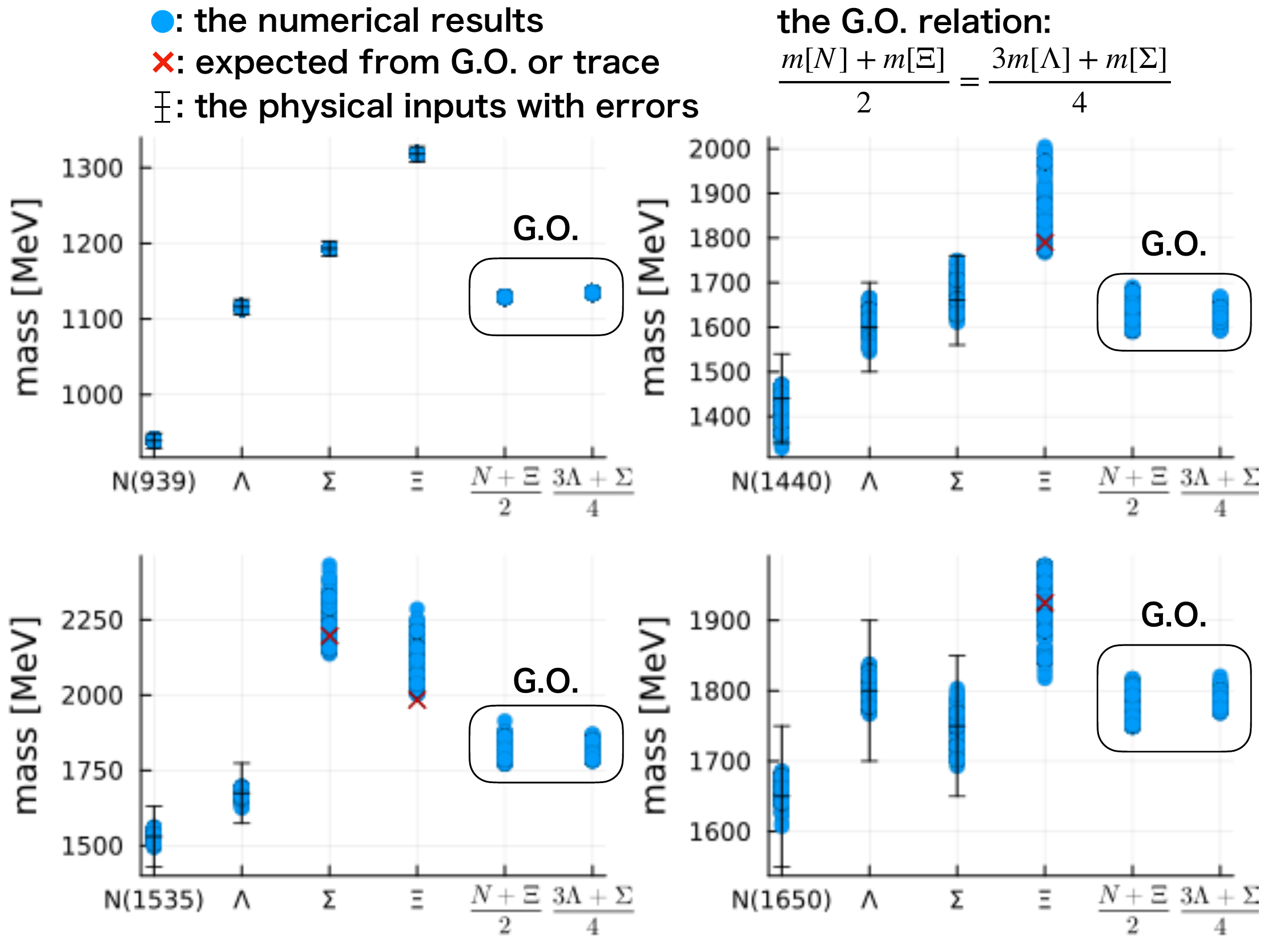}
\caption{
Numerical results of the four octet masses for 
the chiral invariant mass $m_0=800\,\mathrm{MeV}$. 
They almost satisfy the {\GO} ($\Delta_\mathrm{GO}<100\,\mathrm{MeV}$) 
and well reproduce the physical inputs. 
For other values of $m_0$, there are solutions which satisfy the same conditions. 
In this model, the $\Sigma$ baryon in the octet member of $N(1535)$ 
becomes heavier than the others. 
}
\label{fig-massoutput}
\end{figure*}
We summarize the results of fitting in Fig.\ref{fig-massoutput} by showing masses of baryons together with the $\Delta_{\mathrm{GO},i}$.
In this figure, black lines with error bars show the inputs listed in Table~\ref{tab-mass-inputs}, 
and blue points show the best fitted values of masses and $\Delta_{\mathrm{GO},i}$.

Figure~\ref{fig-massoutput} shows that positive baryons with the ground and first excited states are reproduced well.
Meanwhile, in the negative parity channel, 
the mass hierarchy between $\Sigma $ and $\Xi$ states look at odd with the counting based on the strange quark mass.
Although this hierarchy is not rejected by the experiments, we regard it as a problem from the view of naturalness.
\footnote{
We note that there is a two-fold ambiguity in identifying $\Xi$ baryons with negative parity, which we call $\Xi(-)$ below. 
For some parameter choice, we found two solutions in which one of $\Xi(-)$ is identified as an octet-member with $N(1535)$ and another with $N(1650)$.  
In Fig.~\ref{fig-massoutput}, we show one solution which provides  smaller value of $\vert \Delta_{{\rm GO},i} \vert$.
}

%
%
%

\begin{figure*}\centering
\includegraphics[width=0.9\hsize]{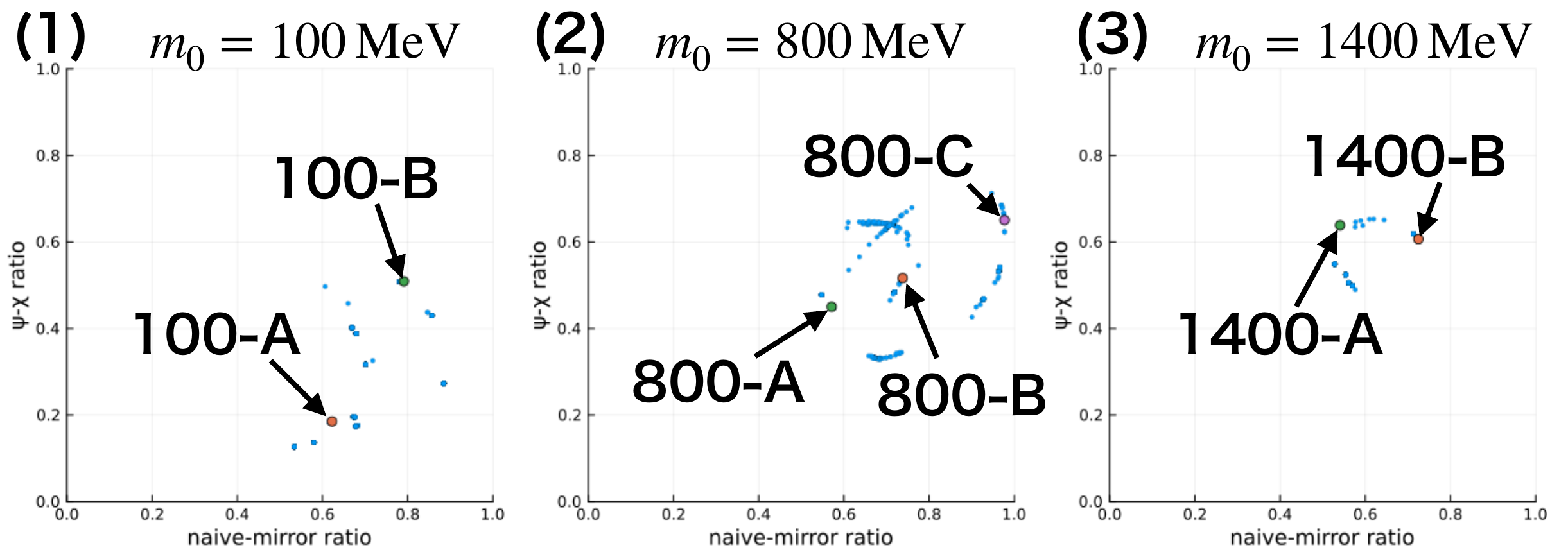}
\caption[]{
%
The some solutions for 
(1) $m_0=100\,\mathrm{MeV}$, 
(2) $m_0=800\,\mathrm{MeV}$, 
and (3) $m_0=1400\,\mathrm{MeV}$. 
The horizontal axis is naive-mirror ratio of $N(939)$. 
0 indicates that all are mirror ($\psi^\mir$ or $\chi^\mir$), 
and 1 indicates that all are naive ($\psi$ or $\chi$). 
The vertical axis is $\psi$-$\chi$ ratio of $N(939)$. 
0 indicates that all are $\chi$ or $\chi^\mir$, 
and 1 indicates that all are $\psi$ or $\psi^\mir$. 
There are seven samples: 
100-A and 100-B in the panel (1), 
800-A, 800-B, and 800-C in the panel (2), 
and 1400-A and 1400-B in the panel (3), 
which are shown in Tab.\ref{tab-samples}. 
}
\label{fig-ratio}
\end{figure*}

We study the mixing structure of the ground-state nucleon, $N(939)$.
In the present analysis, the mass eigenstate for $N(939)$ is expressed as
\begin{align}
B_N^\mathrm{g.s.}
=c_\psi\psi
+c_\chi\chi
+c_\psi^\mir\psi^\mir
+c_\chi^\mir\chi^\mir, 
\end{align}
where 
$c_\psi$, $c_\chi$, $c_\psi^\mir$ and $c_\chi^\mir$ show the ratio of each wave function included in the ground-state nucleon with the normalization of 
$|c_\psi|^2+|c_\chi|+|c_\psi^\mir|^2+|c_\chi^\mir|^2=1$.
For clarifying the mixing structure, we define the ``naive-mirror ratio'' as
\begin{align}
|c_\psi|^2+|c_\chi|^2(=1-|c_\psi^\mir|^2-|c_\chi^\mir|^2)\ , 
\end{align}
and ``$\psi$-$\chi$ ratio'' as 
\begin{align}
|c_\psi|^2+|c_\psi^\mir|^2(=1-|c_\chi|^2-|c_\chi^\mir|^2)\ ,
\end{align}
The results of mixing structure are summarized in Fig.\ref{fig-ratio}
and the couplings used are summarized in Table.\ref{tab-samples}.
It is remarkable that, in most cases, the ``$\psi$-$\chi$ ratio'' is around 50\%, 
which implies that the ground-state nucleon is provided by the maximally mixed state of 
$(3_L\,,\,\bar{3}_R)+(\bar{3}_L\,,\,3_R)$ and $(1_L\,,\,8_R)+(8_L\,,\,1_R)$ representations.
We still note that, for $m_0 = 100$\,MeV, there are some solutions for which $(1_L\,,\,8_R)+(8_L\,,\,1_R)$ representation is dominated.

%
%
%
%
%

\begin{table*}
\caption{
Some sample solutions, 
which correspond to the indicated points 
in Fig.\ref{fig-ratio}. 
$g_{1,2}^\mathrm{s}$ or $g_{1,2}^\mathrm{d}$ 
have values of around 5-10, which is not small. 
This implies the mixing between $\psi^{(\mir)}$ and $\chi^{(\mir)}$ is not small. 
}
\label{tab-samples}
\begin{tabular}{l||cc|ccc|cc}
	& 100-A & 100-B & 800-A & 800-B & 800-C & 1400-A & 1400-B\\
	\hline\hline
	$m_0$ [MeV] & 100 & 100 & 800 & 800 & 800 & 1400 & 1400 \\
	\hline
	$g_1^\mathrm{a}$ ($\psi$-$\psi$) & -12.63 & -8.77 & -3.98 & -8.7 & 3.69 & -1.86 & -3.59 \\
	$g_2^\mathrm{a}$ ($\psi^\mir$-$\psi^\mir$) & 3.35 & -0.04 & -5.26 & -0.75 & -12.46 & -7.13 & -5.5 \\
	$g_1^\mathrm{s}$ ($\psi$-$\chi$) & 7.57 & 9.56 & 2.12 & 7.95 & 7.44 & 5.79 & 0.57 \\
	$g_2^\mathrm{s}$ ($\psi^\mir$-$\chi^\mir$) & -10.97 & -12.44 & 3.78 & 11.19 & -8.95 & 6.88 & 0.82 \\
	$g_1^\mathrm{d}$ ($\psi$-$\chi^\mir$) & -3.83 & 0.01 & 7.09 & -1.04 & -4.78 & -6.14 & -6.58 \\
	$g_2^\mathrm{d}$ ($\chi$-$\psi^\mir$) & -5.1 & 0.66 & -6.17 & -1.47 & -4.15 & 6.19 & 6.89 \\
	$h_1$ ($\psi$-$\psi^\mir$) & 5.88 & -3.38 & 7.35 & 1.33 & 1.17 & 2.35 & 4.25 \\
	$h_2$ ($\psi$-$\psi^\mir$) & -3.73 & 5.61 & -5.64 & -9.51 & -4.01 & -7.24 & -7.39 \\
	$h_3$ ($\chi$-$\chi^\mir$) & -1.6 & 3.56 & -1.65 & 1.51 & 0.78 & -2.42 & -2.36 \\
	$h_4$ ($\chi$-$\chi^\mir$) & -2.85 & 1.4 & 0.33 & -0.18 & -0.53 & -3.7 & -2.71 \\
	\hline
	naive ratio of $N(939)$ & 0.62 & 0.79 & 0.57 & 0.74 & 0.98 & 0.72 & 0.54 \\
	$\psi$ ratio of $N(939)$ & 0.19 & 0.51 & 0.45 & 0.52 & 0.65 & 0.61 & 0.64 \\
\end{tabular}
\end{table*}




\section{Summary and Discussion}\label{sec-summary}

We proposed a systematic way to construct 
models of baryons based on the chiral U(3)$_L\times$U(3)$_R$ symmetry.
The symmetry constraints are far stronger than in models assuming only the SU(3) flavor symmetry,
and the chiral Yukawa interactions appear in very specific ways.
We assume that chiral representations $(3_L\,,\,6_R)$ and $(6_L\,,\,3_R)$ with a bad diquark 
 are heavier than $(3_L\,,,\,\bar{3}_R)$, $(\bar{3}_L\,,\,3_R)$, $(1_L\,,\,8_R)$ and $(8_L\,,\,1_R)$.
We showed that the inclusion of the first order Yukawa interactions for the four representations 
satisfies the Gell-Mann--Okubo mass relation, but cannot reproduce the mass ordering of octet members of the ground states;
the quark graphs convincingly explain why, at the first order, the strange quark mass does not appear to reproduce the correct mass ordering.
Then, we expanded our systematic analyses to the second-order Yukawa interactions,
and showed that the mass ordering problem is cured for the ground state of positive parity baryons.
The state is found to be a maximally mixed state of $(3_L\,,\,\bar{3}_R)+(\bar{3}_L\,,\,3_R)$ and $(1_L\,,\,8_R)+(8_L\,,\,1_R)$ representations.
The results imply that the quark diagrams are very useful to constrain the possible types of Yukawa interactions.

In the present analyses up to the second order,
while the mass ordering in the positive parity ground state is reproduced correctly,
in the negative parity we found the unnatural mass ordering of the ground state;
the mass of $\Sigma$ including a single strange quark is heavier than $\Xi$ with two strange quarks.
Although such ordering is not fully excluded because these two negative parity states have not been confirmed experimentally, 
we feel unlikely that $\Sigma$ is heavier than $\Xi$.
After extensive parameter searches, we have reached somewhat unexpected conclusion that the second order Yukawa interactions are not sufficient.
This sort of difficulties has not been manifest within the analyses for two-flavor models.
Further studies are mandatory.

This work is partially motivated from the hope to saturate the $U(3)_\rl\times U(3)_\rr$ dynamics of baryons
within a few chiral representations, as done for mesons $(\pi,\sigma, \rho, a_1)$ by Weinberg.
We introduced a hierarchy based on good and bad diquarks to pick up chiral representations which are supposed to be important,
but our analyses indicate the necessity to include at least the second order of Yukawa interactions;
the descriptions based on $U(3)_\rl\times U(3)_\rr$ chiral representations are 
much more involved than those requiring only the $SU(3)$-flavor symmetry.
Clearly our analyses need improvement.
Several possibilities are in order:

i) It is possible that the classification of chiral representations based on good and bad diquarks is not very effective.
If this is the case, we need to explicitly include several additional chiral representations, 
such as $(3_L\,,\,6_R)$ and $(6_L\,,\,3_R)$.
The necessity to include baryons with bad diquarks raises questions whether we should manifestly include the decuplet baryons such as $\Delta$
and the interactions with the octet baryons.
This would greatly increase the number of couplings at the tree level.
On the other hand, it is possible that including massive resonances at the leading order
reduces the importance of Yukawa interactions at higher orders.

ii) Another possibility is that the linear realization, even after superposing many chiral representations, 
is not sufficient to explain baryons in vacuum.
If we are indeed required to include infinite number of the Nambu-Goldstone (NG) bosons around baryons,
the non-linear realization is a more natural choice for baryons in vacuum, 
although the description near the chiral restoration should become more complicated.

iii) The extreme limit of infinite number of NG bosons around a baryon
leads to the description of a baryon in the chiral soliton models.
Here a coherent pion cloud represents the baryon charge at the core
in the same way as electric fields around an electron allow us to infer the existence of the electric charge.
If including many NG bosons are indeed essential, 
the physical baryons would be hardly saturated by a few chiral representations.

\begin{acknowledgments}
The work of T.M., B.G., and M.H. was supported in part by JSPS KAKENHI Grant No. 20K03927. 
T.M. and B.G. were also supported by JST SPRING, Grant No. JPMJSP2125.  T.M. and B.G. would like to take this opportunity to thank the
“Interdisciplinary Frontier Next-Generation Researcher Program of the Tokai
Higher Education and Research System.”
T.K. was supported by JSPS KAKENHI Grant No. 23K03377 and by the Graduate Program on Physics for the Universe (GP-PU) at Tohoku university.
\end{acknowledgments}

\bibliography{ref_3fPDM_2022.bib}

\begin{thebibliography}{40}%
\makeatletter
\providecommand \@ifxundefined [1]{%
 \@ifx{#1\undefined}
}%
\providecommand \@ifnum [1]{%
 \ifnum #1\expandafter \@firstoftwo
 \else \expandafter \@secondoftwo
 \fi
}%
\providecommand \@ifx [1]{%
 \ifx #1\expandafter \@firstoftwo
 \else \expandafter \@secondoftwo
 \fi
}%
\providecommand \natexlab [1]{#1}%
\providecommand \enquote  [1]{``#1''}%
\providecommand \bibnamefont  [1]{#1}%
\providecommand \bibfnamefont [1]{#1}%
\providecommand \citenamefont [1]{#1}%
\providecommand \href@noop [0]{\@secondoftwo}%
\providecommand \href [0]{\begingroup \@sanitize@url \@href}%
\providecommand \@href[1]{\@@startlink{#1}\@@href}%
\providecommand \@@href[1]{\endgroup#1\@@endlink}%
\providecommand \@sanitize@url [0]{\catcode `\\12\catcode `\$12\catcode
  `\&12\catcode `\#12\catcode `\^12\catcode `\_12\catcode `\%12\relax}%
\providecommand \@@startlink[1]{}%
\providecommand \@@endlink[0]{}%
\providecommand \url  [0]{\begingroup\@sanitize@url \@url }%
\providecommand \@url [1]{\endgroup\@href {#1}{\urlprefix }}%
\providecommand \urlprefix  [0]{URL }%
\providecommand \Eprint [0]{\href }%
\providecommand \doibase [0]{https://doi.org/}%
\providecommand \selectlanguage [0]{\@gobble}%
\providecommand \bibinfo  [0]{\@secondoftwo}%
\providecommand \bibfield  [0]{\@secondoftwo}%
\providecommand \translation [1]{[#1]}%
\providecommand \BibitemOpen [0]{}%
\providecommand \bibitemStop [0]{}%
\providecommand \bibitemNoStop [0]{.\EOS\space}%
\providecommand \EOS [0]{\spacefactor3000\relax}%
\providecommand \BibitemShut  [1]{\csname bibitem#1\endcsname}%
\let\auto@bib@innerbib\@empty
\bibitem [{\citenamefont {Nambu}\ and\ \citenamefont
  {Jona-Lasinio}(1961{\natexlab{a}})}]{Nambu:1961tp}%
  \BibitemOpen
  \bibfield  {author} {\bibinfo {author} {\bibfnamefont {Y.}~\bibnamefont
  {Nambu}}\ and\ \bibinfo {author} {\bibfnamefont {G.}~\bibnamefont
  {Jona-Lasinio}},\ }\bibfield  {title} {\bibinfo {title} {{Dynamical Model of
  Elementary Particles Based on an Analogy with Superconductivity. 1.}},\
  }\href {https://doi.org/10.1103/PhysRev.122.345} {\bibfield  {journal}
  {\bibinfo  {journal} {Phys. Rev.}\ }\textbf {\bibinfo {volume} {122}},\
  \bibinfo {pages} {345} (\bibinfo {year} {1961}{\natexlab{a}})}\BibitemShut
  {NoStop}%
\bibitem [{\citenamefont {Nambu}\ and\ \citenamefont
  {Jona-Lasinio}(1961{\natexlab{b}})}]{Nambu:1961fr}%
  \BibitemOpen
  \bibfield  {author} {\bibinfo {author} {\bibfnamefont {Y.}~\bibnamefont
  {Nambu}}\ and\ \bibinfo {author} {\bibfnamefont {G.}~\bibnamefont
  {Jona-Lasinio}},\ }\bibfield  {title} {\bibinfo {title} {{Dynamical model of
  elementary particles based on an analogy with superconductivity. II.}},\
  }\href {https://doi.org/10.1103/PhysRev.124.246} {\bibfield  {journal}
  {\bibinfo  {journal} {Phys. Rev.}\ }\textbf {\bibinfo {volume} {124}},\
  \bibinfo {pages} {246} (\bibinfo {year} {1961}{\natexlab{b}})}\BibitemShut
  {NoStop}%
\bibitem [{\citenamefont {Hatsuda}\ and\ \citenamefont
  {Kunihiro}(1994)}]{Hatsuda:1994pi}%
  \BibitemOpen
  \bibfield  {author} {\bibinfo {author} {\bibfnamefont {T.}~\bibnamefont
  {Hatsuda}}\ and\ \bibinfo {author} {\bibfnamefont {T.}~\bibnamefont
  {Kunihiro}},\ }\bibfield  {title} {\bibinfo {title} {{QCD phenomenology based
  on a chiral effective Lagrangian}},\ }\href
  {https://doi.org/10.1016/0370-1573(94)90022-1} {\bibfield  {journal}
  {\bibinfo  {journal} {Phys. Rept.}\ }\textbf {\bibinfo {volume} {247}},\
  \bibinfo {pages} {221} (\bibinfo {year} {1994})},\ \Eprint
  {https://arxiv.org/abs/hep-ph/9401310} {arXiv:hep-ph/9401310} \BibitemShut
  {NoStop}%
\bibitem [{\citenamefont {Weinberg}(1966)}]{Weinberg:1966kf}%
  \BibitemOpen
  \bibfield  {author} {\bibinfo {author} {\bibfnamefont {S.}~\bibnamefont
  {Weinberg}},\ }\bibfield  {title} {\bibinfo {title} {{Pion scattering
  lengths}},\ }\href {https://doi.org/10.1103/PhysRevLett.17.616} {\bibfield
  {journal} {\bibinfo  {journal} {Phys. Rev. Lett.}\ }\textbf {\bibinfo
  {volume} {17}},\ \bibinfo {pages} {616} (\bibinfo {year} {1966})}\BibitemShut
  {NoStop}%
\bibitem [{\citenamefont {Weinberg}(1969)}]{Weinberg:1969hw}%
  \BibitemOpen
  \bibfield  {author} {\bibinfo {author} {\bibfnamefont {S.}~\bibnamefont
  {Weinberg}},\ }\bibfield  {title} {\bibinfo {title} {{Algebraic realizations
  of chiral symmetry}},\ }\href {https://doi.org/10.1103/PhysRev.177.2604}
  {\bibfield  {journal} {\bibinfo  {journal} {Phys. Rev.}\ }\textbf {\bibinfo
  {volume} {177}},\ \bibinfo {pages} {2604} (\bibinfo {year}
  {1969})}\BibitemShut {NoStop}%
\bibitem [{\citenamefont {Weinberg}(1990)}]{Weinberg:1990xn}%
  \BibitemOpen
  \bibfield  {author} {\bibinfo {author} {\bibfnamefont {S.}~\bibnamefont
  {Weinberg}},\ }\bibfield  {title} {\bibinfo {title} {{Mended symmetries}},\
  }\href {https://doi.org/10.1103/PhysRevLett.65.1177} {\bibfield  {journal}
  {\bibinfo  {journal} {Phys. Rev. Lett.}\ }\textbf {\bibinfo {volume} {65}},\
  \bibinfo {pages} {1177} (\bibinfo {year} {1990})}\BibitemShut {NoStop}%
\bibitem [{\citenamefont {Weinberg}(2010)}]{Weinberg:2010bq}%
  \BibitemOpen
  \bibfield  {author} {\bibinfo {author} {\bibfnamefont {S.}~\bibnamefont
  {Weinberg}},\ }\bibfield  {title} {\bibinfo {title} {{Pions in Large-$N$
  Quantum Chromodynamics}},\ }\href
  {https://doi.org/10.1103/PhysRevLett.105.261601} {\bibfield  {journal}
  {\bibinfo  {journal} {Phys. Rev. Lett.}\ }\textbf {\bibinfo {volume} {105}},\
  \bibinfo {pages} {261601} (\bibinfo {year} {2010})},\ \Eprint
  {https://arxiv.org/abs/1009.1537} {arXiv:1009.1537 [hep-ph]} \BibitemShut
  {NoStop}%
\bibitem [{\citenamefont {Coleman}\ \emph {et~al.}(1969)\citenamefont
  {Coleman}, \citenamefont {Wess},\ and\ \citenamefont
  {Zumino}}]{Coleman:1969sm}%
  \BibitemOpen
  \bibfield  {author} {\bibinfo {author} {\bibfnamefont {S.~R.}\ \bibnamefont
  {Coleman}}, \bibinfo {author} {\bibfnamefont {J.}~\bibnamefont {Wess}},\ and\
  \bibinfo {author} {\bibfnamefont {B.}~\bibnamefont {Zumino}},\ }\bibfield
  {title} {\bibinfo {title} {{Structure of phenomenological Lagrangians. 1.}},\
  }\href {https://doi.org/10.1103/PhysRev.177.2239} {\bibfield  {journal}
  {\bibinfo  {journal} {Phys. Rev.}\ }\textbf {\bibinfo {volume} {177}},\
  \bibinfo {pages} {2239} (\bibinfo {year} {1969})}\BibitemShut {NoStop}%
\bibitem [{\citenamefont {Callan}\ \emph {et~al.}(1969)\citenamefont {Callan},
  \citenamefont {Coleman}, \citenamefont {Wess},\ and\ \citenamefont
  {Zumino}}]{Callan:1969sn}%
  \BibitemOpen
  \bibfield  {author} {\bibinfo {author} {\bibfnamefont {C.~G.}\ \bibnamefont
  {Callan}, \bibfnamefont {Jr.}}, \bibinfo {author} {\bibfnamefont {S.~R.}\
  \bibnamefont {Coleman}}, \bibinfo {author} {\bibfnamefont {J.}~\bibnamefont
  {Wess}},\ and\ \bibinfo {author} {\bibfnamefont {B.}~\bibnamefont {Zumino}},\
  }\bibfield  {title} {\bibinfo {title} {{Structure of phenomenological
  Lagrangians. 2.}},\ }\href {https://doi.org/10.1103/PhysRev.177.2247}
  {\bibfield  {journal} {\bibinfo  {journal} {Phys. Rev.}\ }\textbf {\bibinfo
  {volume} {177}},\ \bibinfo {pages} {2247} (\bibinfo {year}
  {1969})}\BibitemShut {NoStop}%
\bibitem [{\citenamefont {Manohar}\ and\ \citenamefont
  {Georgi}(1984)}]{Manohar:1983md}%
  \BibitemOpen
  \bibfield  {author} {\bibinfo {author} {\bibfnamefont {A.}~\bibnamefont
  {Manohar}}\ and\ \bibinfo {author} {\bibfnamefont {H.}~\bibnamefont
  {Georgi}},\ }\bibfield  {title} {\bibinfo {title} {{Chiral Quarks and the
  Nonrelativistic Quark Model}},\ }\href
  {https://doi.org/10.1016/0550-3213(84)90231-1} {\bibfield  {journal}
  {\bibinfo  {journal} {Nucl. Phys. B}\ }\textbf {\bibinfo {volume} {234}},\
  \bibinfo {pages} {189} (\bibinfo {year} {1984})}\BibitemShut {NoStop}%
\bibitem [{\citenamefont {Gasser}\ and\ \citenamefont
  {Leutwyler}(1984)}]{Gasser:1983yg}%
  \BibitemOpen
  \bibfield  {author} {\bibinfo {author} {\bibfnamefont {J.}~\bibnamefont
  {Gasser}}\ and\ \bibinfo {author} {\bibfnamefont {H.}~\bibnamefont
  {Leutwyler}},\ }\bibfield  {title} {\bibinfo {title} {{Chiral Perturbation
  Theory to One Loop}},\ }\href {https://doi.org/10.1016/0003-4916(84)90242-2}
  {\bibfield  {journal} {\bibinfo  {journal} {Annals Phys.}\ }\textbf {\bibinfo
  {volume} {158}},\ \bibinfo {pages} {142} (\bibinfo {year}
  {1984})}\BibitemShut {NoStop}%
\bibitem [{\citenamefont {Gasser}\ and\ \citenamefont
  {Leutwyler}(1985)}]{Gasser:1984gg}%
  \BibitemOpen
  \bibfield  {author} {\bibinfo {author} {\bibfnamefont {J.}~\bibnamefont
  {Gasser}}\ and\ \bibinfo {author} {\bibfnamefont {H.}~\bibnamefont
  {Leutwyler}},\ }\bibfield  {title} {\bibinfo {title} {{Chiral Perturbation
  Theory: Expansions in the Mass of the Strange Quark}},\ }\href
  {https://doi.org/10.1016/0550-3213(85)90492-4} {\bibfield  {journal}
  {\bibinfo  {journal} {Nucl. Phys. B}\ }\textbf {\bibinfo {volume} {250}},\
  \bibinfo {pages} {465} (\bibinfo {year} {1985})}\BibitemShut {NoStop}%
\bibitem [{\citenamefont {Baym}\ \emph {et~al.}(2018)\citenamefont {Baym},
  \citenamefont {Hatsuda}, \citenamefont {Kojo}, \citenamefont {Powell},
  \citenamefont {Song},\ and\ \citenamefont {Takatsuka}}]{Baym:2017whm}%
  \BibitemOpen
  \bibfield  {author} {\bibinfo {author} {\bibfnamefont {G.}~\bibnamefont
  {Baym}}, \bibinfo {author} {\bibfnamefont {T.}~\bibnamefont {Hatsuda}},
  \bibinfo {author} {\bibfnamefont {T.}~\bibnamefont {Kojo}}, \bibinfo {author}
  {\bibfnamefont {P.~D.}\ \bibnamefont {Powell}}, \bibinfo {author}
  {\bibfnamefont {Y.}~\bibnamefont {Song}},\ and\ \bibinfo {author}
  {\bibfnamefont {T.}~\bibnamefont {Takatsuka}},\ }\bibfield  {title} {\bibinfo
  {title} {{From hadrons to quarks in neutron stars: a review}},\ }\href
  {https://doi.org/10.1088/1361-6633/aaae14} {\bibfield  {journal} {\bibinfo
  {journal} {Rept. Prog. Phys.}\ }\textbf {\bibinfo {volume} {81}},\ \bibinfo
  {pages} {056902} (\bibinfo {year} {2018})},\ \Eprint
  {https://arxiv.org/abs/1707.04966} {arXiv:1707.04966 [astro-ph.HE]}
  \BibitemShut {NoStop}%
\bibitem [{\citenamefont {Detar}\ and\ \citenamefont
  {Kunihiro}(1989)}]{Detar:1988kn}%
  \BibitemOpen
  \bibfield  {author} {\bibinfo {author} {\bibfnamefont {C.~E.}\ \bibnamefont
  {Detar}}\ and\ \bibinfo {author} {\bibfnamefont {T.}~\bibnamefont
  {Kunihiro}},\ }\bibfield  {title} {\bibinfo {title} {{Linear $\sigma$ Model
  With Parity Doubling}},\ }\href {https://doi.org/10.1103/PhysRevD.39.2805}
  {\bibfield  {journal} {\bibinfo  {journal} {Phys. Rev. D}\ }\textbf {\bibinfo
  {volume} {39}},\ \bibinfo {pages} {2805} (\bibinfo {year}
  {1989})}\BibitemShut {NoStop}%
\bibitem [{\citenamefont {Jido}\ \emph
  {et~al.}(2000{\natexlab{a}})\citenamefont {Jido}, \citenamefont {Nemoto},
  \citenamefont {Oka},\ and\ \citenamefont {Hosaka}}]{Jido:1998av}%
  \BibitemOpen
  \bibfield  {author} {\bibinfo {author} {\bibfnamefont {D.}~\bibnamefont
  {Jido}}, \bibinfo {author} {\bibfnamefont {Y.}~\bibnamefont {Nemoto}},
  \bibinfo {author} {\bibfnamefont {M.}~\bibnamefont {Oka}},\ and\ \bibinfo
  {author} {\bibfnamefont {A.}~\bibnamefont {Hosaka}},\ }\bibfield  {title}
  {\bibinfo {title} {{Chiral symmetry for positive and negative parity
  nucleons}},\ }\href {https://doi.org/10.1016/S0375-9474(99)00844-1}
  {\bibfield  {journal} {\bibinfo  {journal} {Nucl. Phys. A}\ }\textbf
  {\bibinfo {volume} {671}},\ \bibinfo {pages} {471} (\bibinfo {year}
  {2000}{\natexlab{a}})},\ \Eprint {https://arxiv.org/abs/hep-ph/9805306}
  {arXiv:hep-ph/9805306} \BibitemShut {NoStop}%
\bibitem [{\citenamefont {Jido}\ \emph
  {et~al.}(2000{\natexlab{b}})\citenamefont {Jido}, \citenamefont {Hatsuda},\
  and\ \citenamefont {Kunihiro}}]{Jido:1999hd}%
  \BibitemOpen
  \bibfield  {author} {\bibinfo {author} {\bibfnamefont {D.}~\bibnamefont
  {Jido}}, \bibinfo {author} {\bibfnamefont {T.}~\bibnamefont {Hatsuda}},\ and\
  \bibinfo {author} {\bibfnamefont {T.}~\bibnamefont {Kunihiro}},\ }\bibfield
  {title} {\bibinfo {title} {{Chiral symmetry realization for even parity and
  odd parity baryon resonances}},\ }\href
  {https://doi.org/10.1103/PhysRevLett.84.3252} {\bibfield  {journal} {\bibinfo
   {journal} {Phys. Rev. Lett.}\ }\textbf {\bibinfo {volume} {84}},\ \bibinfo
  {pages} {3252} (\bibinfo {year} {2000}{\natexlab{b}})},\ \Eprint
  {https://arxiv.org/abs/hep-ph/9910375} {arXiv:hep-ph/9910375} \BibitemShut
  {NoStop}%
\bibitem [{\citenamefont {Jido}\ \emph {et~al.}(2001)\citenamefont {Jido},
  \citenamefont {Oka},\ and\ \citenamefont {Hosaka}}]{Jido:2001nt}%
  \BibitemOpen
  \bibfield  {author} {\bibinfo {author} {\bibfnamefont {D.}~\bibnamefont
  {Jido}}, \bibinfo {author} {\bibfnamefont {M.}~\bibnamefont {Oka}},\ and\
  \bibinfo {author} {\bibfnamefont {A.}~\bibnamefont {Hosaka}},\ }\bibfield
  {title} {\bibinfo {title} {{Chiral symmetry of baryons}},\ }\href
  {https://doi.org/10.1143/PTP.106.873} {\bibfield  {journal} {\bibinfo
  {journal} {Prog. Theor. Phys.}\ }\textbf {\bibinfo {volume} {106}},\ \bibinfo
  {pages} {873} (\bibinfo {year} {2001})},\ \Eprint
  {https://arxiv.org/abs/hep-ph/0110005} {arXiv:hep-ph/0110005} \BibitemShut
  {NoStop}%
\bibitem [{\citenamefont {Nagata}\ \emph {et~al.}(2008)\citenamefont {Nagata},
  \citenamefont {Hosaka},\ and\ \citenamefont {Dmitrasinovic}}]{Nagata:2008xf}%
  \BibitemOpen
  \bibfield  {author} {\bibinfo {author} {\bibfnamefont {K.}~\bibnamefont
  {Nagata}}, \bibinfo {author} {\bibfnamefont {A.}~\bibnamefont {Hosaka}},\
  and\ \bibinfo {author} {\bibfnamefont {V.}~\bibnamefont {Dmitrasinovic}},\
  }\bibfield  {title} {\bibinfo {title} {{pi N and pi pi N Couplings of the
  Delta(1232) and its Chiral Partners}},\ }\href
  {https://doi.org/10.1103/PhysRevLett.101.092001} {\bibfield  {journal}
  {\bibinfo  {journal} {Phys. Rev. Lett.}\ }\textbf {\bibinfo {volume} {101}},\
  \bibinfo {pages} {092001} (\bibinfo {year} {2008})},\ \Eprint
  {https://arxiv.org/abs/0804.3185} {arXiv:0804.3185 [hep-ph]} \BibitemShut
  {NoStop}%
\bibitem [{\citenamefont {Gallas}\ \emph {et~al.}(2010)\citenamefont {Gallas},
  \citenamefont {Giacosa},\ and\ \citenamefont {Rischke}}]{Gallas:2009qp}%
  \BibitemOpen
  \bibfield  {author} {\bibinfo {author} {\bibfnamefont {S.}~\bibnamefont
  {Gallas}}, \bibinfo {author} {\bibfnamefont {F.}~\bibnamefont {Giacosa}},\
  and\ \bibinfo {author} {\bibfnamefont {D.~H.}\ \bibnamefont {Rischke}},\
  }\bibfield  {title} {\bibinfo {title} {{Vacuum phenomenology of the chiral
  partner of the nucleon in a linear sigma model with vector mesons}},\ }\href
  {https://doi.org/10.1103/PhysRevD.82.014004} {\bibfield  {journal} {\bibinfo
  {journal} {Phys. Rev. D}\ }\textbf {\bibinfo {volume} {82}},\ \bibinfo
  {pages} {014004} (\bibinfo {year} {2010})},\ \Eprint
  {https://arxiv.org/abs/0907.5084} {arXiv:0907.5084 [hep-ph]} \BibitemShut
  {NoStop}%
\bibitem [{\citenamefont {Gallas}\ and\ \citenamefont
  {Giacosa}(2014)}]{Gallas:2013ipa}%
  \BibitemOpen
  \bibfield  {author} {\bibinfo {author} {\bibfnamefont {S.}~\bibnamefont
  {Gallas}}\ and\ \bibinfo {author} {\bibfnamefont {F.}~\bibnamefont
  {Giacosa}},\ }\bibfield  {title} {\bibinfo {title} {{Mirror versus naive
  assignment in chiral models for the nucleon}},\ }\href
  {https://doi.org/10.1142/S0217751X14500985} {\bibfield  {journal} {\bibinfo
  {journal} {Int. J. Mod. Phys. A}\ }\textbf {\bibinfo {volume} {29}},\
  \bibinfo {pages} {1450098} (\bibinfo {year} {2014})},\ \Eprint
  {https://arxiv.org/abs/1308.4817} {arXiv:1308.4817 [hep-ph]} \BibitemShut
  {NoStop}%
\bibitem [{\citenamefont {Minamikawa}\ \emph
  {et~al.}(2021{\natexlab{a}})\citenamefont {Minamikawa}, \citenamefont
  {Kojo},\ and\ \citenamefont {Harada}}]{Minamikawa:2020jfj}%
  \BibitemOpen
  \bibfield  {author} {\bibinfo {author} {\bibfnamefont {T.}~\bibnamefont
  {Minamikawa}}, \bibinfo {author} {\bibfnamefont {T.}~\bibnamefont {Kojo}},\
  and\ \bibinfo {author} {\bibfnamefont {M.}~\bibnamefont {Harada}},\
  }\bibfield  {title} {\bibinfo {title} {{Quark-hadron crossover equations of
  state for neutron stars: constraining the chiral invariant mass in a parity
  doublet model}},\ }\href {https://doi.org/10.1103/PhysRevC.103.045205}
  {\bibfield  {journal} {\bibinfo  {journal} {Phys. Rev. C}\ }\textbf {\bibinfo
  {volume} {103}},\ \bibinfo {pages} {045205} (\bibinfo {year}
  {2021}{\natexlab{a}})},\ \Eprint {https://arxiv.org/abs/2011.13684}
  {arXiv:2011.13684 [nucl-th]} \BibitemShut {NoStop}%
\bibitem [{\citenamefont {Minamikawa}\ \emph
  {et~al.}(2021{\natexlab{b}})\citenamefont {Minamikawa}, \citenamefont
  {Kojo},\ and\ \citenamefont {Harada}}]{Minamikawa:2021fln}%
  \BibitemOpen
  \bibfield  {author} {\bibinfo {author} {\bibfnamefont {T.}~\bibnamefont
  {Minamikawa}}, \bibinfo {author} {\bibfnamefont {T.}~\bibnamefont {Kojo}},\
  and\ \bibinfo {author} {\bibfnamefont {M.}~\bibnamefont {Harada}},\
  }\bibfield  {title} {\bibinfo {title} {{Chiral condensates for neutron stars
  in hadron-quark crossover: From a parity doublet nucleon model to a
  Nambu\textendash{}Jona-Lasinio quark model}},\ }\href
  {https://doi.org/10.1103/PhysRevC.104.065201} {\bibfield  {journal} {\bibinfo
   {journal} {Phys. Rev. C}\ }\textbf {\bibinfo {volume} {104}},\ \bibinfo
  {pages} {065201} (\bibinfo {year} {2021}{\natexlab{b}})},\ \Eprint
  {https://arxiv.org/abs/2107.14545} {arXiv:2107.14545 [nucl-th]} \BibitemShut
  {NoStop}%
\bibitem [{\citenamefont {Gao}\ \emph {et~al.}(2022)\citenamefont {Gao},
  \citenamefont {Minamikawa}, \citenamefont {Kojo},\ and\ \citenamefont
  {Harada}}]{Gao:2022klm}%
  \BibitemOpen
  \bibfield  {author} {\bibinfo {author} {\bibfnamefont {B.}~\bibnamefont
  {Gao}}, \bibinfo {author} {\bibfnamefont {T.}~\bibnamefont {Minamikawa}},
  \bibinfo {author} {\bibfnamefont {T.}~\bibnamefont {Kojo}},\ and\ \bibinfo
  {author} {\bibfnamefont {M.}~\bibnamefont {Harada}},\ }\bibfield  {title}
  {\bibinfo {title} {{Impacts of the U(1)A anomaly on nuclear and neutron star
  equation~of state based on a parity doublet model}},\ }\href
  {https://doi.org/10.1103/PhysRevC.106.065205} {\bibfield  {journal} {\bibinfo
   {journal} {Phys. Rev. C}\ }\textbf {\bibinfo {volume} {106}},\ \bibinfo
  {pages} {065205} (\bibinfo {year} {2022})},\ \Eprint
  {https://arxiv.org/abs/2207.05970} {arXiv:2207.05970 [nucl-th]} \BibitemShut
  {NoStop}%
\bibitem [{\citenamefont {Minamikawa}\ \emph {et~al.}(2023)\citenamefont
  {Minamikawa}, \citenamefont {Gao}, \citenamefont {Kojo},\ and\ \citenamefont
  {Harada}}]{Minamikawa:2023eky}%
  \BibitemOpen
  \bibfield  {author} {\bibinfo {author} {\bibfnamefont {T.}~\bibnamefont
  {Minamikawa}}, \bibinfo {author} {\bibfnamefont {B.}~\bibnamefont {Gao}},
  \bibinfo {author} {\bibfnamefont {T.}~\bibnamefont {Kojo}},\ and\ \bibinfo
  {author} {\bibfnamefont {M.}~\bibnamefont {Harada}},\ }\bibfield  {title}
  {\bibinfo {title} {{Chiral Restoration of Nucleons in Neutron Star Matter:
  Studies Based on a Parity Doublet Model}},\ }\href
  {https://doi.org/10.3390/sym15030745} {\bibfield  {journal} {\bibinfo
  {journal} {Symmetry}\ }\textbf {\bibinfo {volume} {15}},\ \bibinfo {pages}
  {745} (\bibinfo {year} {2023})},\ \Eprint {https://arxiv.org/abs/2302.00825}
  {arXiv:2302.00825 [nucl-th]} \BibitemShut {NoStop}%
\bibitem [{\citenamefont {Chen}\ \emph {et~al.}(2010)\citenamefont {Chen},
  \citenamefont {Dmitrasinovic},\ and\ \citenamefont {Hosaka}}]{Chen:2009sf}%
  \BibitemOpen
  \bibfield  {author} {\bibinfo {author} {\bibfnamefont {H.-X.}\ \bibnamefont
  {Chen}}, \bibinfo {author} {\bibfnamefont {V.}~\bibnamefont
  {Dmitrasinovic}},\ and\ \bibinfo {author} {\bibfnamefont {A.}~\bibnamefont
  {Hosaka}},\ }\bibfield  {title} {\bibinfo {title} {{Baryon fields with
  U(L)(3) X U(R)(3) chiral symmetry II: Axial currents of nucleons and
  hyperons}},\ }\href {https://doi.org/10.1103/PhysRevD.81.054002} {\bibfield
  {journal} {\bibinfo  {journal} {Phys. Rev. D}\ }\textbf {\bibinfo {volume}
  {81}},\ \bibinfo {pages} {054002} (\bibinfo {year} {2010})},\ \Eprint
  {https://arxiv.org/abs/0912.4338} {arXiv:0912.4338 [hep-ph]} \BibitemShut
  {NoStop}%
\bibitem [{\citenamefont {Chen}\ \emph {et~al.}(2011)\citenamefont {Chen},
  \citenamefont {Dmitrasinovic},\ and\ \citenamefont {Hosaka}}]{Chen:2010ba}%
  \BibitemOpen
  \bibfield  {author} {\bibinfo {author} {\bibfnamefont {H.-X.}\ \bibnamefont
  {Chen}}, \bibinfo {author} {\bibfnamefont {V.}~\bibnamefont
  {Dmitrasinovic}},\ and\ \bibinfo {author} {\bibfnamefont {A.}~\bibnamefont
  {Hosaka}},\ }\bibfield  {title} {\bibinfo {title} {{Baryon Fields with
  $U_L(3) times U_R(3)$ Chiral Symmetry III: Interactions with Chiral
  $(3,\bar{3})+ (\bar{3},3)$ Spinless Mesons}},\ }\href
  {https://doi.org/10.1103/PhysRevD.83.014015} {\bibfield  {journal} {\bibinfo
  {journal} {Phys. Rev. D}\ }\textbf {\bibinfo {volume} {83}},\ \bibinfo
  {pages} {014015} (\bibinfo {year} {2011})},\ \Eprint
  {https://arxiv.org/abs/1009.2422} {arXiv:1009.2422 [hep-ph]} \BibitemShut
  {NoStop}%
\bibitem [{\citenamefont {Chen}\ \emph {et~al.}(2012)\citenamefont {Chen},
  \citenamefont {Dmitrasinovic},\ and\ \citenamefont {Hosaka}}]{Chen:2011rh}%
  \BibitemOpen
  \bibfield  {author} {\bibinfo {author} {\bibfnamefont {H.-X.}\ \bibnamefont
  {Chen}}, \bibinfo {author} {\bibfnamefont {V.}~\bibnamefont
  {Dmitrasinovic}},\ and\ \bibinfo {author} {\bibfnamefont {A.}~\bibnamefont
  {Hosaka}},\ }\bibfield  {title} {\bibinfo {title} {{$mathrm{Baryons with}$
  $U_L(3) \times U_R(3)$ Chiral Symmetry IV: Interactions with Chiral (8,1)
  $\oplus$ (1,8) Vector and Axial-vector Mesons and Anomalous Magnetic
  Moments}},\ }\href {https://doi.org/10.1103/PhysRevC.85.055205} {\bibfield
  {journal} {\bibinfo  {journal} {Phys. Rev. C}\ }\textbf {\bibinfo {volume}
  {85}},\ \bibinfo {pages} {055205} (\bibinfo {year} {2012})},\ \Eprint
  {https://arxiv.org/abs/1109.3130} {arXiv:1109.3130 [hep-ph]} \BibitemShut
  {NoStop}%
\bibitem [{\citenamefont {Schramm}\ \emph {et~al.}(2016)\citenamefont
  {Schramm}, \citenamefont {Dexheimer},\ and\ \citenamefont
  {Negreiros}}]{Schramm:2015hba}%
  \BibitemOpen
  \bibfield  {author} {\bibinfo {author} {\bibfnamefont {S.}~\bibnamefont
  {Schramm}}, \bibinfo {author} {\bibfnamefont {V.}~\bibnamefont {Dexheimer}},\
  and\ \bibinfo {author} {\bibfnamefont {R.}~\bibnamefont {Negreiros}},\
  }\bibfield  {title} {\bibinfo {title} {{Modelling Hybrid Stars in
  Quark-Hadron Approaches}},\ }\href
  {https://doi.org/10.1140/epja/i2016-16014-5} {\bibfield  {journal} {\bibinfo
  {journal} {Eur. Phys. J. A}\ }\textbf {\bibinfo {volume} {52}},\ \bibinfo
  {pages} {14} (\bibinfo {year} {2016})},\ \Eprint
  {https://arxiv.org/abs/1508.04699} {arXiv:1508.04699 [nucl-th]} \BibitemShut
  {NoStop}%
\bibitem [{\citenamefont {Nishihara}\ and\ \citenamefont
  {Harada}(2015)}]{Nishihara:2015fka}%
  \BibitemOpen
  \bibfield  {author} {\bibinfo {author} {\bibfnamefont {H.}~\bibnamefont
  {Nishihara}}\ and\ \bibinfo {author} {\bibfnamefont {M.}~\bibnamefont
  {Harada}},\ }\bibfield  {title} {\bibinfo {title} {{Extended
  Goldberger-Treiman relation in a three-flavor parity doublet model}},\ }\href
  {https://doi.org/10.1103/PhysRevD.92.054022} {\bibfield  {journal} {\bibinfo
  {journal} {Phys. Rev. D}\ }\textbf {\bibinfo {volume} {92}},\ \bibinfo
  {pages} {054022} (\bibinfo {year} {2015})},\ \Eprint
  {https://arxiv.org/abs/1506.07956} {arXiv:1506.07956 [hep-ph]} \BibitemShut
  {NoStop}%
\bibitem [{\citenamefont {Dmitra\v{s}inovi\'c}\ \emph
  {et~al.}(2016)\citenamefont {Dmitra\v{s}inovi\'c}, \citenamefont {Chen},\
  and\ \citenamefont {Hosaka}}]{Dmitrasinovic:2016hup}%
  \BibitemOpen
  \bibfield  {author} {\bibinfo {author} {\bibfnamefont {V.}~\bibnamefont
  {Dmitra\v{s}inovi\'c}}, \bibinfo {author} {\bibfnamefont {H.-X.}\
  \bibnamefont {Chen}},\ and\ \bibinfo {author} {\bibfnamefont
  {A.}~\bibnamefont {Hosaka}},\ }\bibfield  {title} {\bibinfo {title} {{Baryon
  fields with UL(3)\texttimes{}UR(3) chiral symmetry. V. Pion-nucleon and
  kaon-nucleon \ensuremath{\Sigma} terms}},\ }\href
  {https://doi.org/10.1103/PhysRevC.93.065208} {\bibfield  {journal} {\bibinfo
  {journal} {Phys. Rev. C}\ }\textbf {\bibinfo {volume} {93}},\ \bibinfo
  {pages} {065208} (\bibinfo {year} {2016})},\ \Eprint
  {https://arxiv.org/abs/1812.03414} {arXiv:1812.03414 [hep-ph]} \BibitemShut
  {NoStop}%
\bibitem [{\citenamefont {Sasaki}(2018)}]{Sasaki:2017glk}%
  \BibitemOpen
  \bibfield  {author} {\bibinfo {author} {\bibfnamefont {C.}~\bibnamefont
  {Sasaki}},\ }\bibfield  {title} {\bibinfo {title} {{Parity doubling of
  baryons in a chiral approach with three flavors}},\ }\href
  {https://doi.org/10.1016/j.nuclphysa.2018.01.004} {\bibfield  {journal}
  {\bibinfo  {journal} {Nucl. Phys. A}\ }\textbf {\bibinfo {volume} {970}},\
  \bibinfo {pages} {388} (\bibinfo {year} {2018})},\ \Eprint
  {https://arxiv.org/abs/1707.05081} {arXiv:1707.05081 [hep-ph]} \BibitemShut
  {NoStop}%
\bibitem [{\citenamefont {Shivam}\ and\ \citenamefont
  {Kumar}(2019)}]{Shivam:2019cmw}%
  \BibitemOpen
  \bibfield  {author} {\bibinfo {author} {\bibnamefont {Shivam}}\ and\ \bibinfo
  {author} {\bibfnamefont {A.}~\bibnamefont {Kumar}},\ }\bibfield  {title}
  {\bibinfo {title} {{Possibility of $ \rho$ meson condensation in neutron
  stars: Unified approach of chiral SU(3) model and QCD sum rules}},\ }\href
  {https://doi.org/10.1140/epjp/i2019-12945-x} {\bibfield  {journal} {\bibinfo
  {journal} {Eur. Phys. J. Plus}\ }\textbf {\bibinfo {volume} {134}},\ \bibinfo
  {pages} {592} (\bibinfo {year} {2019})},\ \Eprint
  {https://arxiv.org/abs/1905.13184} {arXiv:1905.13184 [nucl-th]} \BibitemShut
  {NoStop}%
\bibitem [{\citenamefont {Motornenko}\ \emph {et~al.}(2021)\citenamefont
  {Motornenko}, \citenamefont {Steinheimer}, \citenamefont {Vovchenko},
  \citenamefont {Schramm},\ and\ \citenamefont
  {Stoecker}}]{Motornenko:2020vqm}%
  \BibitemOpen
  \bibfield  {author} {\bibinfo {author} {\bibfnamefont {A.}~\bibnamefont
  {Motornenko}}, \bibinfo {author} {\bibfnamefont {J.}~\bibnamefont
  {Steinheimer}}, \bibinfo {author} {\bibfnamefont {V.}~\bibnamefont
  {Vovchenko}}, \bibinfo {author} {\bibfnamefont {S.}~\bibnamefont {Schramm}},\
  and\ \bibinfo {author} {\bibfnamefont {H.}~\bibnamefont {Stoecker}},\
  }\bibfield  {title} {\bibinfo {title} {{QCD equation of state at vanishing
  and high baryon density: Chiral Mean Field model}},\ }\href
  {https://doi.org/10.1016/j.nuclphysa.2020.121836} {\bibfield  {journal}
  {\bibinfo  {journal} {Nucl. Phys. A}\ }\textbf {\bibinfo {volume} {1005}},\
  \bibinfo {pages} {121836} (\bibinfo {year} {2021})},\ \Eprint
  {https://arxiv.org/abs/2002.01217} {arXiv:2002.01217 [hep-ph]} \BibitemShut
  {NoStop}%
\bibitem [{\citenamefont {Jaffe}(1977{\natexlab{a}})}]{Jaffe:1976ig}%
  \BibitemOpen
  \bibfield  {author} {\bibinfo {author} {\bibfnamefont {R.~L.}\ \bibnamefont
  {Jaffe}},\ }\bibfield  {title} {\bibinfo {title} {{Multi-Quark Hadrons. 1.
  The Phenomenology of (2 Quark 2 anti-Quark) Mesons}},\ }\href
  {https://doi.org/10.1103/PhysRevD.15.267} {\bibfield  {journal} {\bibinfo
  {journal} {Phys. Rev. D}\ }\textbf {\bibinfo {volume} {15}},\ \bibinfo
  {pages} {267} (\bibinfo {year} {1977}{\natexlab{a}})}\BibitemShut {NoStop}%
\bibitem [{\citenamefont {Jaffe}(1977{\natexlab{b}})}]{Jaffe:1976ih}%
  \BibitemOpen
  \bibfield  {author} {\bibinfo {author} {\bibfnamefont {R.~L.}\ \bibnamefont
  {Jaffe}},\ }\bibfield  {title} {\bibinfo {title} {{Multi-Quark Hadrons. 2.
  Methods}},\ }\href {https://doi.org/10.1103/PhysRevD.15.281} {\bibfield
  {journal} {\bibinfo  {journal} {Phys. Rev. D}\ }\textbf {\bibinfo {volume}
  {15}},\ \bibinfo {pages} {281} (\bibinfo {year}
  {1977}{\natexlab{b}})}\BibitemShut {NoStop}%
\bibitem [{\citenamefont {Rapp}\ \emph {et~al.}(1998)\citenamefont {Rapp},
  \citenamefont {Sch\"afer}, \citenamefont {Shuryak},\ and\ \citenamefont
  {Velkovsky}}]{Rapp:1997zu}%
  \BibitemOpen
  \bibfield  {author} {\bibinfo {author} {\bibfnamefont {R.}~\bibnamefont
  {Rapp}}, \bibinfo {author} {\bibfnamefont {T.}~\bibnamefont {Sch\"afer}},
  \bibinfo {author} {\bibfnamefont {E.~V.}\ \bibnamefont {Shuryak}},\ and\
  \bibinfo {author} {\bibfnamefont {M.}~\bibnamefont {Velkovsky}},\ }\bibfield
  {title} {\bibinfo {title} {{Diquark Bose condensates in high density matter
  and instantons}},\ }\href {https://doi.org/10.1103/PhysRevLett.81.53}
  {\bibfield  {journal} {\bibinfo  {journal} {Phys. Rev. Lett.}\ }\textbf
  {\bibinfo {volume} {81}},\ \bibinfo {pages} {53} (\bibinfo {year} {1998})},\
  \Eprint {https://arxiv.org/abs/hep-ph/9711396} {arXiv:hep-ph/9711396}
  \BibitemShut {NoStop}%
\bibitem [{\citenamefont {Jaffe}\ and\ \citenamefont
  {Wilczek}(2003)}]{Jaffe:2003sg}%
  \BibitemOpen
  \bibfield  {author} {\bibinfo {author} {\bibfnamefont {R.~L.}\ \bibnamefont
  {Jaffe}}\ and\ \bibinfo {author} {\bibfnamefont {F.}~\bibnamefont
  {Wilczek}},\ }\bibfield  {title} {\bibinfo {title} {{Diquarks and exotic
  spectroscopy}},\ }\href {https://doi.org/10.1103/PhysRevLett.91.232003}
  {\bibfield  {journal} {\bibinfo  {journal} {Phys. Rev. Lett.}\ }\textbf
  {\bibinfo {volume} {91}},\ \bibinfo {pages} {232003} (\bibinfo {year}
  {2003})},\ \Eprint {https://arxiv.org/abs/hep-ph/0307341}
  {arXiv:hep-ph/0307341} \BibitemShut {NoStop}%
\bibitem [{\citenamefont {De~Rujula}\ \emph {et~al.}(1975)\citenamefont
  {De~Rujula}, \citenamefont {Georgi},\ and\ \citenamefont
  {Glashow}}]{DeRujula:1975qlm}%
  \BibitemOpen
  \bibfield  {author} {\bibinfo {author} {\bibfnamefont {A.}~\bibnamefont
  {De~Rujula}}, \bibinfo {author} {\bibfnamefont {H.}~\bibnamefont {Georgi}},\
  and\ \bibinfo {author} {\bibfnamefont {S.~L.}\ \bibnamefont {Glashow}},\
  }\bibfield  {title} {\bibinfo {title} {{Hadron Masses in a Gauge Theory}},\
  }\href {https://doi.org/10.1103/PhysRevD.12.147} {\bibfield  {journal}
  {\bibinfo  {journal} {Phys. Rev. D}\ }\textbf {\bibinfo {volume} {12}},\
  \bibinfo {pages} {147} (\bibinfo {year} {1975})}\BibitemShut {NoStop}%
\bibitem [{\citenamefont {Georgi}(1982)}]{georgi1982lie}%
  \BibitemOpen
  \bibfield  {author} {\bibinfo {author} {\bibfnamefont {H.}~\bibnamefont
  {Georgi}},\ }\href {https://books.google.co.jp/books?id=5ex644CLpZcC} {\emph
  {\bibinfo {title} {Lie Algebras In Particle Physics: From Isospin To Unified
  Theories}}},\ Advanced Book Program\ (\bibinfo  {publisher} {Basic Books},\
  \bibinfo {year} {1982})\BibitemShut {NoStop}%
\bibitem [{\citenamefont {Workman}\ \emph {et~al.}(2022)\citenamefont {Workman}
  \emph {et~al.}}]{Workman:2022ynf}%
  \BibitemOpen
  \bibfield  {author} {\bibinfo {author} {\bibfnamefont {R.~L.}\ \bibnamefont
  {Workman}} \emph {et~al.} (\bibinfo {collaboration} {Particle Data Group}),\
  }\bibfield  {title} {\bibinfo {title} {{Review of Particle Physics}},\ }\href
  {https://doi.org/10.1093/ptep/ptac097} {\bibfield  {journal} {\bibinfo
  {journal} {PTEP}\ }\textbf {\bibinfo {volume} {2022}},\ \bibinfo {pages}
  {083C01} (\bibinfo {year} {2022})}\BibitemShut {NoStop}%
\end{thebibliography}%

\end{document}